\documentclass[12pt]{article}
\usepackage{times}
\usepackage{geometry}
\usepackage[utf8]{inputenc}
\usepackage{enumitem,amssymb}
\usepackage{ragged2e}
\newlist{thematic}{itemize}{8}
\setlist[thematic]{label=$\square$}
\usepackage{pifont}

\usepackage[compress, numbers]{natbib}
\usepackage{aas_macros}
\usepackage[hidelinks]{hyperref}
\usepackage{graphicx,color,rotating}
\usepackage[svgnames]{xcolor}
\usepackage[labelfont={color=RoyalBlue,bf}]{caption}
\usepackage{lineno,amssymb}
\usepackage[T1]{fontenc}

\begin{document}
\thispagestyle{empty}
{\RaggedRight
\huge
Astro2020 Science White Paper \linebreak

Joint Gravitational Wave and Electromagnetic Astronomy with LIGO and LSST in the 2020's \linebreak
\normalsize

\noindent \textbf{Thematic Areas:} \hspace*{60pt} $\square$ Planetary Systems \hspace*{10pt} $\square$ Star and Planet Formation \hspace*{20pt}\linebreak
\makebox[0pt][l]{$\square$}\raisebox{.15ex}{\hspace{0.1em}$\checkmark$} Formation and Evolution of Compact Objects \hspace*{29pt} \makebox[0pt][l]{$\square$}\raisebox{.15ex}{\hspace{0.1em}$\checkmark$} Cosmology and Fundamental Physics \linebreak
  $\square$  Stars and Stellar Evolution \hspace*{5pt} $\square$ Resolved Stellar Populations and their Environments \hspace*{40pt} \linebreak
  $\square$    Galaxy Evolution   \hspace*{49pt} \makebox[0pt][l]{$\square$}\raisebox{.15ex}{\hspace{0.1em}$\checkmark$}             Multi-Messenger Astronomy and Astrophysics \hspace*{65pt} \linebreak
  
\textbf{Principal Author:}

Name: Philip S. Cowperthwaite
 \linebreak						
Institution:  Carnegie Observatories
 \linebreak
Email: pcowperthwaite@carnegiescience.edu
 \linebreak
Phone: 1-626-304-0205
 \linebreak
 
\textbf{Co-authors:} Hsin-Yu Chen (Harvard/BHI), Ben Margalit (Columbia), Raffaella Margutti (Northwestern), Morgan May (Columbia), Brian Metzger (Columbia), and Chris Pankow (Northwestern).
  \linebreak
  
\textbf{Abstract  (optional):} The blossoming field of joint gravitational wave and electromagnetic (GW-EM) astronomy is one of the most promising in astronomy. The first, and only, joint GW-EM event GW170817 provided remarkable science returns that still continue to this day. Continued growth in this field requires increasing the sample size of joint GW-EM detections. In this white paper, we outline the case for using some percentage of LSST survey time for dedicated target-of-opportunity follow up of GW triggers in order to efficiently and rapidly identify optical counterparts. We show that the timeline for the LSST science survey is well matched to the planned improvements to ground based GW detectors in the next decade. LSST will become particularly crucial in the later half of the 2020s as more and more distant GW sources are detected. Lastly, we highlight some of the key science goals that can be addressed by a large sample of joint GW-EM detections.

}
\clearpage
\setcounter{page}{1}

\begin{figure}
\centering
\begin{minipage}[c]{0.70\textwidth}
{\includegraphics[width=\textwidth]{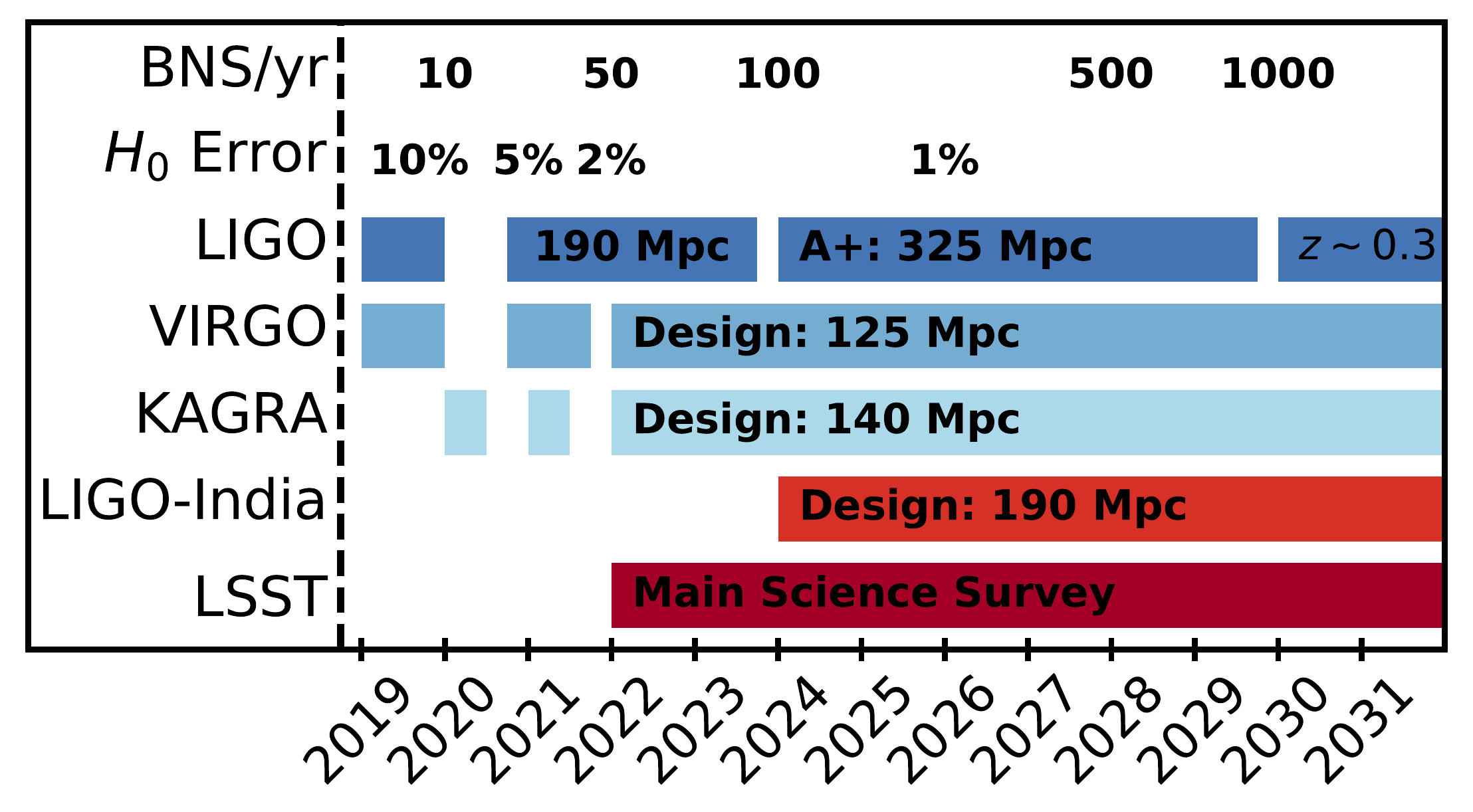}}
\end{minipage}\hfill
\begin{minipage}[c]{0.28\textwidth}
\caption{Example timeline showing the favorable overlap between the LSST science survey and increased GW detector sensitivity. The average BNS detection horizon is given for each GW facility. Example improvements in key joint GW-EM science are shown for reference.}
\label{fig:timeline}
\end{minipage}
\vspace{-0.15in}
\end{figure}

\section{Introduction}{\label{sec:intro}}
The direct detection of gravitational waves (GW) from astrophysical sources has ushered in an exciting new era of astronomy. The GW data from sources such as the inspiral and merger of compact object binaries comprised of black holes (BH) and/or neutron stars (NS) provides unparalleled insight into the component masses and spins, binary parameters (e.g., inclination, eccentricity), and the strong gravity regime of general relativity. However, the true power of GW detections become apparent when combined with traditional electromagnetic (EM) observations. The discovery of an EM counterpart provides improved localization and distance information, host galaxy identification and demographics, breaks modeling degeneracies (e.g., distance vs. inclination), and allows modeling of the merger hydrodynamics.

The first such joint detection, GW170817~\citep{2017PhRvL.119p1101A}, was an exceptional event. It remains the loudest signal detected by GW instruments from the first two observing runs. Its proximity (40 Mpc) allowed not only rapid counterpart identification and a bevy of electromagnetic observations~\citep{2017ApJ...848L..12A}, but also a number of joint GW-EM studies. The GW data provided the first measurements of NS tidal deformation and subsequent equation of state constraints~\citep{2018PhRvL.121p1101A}, constraints on non-linear tides in NS~\citep{2019PhRvL.122f1104A}, constraints on the formation of the progenitor system~\citep{2017ApJ...850L..40A}, and new tests of general relativity~\citep{2018arXiv181100364T}. Joint GW-EM studies provided an independent measurement of the Hubble constant~\citep{2017Natur.551...85A}, 
placed complementary constraints on the NS equation of state \citep{Margalit&Metzger17,Bauswein+17,Radice+18,Coughlin+18},
the fate of the post-merger object~\citep{2018ApJ...856..101M}, and prompted reanalysis and questions about the distribution and formation of binary neutron stars in the galaxy~\citep{2018ApJ...866...60P,2019arXiv190203300F}.

Building upon the tremendous success of GW170817 requires the continued support for on-going improvements to both the network of GW detectors and next-generation EM facilities. As the network of ground-based GW detectors continues to grow more sensitive (\autoref{sec:GWDet}) wide-field optical telescopes such as the Large Synoptic Survey Telescope \citep[LSST\footnote{\url{www.lsst.org}};][]{2008arXiv0805.2366I} will be crucial for detecting distant EM counterparts. In this white paper we discuss the landscape of joint GW-EM science in the next decade with a focus on target of opportunity follow-up with LSST to identify a large sample of EM counterparts (\autoref{sec:LSST}), as well as briefly introducing some of the major science questions such a sample can help address (\autoref{sec:examples}).

\section{Gravitational-Wave Detectors in the Next Decade}{\label{sec:GWDet}}
Gravitational-wave networks are expected to both expand in number and increase in sensitivity over the next 5--10 years~\citep{2018LRR....21....3A}. The first two observing runs included both of the LIGO interferometers~\citep{2015CQGra..32g4001L,2016PhRvL.116m1103A} as well as the French-Italian Virgo~\citep{2012JInst...7.3012A} interferometer. The third observing run will likely see the inclusion of a fourth site in Japan \citep[KAGRA;][]{2019NatAs...3...35K}, and a fifth site in India \citep[LIGO-India;][]{ligoindia} is in the planning and construction stages with a 2024 operations date. By the middle of the decade, a five site, world-wide network will be operational, with a binary neutron star (BNS) merger detection horizon of approximately 400 Mpc. Current observational estimates of merger event rates for these systems range over 100 -- 4000 Gpc$^{-3}$ yr${-1}$~\citep{2015MNRAS.448..928K,2019ApJ...870...71P,2018arXiv181112907T}. Taking the median of current estimates and the detection range quoted above, a year of GW network operations would yield a few tens of events. As sensitivity improves, BNS localizations are expected to shrink by at least an order of magnitude; from hundreds to tens of square degrees on the sky~\citep{2016LRR....19....1A,2016ApJ...825..116F,pankow2018,2018CQGra..35j5002F}.

Plans for going beyond the design sensitivity of second generation instruments include the A+ upgrade~\citep{LIGOT1800042} for the LIGO instruments, as well as proposed third generation instruments such as LIGO Voyager and the Einstein telescope~\citep{2010CQGra..27s4002P,2017CQGra..34d4001A}. The surveyed volumes for these instruments are at the Gpc${}^3$ scale, and will allow for hundreds to thousands of BNS detections. This sample will provide exquisitely precise measurements of the GW emission across the entire accessible bandwidth as well as commensurate contributions to our understanding of a variety of fundamental open questions~\citep{2012CQGra..29l4013S,2014JPhCS.484a2008V}. 


\section{Targeted Follow-Up of GW Events with LSST}{\label{sec:LSST}}

While GW170817 was detected with relative ease due to it's small sky localization ($\sim28$~deg$^2$) and luminosity distance ($D_L\sim40$~Mpc), future detections may not be so straight-forward. This is particularly true as the sensitivity horizon of GW detectors pushes out to hundreds of Mpc or Gpc scales in the case of NS-BH mergers. In this regime, LSST is the instrument of choice due to its unmatched combination of wide field of view (9.6 deg$^2$), depth and speed ($g\sim25$~mag in 30s). Thus, LSST will be able to efficiently observe GW localization regions \citep[20--80 deg$^2$;][]{hsinyu2017,pankow2018} and rapidly identify EM counterparts. The operations timescale for LSST is well matched to this goal with full science observations expected to begin in 2022 and continuing for 10 years. 

The present LSST operations plan sets aside 5$\%$ of survey time for mini-surveys that require a special cadence. It is imperative that some fraction of this allocation go to target-of-opportunity observations in response to GW events. Such triggered observations are essential to ensure a large sample of joint GW-EM detections as the primary survey cadence is insufficient for the efficient detection of rapidly evolving transients \citep[e.g.,][]{Cowp+18LSST,Scolnic+18,MarguttiTOO}. In addition to growing the sample of joint GW-EM detections, targeted observations with LSST have additional benefits such as the potential for detecting the optical counterpart at very early times. Such observations can provide key insights into the nature of the early-time emission \citep[e.g.,][]{Metzger+15,Piro2018}. This is particularly interesting in cases where strong BNS signals provide an ``early-warning''~\citep{2012ApJ...748..136C,2018PhRvD..97l3014C} potentially allowing observations to begin before the final merger occurs~\citep{2018ApJ...855L..23A}. One challenge facing these deep follow-up observations is the presence of a background of transients unrelated to the GW event, particularly as observations increase in depth. However, these effects can be mitigated thanks to the unique temporal and spectral behavior of kilonovae \citep{CowpBerger15,Cowp+18LSST,Cowp+18Cont}.



\section{Examples of Joint GW-EM Science}{\label{sec:examples}}

The field of joint GW-EM astronomy is rich with opportunity to study the Universe in new and exciting ways previously inaccessible to studies using just traditional EM observations. Here we briefly discuss just a fraction of these exciting possibilities:

\subsection{The Growth of Joint GW-EM Cosmology}{\label{sec:cosmology}}

The tension between local measurements of H$_0$ and CMB measurements by Planck is now at 3.8 $\sigma$ (99.9$\%$) \citep{riess2018} due to an improvement in the distance ladder using Gaia parallax and HST photometry. In light of this increasing tension, an independent measurement of $H_0$ is desirable as a potential resolution. Such an independent measurement is made possible by joint GW-EM science whereby GW signals provide a direct measurement of the luminosity distance of the GW sources and the EM counterpart provide a unique determination of the host galaxy and thus the redshift of the source. Combining these two pieces of information, one can measure cosmological parameters, including $H_0$~\citep{1986Natur.323..310S,2005ApJ...629...15H}. The value of $H_0$ measured in this fashion is completely independent of other cosmological experiments, with its own unique systematics. 

\begin{figure}
\centering
\raisebox{0.06\height}{\includegraphics[width=0.49\textwidth]{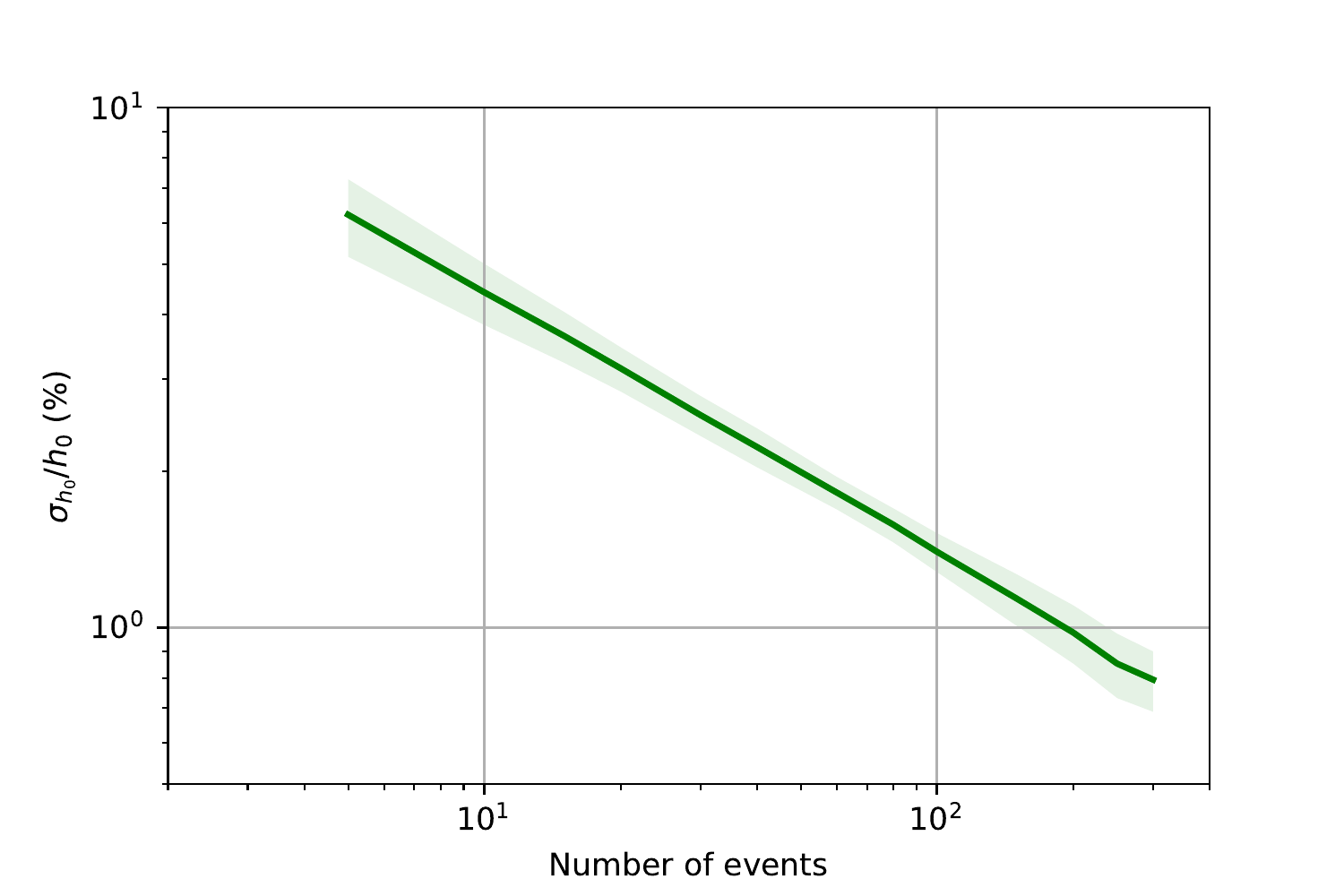}}\hspace{-0.1in}
{\includegraphics[width=0.49\textwidth]{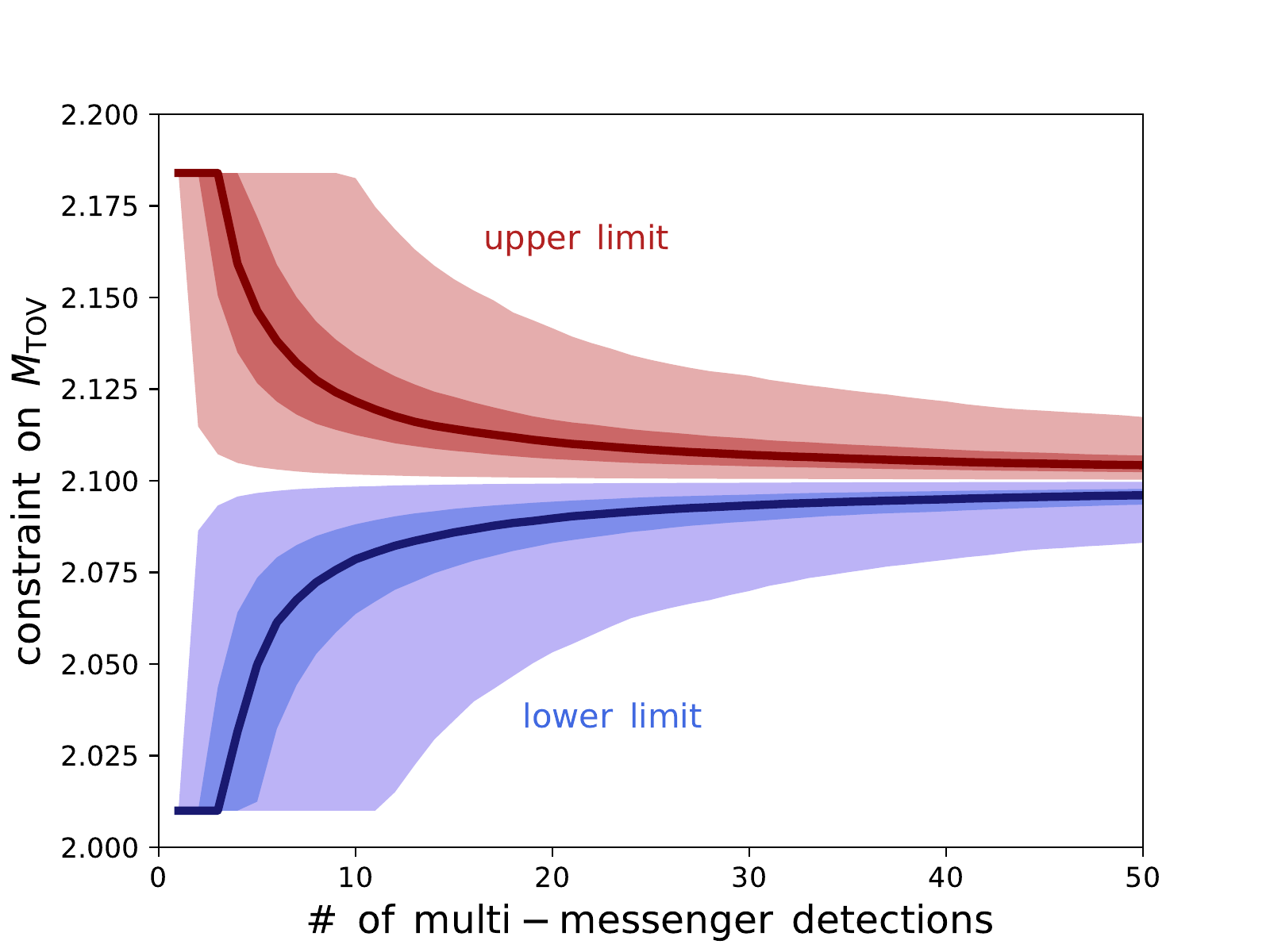}}
\caption{{\it Left Panel:} Hubble constant precision as a function of the number of joint GW-EM detections. 
         {\it Right Panel:} Constraints on maximum neutron star mass, $M_{\rm TOV}$ (here assumed to be $M_{\rm TOV} = 2.1M_\odot$), as a function of the number of joint GW-EM detections.}
\label{fig:evolution}
\vspace{-0.15in}
\end{figure}

The first $H_0$ measurement using this technique was made following the detection of GW170817 and its associated optical counterpart~\citep{2017Natur.551...85A,Guidorzi+17}. As more GW-EM joint detections are made the precision of our $H_0$ measurement will continue to increase significantly (\autoref{fig:evolution}). It is expected that within the decade we will have a sufficient sample size to make a percent-level determination of $H_0$ which can potentially resolve the current tension~\citep{2018Natur.562..545C}. In addition to BNS mergers, NS-BH mergers provide another potential candidate for GW-EM $H_0$ measurements. The precession of NS-BH binaries allow for a more accurate determination of the luminosity distance and therefore an even more precise $H_0$ measurement, assuming that an EM counterpart can be identified~\citep{2018PhRvL.121b1303V}. As the sensitivity of GW detectors continues to improve, GW sources at higher redshifts may eventually provide the leverage necessary to constrain other cosmological parameters through measurements of the distance-redshift relation beyond the local volume.

\subsection{Unveiling the Sites of Cosmic $r$-process Production}{\label{sed:rprocess}}

Roughly half of the elements heavier than the iron group are produced by the rapid capture of neutrons onto lighter seed nuclei \citep[the so-called $r$-process; e.g.,][]{Burbidge+57}.  However, the astrophysical site or sites where the $r$-process takes place has long been debated, with core collapse supernovae \citep[e.g.,][]{Woosley+94} and binary NS mergers \citep{Lattimer&Schramm74} being traditionally favored contenders.  The kilonova transient following GW170817 was very likely powered by the radioactive decay of freshly synthesized $r$-process nuclei, providing the first direct evidence for its origin.  Furthermore, given the large quantity of $r$-process material (of at least several percent of a solar mass) inferred from the light curve modeling, and the volumetric rate of BNS mergers implied by Advanced LIGO, one is already led to conclude that NS mergers are major, if not dominant, $r$-process sources in the Galaxy \citep[e.g.,][]{Cowperthwaite+17,Drout+17}.  Nevertheless, given just a single event, it is not known whether the $r$-process yield of GW170817 was in fact "typical".  Furthermore, other astrophysical events \citep[e.g.~collapsars;][]{Siegel+17} may contribute to the $r$-process. 

A future sample of kilonovae discovered by LSST following triggers from LIGO/Virgo will provide much better statistics on the diversity of $r$-process production from NS mergers.  The colors and spectra of the kilonovae provide key information on the detailed composition of the ejecta \citep[e.g.,][]{Kasen+17}, particularly the presence or absence of heavy lanthanide elements (atomic mass number $A \gtrsim 140$), which may vary from event to event, depending on the neutron richness (electron fraction) of the merger ejecta \citep[e.g.,][]{Metzger&Fernandez14}.  The abundance patterns of individual $r$-process "pollution events" are measured in the spectra of metal-poor stars in the galactic halo \citep[e.g.,][]{Honda+06} or in dwarf galaxies orbiting the Milky Way \citep[e.g.,][]{Ji+16}.  Searching for similar diversity in the the $r$-process signatures of kilonovae would help determine whether binary NS mergers were present and contributing to chemical enrichment in the early universe \citep[e.g.,][]{Horowitz+18}.  

\subsection{New Constraints on the Neutron Star Equation of State}{\label{sec:NSEOS}}

Despite significant efforts, the equation of state (EOS) of dense neutron-rich matter is fundamentally unknown.  This property is key to understanding the physics of matter at extreme densities and the properties of NSs.  The EOS is also essential input to numerical simulations of NS mergers and core-collapse SNe.  However, most properties of the EOS are inaccessible to terrestrial laboratory experiments and currently incalculable from first principles QCD. As such, astrophysical observations of NSs remain the most promising avenue of probing the regime of dense matter.

The GW signal of binary NS mergers alone can provide important information about the radii of NSs \citep{2017PhRvL.119p1101A,De+18,Raithel+18,Most+18,2018PhRvL.121p1101A} --- a proxy for the pressure at $\sim$twice nuclear saturation density \citep{Lattimer&Prakash01}, however probing the EOS at even higher densities in the NS core remains difficult \citep[though see e.g.][]{Bauswein+14}. Joint GW-EM observations of NS mergers (particularly the kilonova and gamma-ray burst emission) offer a fresh opportunity to probe this regime by using the electromagnetic signature to infer the fate of the merger remnant, e.g. when and whether a black hole forms.  This approach was applied to GW170817, leading to stringent new constraints on the maximal mass of (cold, non-rotating) NSs \citep{Margalit&Metzger17,Shibata+17,Rezzolla+18,Ruiz+18} as well as lower limits on the NS radius \citep{Bauswein+17,Radice+18,Coughlin+18}.

A larger statistical sample of joint GW-EM binary NS merger observations will further strengthen these constraints and provide invaluable insight into the behavior of matter at the highest densities.
Figure \ref{fig:evolution} shows the predicted constraints (upper/lower limits) on the maximum NS mass, $M_{\rm TOV}$, as a function of the number of well-characterized joint GW+EM binary NS merger detections, simulated for a binary NS mass distribution drawn from the Galactic population. Future LSST discoveries of GW events, accompanied by dedicated ground or space-based EM follow-up efforts to ascertain the remnant fates, thus represents a powerful new probe on constraining the NS EOS and the only current method which can provide an {\it upper limit} on $M_{\rm TOV}$.

In addition to NS-NS mergers, the EM follow-up of mergers of BH-NS binaries can provide constraints on the NS radius and the properties of the BH complementary to those obtained through the GW signal alone.  This is driven by the fact that the quantity of ejected mass (which can be measured from the kilonova emission) in BH-NS merger depends sensitively on the NS radius and the size of the BH event horizon \citep[e.g.~][]{Foucart12}. The presence or absence of luminous EM counterpart in a population of BH-NS merger, in addition to providing constraints on the NS radius given the GW-inferred mass and spin of the BH, is sensitive to whether the BH event horizon has the properties predicted by General Relativity.

\subsection{Additional Science Drivers}{\label{sec:other}}
In this white paper, we have touched on only a small fraction of the GW-EM science possible in the next decade. In reality, the potential for new discoveries is vast. This includes as of yet undetected GW sources, such as Local Group supernova~\citep{2016PhRvD..94j2001A} and neutron-star black-hole mergers~\citep{2010CQGra..27q3001A,2018arXiv181112907T} which may also begin to emerge. In both cases, GW detectors provide complementary facilities to study complex physical phenomena such as accretion disks, jet physics, and the collapse of massive stars. Additionally, binary black hole (BBH) mergers are expected to form the majority (up to hundreds per year) of GW event catalogs. While not anticipated to be EM active, tentative links to gamma-ray emission have been made~\citep{2016ApJ...826L...6C}, and further detections will solidify or rule out these claims as well as constrain other models. This broad range of science possibilities means that LSST should remain available and flexible enough to respond to unexpected triggers.

\section{Outlook For The Next Decade}{\label{sec:conclusions}}
We have presented a wide range of possible science goals using joint GW-EM astronomy that are achievable in the next decade. The common thread between all of these goals is that they require the dedicated allocation of resources. Specifically, they all require a large sample of joint GW-EM detections that enable deep statistical studies of both GW and EM sources. As outlined in \autoref{sec:LSST}, the use of GW triggered target of opportunity observations with LSST is the most efficient and effective approach to building up such a sample. We therefore reiterate our recommendation that significant effort and telescope time be dedicated to this task. This includes an appropriate fraction of LSST survey time along with the development of rapid response data brokers fine-tuned for GW follow-up. Ideally, LSST should also strive to interact with other telescopes \citep[e.g., ELTs;][]{chornockWP,grahamWP} to ensure that we obtain rich data sets spanning the entire EM spectrum. Data from these programs should be made promptly available to the entire community. Joint GW-EM astronomy has the potential to truly revolutionize the way in which we observe and understand the Universe, but only with the proper allocation of resources and opportunities.

\pagebreak
{\RaggedRight \small
\bibliographystyle{yahapj}
\let\oldbibliography\thebibliography
\renewcommand{\thebibliography}[1]{\oldbibliography{#1}
\setlength{\itemsep}{0pt}} 
\bibliography{references}

\begin{thebibliography}{}
\providecommand\natexlab[1]{#1}
\providecommand\JournalTitle[1]{#1}

\bibitem[{{Abadie} {et~al.}(2010){Abadie}, {Abbott}, {Abbott}, {Abernathy},
  {Accadia}, {Acernese}, {Adams}, {Adhikari}, {Ajith}, {Allen}, {Allen},
  {Amador Ceron}, {Amin}, {Anderson}, {Anderson}, {Antonucci}, {Aoudia},
  {Arain}, {Araya}, {Aronsson}, {Arun}, {Aso}, {Aston}, {Astone}, {Atkinson},
  {Aufmuth}, {Aulbert}, {Babak}, {Baker}, {Ballardin}, {Ballmer}, {Barker},
  {Barnum}, {Barone}, {Barr}, {Barriga}, {Barsotti}, {Barsuglia}, {Barton},
  {Bartos}, {Bassiri}, {Bastarrika}, {Bauchrowitz}, {Bauer}, {Behnke}, {Beker},
  {Belczynski}, {Benacquista}, {Bertolini}, {Betzwieser}, {Beveridge},
  {Beyersdorf}, {Bigotta}, {Bilenko}, {Billingsley}, {Birch}, {Birindelli},
  {Biswas}, {Bitossi}, {Bizouard}, {Black}, {Blackburn}, {Blackburn}, {Blair},
  {Bland}, {Blom}, {Blomberg}, {Boccara}, {Bock}, {Bodiya}, {Bondarescu},
  {Bondu}, {Bonelli}, {Bork}, {Born}, {Bose}, {Bosi}, {Boyle}, {Braccini},
  {Bradaschia}, {Brady}, {Braginsky}, {Brau}, {Breyer}, {Bridges}, {Brillet},
  {Brinkmann}, {Brisson}, {Britzger}, {Brooks}, {Brown}, {Budzy{\'n}ski},
  {Bulik}, {Bulten}, {Buonanno}, {Burguet-Castell}, {Burmeister}, {Buskulic},
  {Byer}, {Cadonati}, {Cagnoli}, {Calloni}, {Camp}, {Campagna}, {Campsie},
  {Cannizzo}, {Cannon}, {Canuel}, {Cao}, {Capano}, {Carbognani}, {Caride},
  {Caudill}, {Cavagli{\`a}}, {Cavalier}, {Cavalieri}, {Cella}, {Cepeda},
  {Cesarini}, {Chalermsongsak}, {Chalkley}, {Charlton}, {Chassand e Mottin},
  {Chelkowski}, {Chen}, {Chincarini}, {Christensen}, {Chua}, {Chung}, {Clark},
  {Clark}, {Clayton}, {Cleva}, {Coccia}, {Colacino}, {Colas}, {Colla},
  {Colombini}, {Conte}, {Cook}, {Corbitt}, {Corda}, {Cornish}, {Corsi},
  {Costa}, {Coulon}, {Coward}, {Coyne}, {Creighton}, {Creighton}, {Cruise},
  {Culter}, {Cumming}, {Cunningham}, {Cuoco}, {Dahl}, {Danilishin},
  {Dannenberg}, {D'Antonio}, {Danzmann}, {Dari}, {Das}, {Dattilo}, {Daudert},
  {Davier}, {Davies}, {Davis}, {Daw}, {Day}, {Dayanga}, {De Rosa}, {DeBra},
  {Degallaix}, {del Prete}, {Dergachev}, {DeRosa}, {DeSalvo}, {Devanka},
  {Dhurandhar}, {Di Fiore}, {Di Lieto}, {Di Palma}, {Emilio}, {Di Virgilio},
  {D{\'\i}az}, {Dietz}, {Donovan}, {Dooley}, {Doomes}, {Dorsher}, {Douglas},
  {Drago}, {Drever}, {Driggers}, {Dueck}, {Dumas}, {Eberle}, {Edgar},
  {Edwards}, {Effler}, {Ehrens}, {Engel}, {Etzel}, {Evans}, {Evans}, {Fafone},
  {Fairhurst}, {Fan}, {Farr}, {Fazi}, {Fehrmann}, {Feldbaum}, {Ferrante},
  {Fidecaro}, {Finn}, {Fiori}, {Flaminio}, {Flanigan}, {Flasch}, {Foley},
  {Forrest}, {Forsi}, {Fotopoulos}, {Fournier}, {Franc}, {Frasca}, {Frasconi},
  {Frede}, {Frei}, {Frei}, {Freise}, {Frey}, {Fricke}, {Friedrich},
  {Fritschel}, {Frolov}, {Fulda}, {Fyffe}, {Gammaitoni}, {Garofoli}, {Garufi},
  {Gemme}, {Genin}, {Gennai}, {Gholami}, {Ghosh}, {Giaime}, {Giampanis},
  {Giardina}, {Giazotto}, {Gill}, {Goetz}, {Goggin}, {Gonz{\'a}lez},
  {Gorodetsky}, {Go{\ss}ler}, {Gouaty}, {Graef}, {Granata}, {Grant}, {Gras},
  {Gray}, {Greenhalgh}, {Gretarsson}, {Greverie}, {Grosso}, {Grote},
  {Grunewald}, {Guidi}, {Gustafson}, {Gustafson}, {Hage}, {Hall}, {Hallam},
  {Hammer}, {Hammond}, {Hanks}, {Hanna}, {Hanson}, {Harms}, {Harry}, {Harry},
  {Harstad}, {Haughian}, {Hayama}, {Heefner}, {Heitmann}, {Hello}, {Heng},
  {Heptonstall}, {Hewitson}, {Hild}, {Hirose}, {Hoak}, {Hodge}, {Holt},
  {Hosken}, {Hough}, {Howell}, {Hoyland}, {Huet}, {Hughey}, {Husa}, {Huttner},
  {Huynh-Dinh}, {Ingram}, {Inta}, {Isogai}, {Ivanov}, {Jaranowski}, {Johnson},
  {Jones}, {Jones}, {Jones}, {Ju}, {Kalmus}, {Kalogera}, {Kandhasamy},
  {Kanner}, {Katsavounidis}, {Kawabe}, {Kawamura}, {Kawazoe}, {Kells},
  {Keppel}, {Khalaidovski}, {Khalili}, {Khazanov}, {Kim}, {Kim}, {King},
  {Kinzel}, {Kissel}, {Klimenko}, {Kondrashov}, {Kopparapu}, {Koranda},
  {Kowalska}, {Kozak}, {Krause}, {Kringel}, {Krishnamurthy}, {Krishnan},
  {Kr{\'o}lak}, {Kuehn}, {Kullman}, {Kumar}, {Kwee}, {Landry}, {Lang}, {Lantz},
  {Lastzka}, {Lazzarini}, {Leaci}, {Leong}, {Leonor}, {Leroy}, {Letendre},
  {Li}, {Li}, {Lin}, {Lindquist}, {Lockerbie}, {Lodhia}, {Lorenzini},
  {Loriette}, {Lormand}, {Losurdo}, {Lu}, {Luan}, {Lubi{\'n}ski}, {LIGO
  Scientific Collaboration}, \& {Virgo Collaboration}}]{2010CQGra..27q3001A}
{Abadie}, J., {Abbott}, B.~P., {Abbott}, R., {et~al.} 2010,
  \href{http://dx.doi.org/10.1088/0264-9381/27/17/173001}{\JournalTitle{Classical
  and Quantum Gravity}, 27, 173001}

\bibitem[{{Abbott} {et~al.}(2016{\natexlab{a}}){Abbott}, {Abbott}, {Abbott},
  {Abernathy}, {Acernese}, {Ackley}, {Adams}, {Adams}, {Addesso}, {Adhikari},
  {Adya}, {Affeldt}, {Agathos}, {Agatsuma}, {Aggarwal}, {Aguiar}, {Aiello},
  {Ain}, {Ajith}, {Allen}, {Allocca}, {Altin}, {Anderson}, {Anderson}, {Arai},
  {Araya}, {Arceneaux}, {Areeda}, {Arnaud}, {Arun}, {Ascenzi}, {Ashton}, {Ast},
  {Aston}, {Astone}, {Aufmuth}, {Aulbert}, {Babak}, {Bacon}, {Bader}, {Baker},
  {Baldaccini}, {Ballardin}, {Ballmer}, {Barayoga}, {Barclay}, {Barish},
  {Barker}, {Barone}, {Barr}, {Barsotti}, {Barsuglia}, {Barta}, {Bartlett},
  {Bartos}, {Bassiri}, {Basti}, {Batch}, {Baune}, {Bavigadda}, {Bazzan},
  {Behnke}, {Bejger}, {Bell}, {Bell}, {Berger}, {Bergman}, {Bergmann}, {Berry},
  {Bersanetti}, {Bertolini}, {Betzwieser}, {Bhagwat}, {Bhand are}, {Bilenko},
  {Billingsley}, {Birch}, {Birney}, {Biscans}, {Bisht}, {Bitossi}, {Biwer},
  {Bizouard}, {Blackburn}, {Blair}, {Blair}, {Blair}, {Bloemen}, {Bock},
  {Bodiya}, {Boer}, {Bogaert}, {Bogan}, {Bohe}, {Bojtos}, {Bond}, {Bondu},
  {Bonnand}, {Boom}, {Bork}, {Boschi}, {Bose}, {Bouffanais}, {Bozzi},
  {Bradaschia}, {Brady}, {Braginsky}, {Branchesi}, {Brau}, {Briant}, {Brillet},
  {Brinkmann}, {Brisson}, {Brockill}, {Brooks}, {Brown}, {Brown}, {Brown},
  {Buchanan}, {Buikema}, {Bulik}, {Bulten}, {Buonanno}, {Buskulic}, {Buy},
  {Byer}, {Cadonati}, {Cagnoli}, {Cahillane}, {Calder{\'o}n Bustillo},
  {Callister}, {Calloni}, {Camp}, {Cannon}, {Cao}, {Capano}, {Capocasa},
  {Carbognani}, {Caride}, {Casanueva Diaz}, {Casentini}, {Caudill},
  {Cavagli{\`a}}, {Cavalier}, {Cavalieri}, {Cella}, {Cepeda}, {Cerboni
  Baiardi}, {Cerretani}, {Cesarini}, {Chakraborty}, {Chalermsongsak},
  {Chamberlin}, {Chan}, {Chao}, {Charlton}, {Chassande-Mottin}, {Chen}, {Chen},
  {Cheng}, {Chincarini}, {Chiummo}, {Cho}, {Cho}, {Chow}, {Christensen}, {Chu},
  {Chua}, {Chung}, {Ciani}, {Clara}, {Clark}, {Cleva}, {Coccia}, {Cohadon},
  {Colla}, {Collette}, {Cominsky}, {Constancio}, {Conte}, {Conti}, {Cook},
  {Corbitt}, {Cornish}, {Corpuz}, {Corsi}, {Cortese}, {Costa}, {Coughlin},
  {Coughlin}, {Coulon}, {Countryman}, {Couvares}, {Coward}, {Cowart}, {Coyne},
  {Coyne}, {Craig}, {Creighton}, {Cripe}, {Crowder}, {Cumming}, {Cunningham},
  {Cuoco}, {Dal Canton}, {Danilishin}, {D'Antonio}, {Danzmann}, {Darman},
  {Dattilo}, {Dave}, {Daveloza}, {Davier}, {Davies}, {Daw}, {Day}, {DeBra},
  {Debreczeni}, {Degallaix}, {De Laurentis}, {Del{\'e}glise}, {Del Pozzo},
  {Denker}, {Dent}, {Dergachev}, {De Rosa}, {DeRosa}, {DeSalvo}, {Dhurandhar},
  {D{\'\i}az}, {Di Fiore}, {Di Giovanni}, {Di Girolamo}, {Di Lieto}, {Di Pace},
  {Di Palma}, {Di Virgilio}, {Dojcinoski}, {Dolique}, {Donovan}, {Dooley},
  {Doravari}, {Douglas}, {Downes}, {Drago}, {Drever}, {Driggers}, {Du},
  {Ducrot}, {Dwyer}, {Edo}, {Edwards}, {Effler}, {Eggenstein}, {Ehrens},
  {Eichholz}, {Eikenberry}, {Engels}, {Essick}, {Etzel}, {Evans}, {Evans},
  {Everett}, {Factourovich}, {Fafone}, {Fair}, {Fairhurst}, {Fan}, {Fang},
  {Farinon}, {Farr}, {Farr}, {Favata}, {Fays}, {Fehrmann}, {Fejer}, {Ferrante},
  {Ferreira}, {Ferrini}, {Fidecaro}, {Fiori}, {Fiorucci}, {Fisher}, {Flaminio},
  {Fletcher}, {Fournier}, {Frasca}, {Frasconi}, {Frei}, {Freise}, {Frey},
  {Frey}, {Fricke}, {Fritschel}, {Frolov}, {Fulda}, {Fyffe}, {Gabbard}, {Gair},
  {Gammaitoni}, {Gaonkar}, {Garufi}, {Gaur}, {Gehrels}, {Gemme}, {Genin},
  {Gennai}, {George}, {Gergely}, {Germain}, {Ghosh}, {Ghosh}, {Giaime},
  {Giardina}, {Giazotto}, {Gill}, {Glaefke}, {Goetz}, {Goetz}, {Gondan},
  {Gonz{\'a}lez}, {Gonzalez Castro}, {Gopakumar}, {Gordon}, {Gorodetsky},
  {Gossan}, {Gosselin}, {Gouaty}, {Grado}, {Graef}, {Graff}, {Granata},
  {Grant}, {Gras}, {Gray}, {Greco}, {Green}, {Groot}, {Grote}, {Grunewald},
  {Guidi}, {Guo}, {Gupta}, {Gupta}, {Gushwa}, {Gustafson}, {Gustafson},
  {Hacker}, {Hall}, {Hall}, {Hammond}, {Haney}, {Hanke}, {Hanks}, {Hanna},
  {Hannam}, {Hanson}, {Hardwick}, {Harms}, {Harry}, {Harry}, {Hart}, {Hartman},
  {Haster}, {Haughian}, {Heidmann}, {Heintze}, {Heitmann}, {Hello}, {Hemming},
  {Hendry}, {Heng}, {Hennig}, {Heptonstall}, {Heurs}, {Hild}, {Hoak}, {Hodge},
  {Hofman}, {Hollitt}, {Holt}, {Holz}, {Hopkins}, {Hosken}, {Hough}, {Houston},
  {Howell}, {Hu}, {Huang}, {Huerta}, {Huet}, {Hughey}, {Husa}, {Huttner},
  {Huynh-Dinh}, {Idrisy}, {Indik}, {Ingram}, {Inta}, {Isa}, {Isac}, {Isi},
  {Islas}, {Isogai}, {Iyer}, {Izumi}, {Jacqmin}, {Jang}, {Jani}, {Jaranowski},
  {Jawahar}, {Jim{\'e}nez-Forteza}, {Johnson}, {Jones}, {Jones}, {Jonker},
  {Ju}, {Haris}, {Kalaghatgi}, {Kalmus}, {Kalogera}, {Kamaretsos},
  {Kandhasamy}, {Kang}, {Kanner}, {Karki}, {Kasprzack}, {Katsavounidis},
  {Katzman}, {Kaufer}, {Kaur}, {Kawabe}, {Kawazoe}, {K{\'e}f{\'e}lian}, {Kehl},
  {Keitel}, {Kelley}, {Kells}, {Kennedy}, {Key}, {Khalaidovski}, {Khalili},
  {Khan}, {Khan}, {Khan}, {Khazanov}, {Kijbunchoo}, {Kim}, {Kim}, {Kim}, {Kim},
  {Kim}, {Kim}, {King}, {King}, {Kinzel}, {Kissel}, {Kleybolte}, {Klimenko},
  {Koehlenbeck}, {Kokeyama}, {Koley}, {Kondrashov}, {Kontos}, {Korobko},
  {Korth}, {Kowalska}, {Kozak}, {Kringel}, {Krishnan}, {Kr{\'o}lak}, {Krueger},
  {Kuehn}, {Kumar}, {Kuo}, {Kutynia}, {Lackey}, {Landry}, {Lange}, {Lantz},
  {Lasky}, {Lazzarini}, {Lazzaro}, {Leaci}, {Leavey}, {Lebigot}, {Lee}, {Lee},
  {Lee}, {Lee}, {Lenon}, {Leonardi}, {Leong}, {Leroy}, {Letendre}, {Levin},
  {Levine}, {Li}, {Libson}, {Littenberg}, {Lockerbie}, {Loew}, {Logue},
  {Lombardi}, {Lord}, {Lorenzini}, {Loriette}, {Lormand}, {Losurdo}, {Lough},
  {L{\"u}ck}, {Lundgren}, {Luo}, {Lynch}, {Ma}, {MacDonald}, {Machenschalk},
  {MacInnis}, {Macleod}, {Maga{\~n}a-Sandoval}, {Magee}, {Mageswaran},
  {Majorana}, {Maksimovic}, {Malvezzi}, {Man}, {Mandel}, {Mandic}, {Mangano},
  {Mansell}, {Manske}, {Mantovani}, {Marchesoni}, {Marion}, {M{\'a}rka},
  {M{\'a}rka}, {Markosyan}, {Maros}, {Martelli}, {Martellini}, {Martin},
  {Martin}, {Martynov}, {Marx}, {Mason}, {Masserot}, {Massinger}, {Masso-Reid},
  {Mastrogiovanni}, {Matichard}, {Matone}, {Mavalvala}, {Mazumder}, {Mazzolo},
  {McCarthy}, {McClelland }, {McCormick}, {McGuire}, {McIntyre}, {McIver},
  {McManus}, {McWilliams}, {Meacher}, {Meadors}, {Meidam}, {Melatos},
  {Mendell}, {Mendoza-Gandara}, {Mercer}, {Merilh}, {Merzougui}, {Meshkov},
  {Messenger}, {Messick}, {Metzdorff}, {Meyers}, {Mezzani}, {Miao}, {Michel},
  {Middleton}, {Mikhailov}, {Milano}, {Miller}, {Miller}, {Millhouse},
  {Minenkov}, {Ming}, {Mirshekari}, {Mishra}, {Mitra}, {Mitrofanov},
  {Mitselmakher}, {Mittleman}, {Moggi}, {Mohan}, {Mohapatra}, {Montani},
  {Moore}, {Moore}, {Moraru}, {Moreno}, {Morriss}, {Mossavi}, {Mours},
  {Mow-Lowry}, {Mueller}, {Mueller}, {Muir}, {Mukherjee}, {Mukherjee},
  {Mukherjee}, {Mukund}, {Mullavey}, {Munch}, {Murphy}, {Murray}, {Mytidis},
  {Nardecchia}, {Naticchioni}, {Nayak}, {Necula}, {Nedkova}, {Nelemans},
  {Neri}, {Neunzert}, {Newton}, {Nguyen}, {Nielsen}, {Nissanke}, {Nitz},
  {Nocera}, {Nolting}, {Normandin}, {Nuttall}, {Oberling}, {Ochsner}, {O'Dell},
  {Oelker}, {Ogin}, {Oh}, {Oh}, {Ohme}, {Oliver}, {Oppermann}, {Oram},
  {O'Reilly}, {O'Shaughnessy}, {Ott}, {Ottaway}, {Ottens}, {Overmier}, {Owen},
  {Pai}, {Pai}, {Palamos}, {Palashov}, {Palomba}, {Pal-Singh}, {Pan}, {Pankow},
  {Pannarale}, {Pant}, {Paoletti}, {Paoli}, {Papa}, {Paris}, {Parker},
  {Pascucci}, {Pasqualetti}, {Passaquieti}, {Passuello}, {Patricelli},
  {Patrick}, {Pearlstone}, {Pedraza}, {Pedurand}, {Pekowsky}, {Pele}, {Penn},
  {Pereira}, {Perreca}, {Phelps}, {Piccinni}, {Pichot}, {Piergiovanni},
  {Pierro}, {Pillant}, {Pinard}, {Pinto}, {Pitkin}, {Poggiani}, {Popolizio},
  {Post}, {Powell}, {Prasad}, {Predoi}, {Premachand ra}, {Prestegard}, {Price},
  {Prijatelj}, {Principe}, {Privitera}, {Prix}, {Prodi}, {Prokhorov},
  {Puncken}, {Punturo}, {Puppo}, {P{\"u}rrer}, {Qi}, {Qin}, {Quetschke},
  {Quintero}, {Quitzow-James}, {Raab}, {Rabeling}, {Radkins}, {Raffai}, {Raja},
  {Rakhmanov}, {Rapagnani}, {Raymond}, {Razzano}, {Re}, {Read}, {Reed},
  {Regimbau}, {Rei}, {Reid}, {Reitze}, {Rew}, {Ricci}, {Riles}, {Robertson},
  {Robie}, {Robinet}, {Rocchi}, {Rolland}, {Rollins}, {Roma}, {Romano},
  {Romano}, {Romanov}, {Romie}, {Rosi{\'n}ska}, {Rowan}, {R{\"u}diger},
  {Ruggi}, {Ryan}, {Sachdev}, {Sadecki}, {Sadeghian}, {Salconi}, {Saleem},
  {Salemi}, {Samajdar}, {Sammut}, {Sanchez}, {Sand berg}, {Sandeen}, {Sanders},
  {Santamaria}, {Sassolas}, {Sathyaprakash}, {Saulson}, {Sauter}, {Savage},
  {Sawadsky}, {Schale}, {Schilling}, {Schmidt}, {Schmidt}, {Schnabel},
  {Schofield}, {Sch{\"o}nbeck}, {Schreiber}, {Schuette}, {Schutz}, {Scott},
  {Scott}, {Sellers}, {Sentenac}, {Sequino}, {Sergeev}, {Serna}, {Setyawati},
  {Sevigny}, {Shaddock}, {Shahriar}, {Shaltev}, {Shao}, {Shapiro}, {Shawhan},
  {Sheperd}, {Shoemaker}, {Shoemaker}, {Siellez}, {Siemens}, {Sieniawska},
  {Sigg}, {Silva}, {Simakov}, {Singer}, {Singer}, {Singh}, {Singh}, {Singhal},
  {Sintes}, {Slagmolen}, {Smith}, {Smith}, {Smith}, {Son}, {Sorazu},
  {Sorrentino}, {Souradeep}, {Srivastava}, {Staley}, {Steinke}, {Steinlechner},
  {Steinlechner}, {Steinmeyer}, {Stephens}, {Stone}, {Strain}, {Straniero},
  {Stratta}, {Strauss}, {Strigin}, {Sturani}, {Stuver}, {Summerscales}, {Sun},
  {Sutton}, {Swinkels}, {Szczepa{\'n}czyk}, {Tacca}, {Talukder}, {Tanner},
  {T{\'a}pai}, {Tarabrin}, {Taracchini}, {Taylor}, {Theeg},
  {Thirugnanasambandam}, {Thomas}, {Thomas}, {Thomas}, {Thorne}, {Thorne},
  {Thrane}, {Tiwari}, {Tiwari}, {Tokmakov}, {Tomlinson}, {Tonelli}, {Torres},
  {Torrie}, {T{\"o}yr{\"a}}, {Travasso}, {Traylor}, {Trifir{\`o}}, {Tringali},
  {Trozzo}, {Tse}, {Turconi}, {Tuyenbayev}, {Ugolini}, {Unnikrishnan}, {Urban},
  {Usman}, {Vahlbruch}, {Vajente}, {Valdes}, {van Bakel}, {van Beuzekom}, {van
  den Brand}, {Van Den Broeck}, {Vander-Hyde}, {van der Schaaf}, {van
  Heijningen}, {van Veggel}, {Vardaro}, {Vass}, {Vas{\'u}th}, {Vaulin},
  {Vecchio}, {Vedovato}, {Veitch}, {Veitch}, {Venkateswara}, {Verkindt},
  {Vetrano}, {Vicer{\'e}}, {Vinciguerra}, {Vine}, {Vinet}, {Vitale}, {Vo},
  {Vocca}, {Vorvick}, {Voss}, {Vousden}, {Vyatchanin}, {Wade}, {Wade}, {Wade},
  {Walker}, {Wallace}, {Walsh}, {Wang}, {Wang}, {Wang}, {Wang}, {Wang}, {Ward},
  {Warner}, {Was}, {Weaver}, {Wei}, {Weinert}, {Weinstein}, {Weiss}, {Welborn},
  {Wen}, {We{\ss}els}, {Westphal}, {Wette}, {Whelan}, {Whitcomb}, {White},
  {Whiting}, {Williams}, {Williamson}, {Willis}, {Willke}, {Wimmer}, {Winkler},
  {Wipf}, {Wittel}, {Woan}, {Worden}, {Wright}, {Wu}, {Yablon}, {Yam},
  {Yamamoto}, {Yancey}, {Yap}, {Yu}, {Yvert}, {Zadro{\.Z}ny}, {Zangrando},
  {Zanolin}, {Zendri}, {Zevin}, {Zhang}, {Zhang}, {Zhang}, {Zhang}, {Zhao},
  {Zhou}, {Zhou}, {Zhu}, {Zucker}, {Zuraw}, {Zweizig}, {LIGO Scientific
  Collaboration}, \& {Virgo Collaboration}}]{2016PhRvD..94j2001A}
{Abbott}, B.~P., {Abbott}, R., {Abbott}, T.~D., {et~al.} 2016{\natexlab{a}},
  \href{http://dx.doi.org/10.1103/PhysRevD.94.102001}{\JournalTitle{\prd}, 94,
  102001}

\bibitem[{{Abbott} {et~al.}(2016{\natexlab{b}}){Abbott}, {Abbott}, {Abbott},
  {Abernathy}, {Acernese}, {Ackley}, {Adams}, {Adams}, {Addesso}, {Adhikari},
  {Adya}, {Affeldt}, {Agathos}, {Agatsuma}, {Aggarwal}, {Aguiar}, {Aiello},
  {Ain}, {Ajith}, {Allen}, {Allocca}, {Altin}, {Anderson}, {Anderson}, {Arai},
  {Araya}, {Arceneaux}, {Areeda}, {Arnaud}, {Arun}, {Ascenzi}, {Ashton}, {Ast},
  {Aston}, {Astone}, {Aufmuth}, {Aulbert}, {Babak}, {Bacon}, {Bader}, {Baker},
  {Baldaccini}, {Ballardin}, {Ballmer}, {Barayoga}, {Barclay}, {Barish},
  {Barker}, {Barone}, {Barr}, {Barsotti}, {Barsuglia}, {Barta}, {Bartlett},
  {Bartos}, {Bassiri}, {Basti}, {Batch}, {Baune}, {Bavigadda}, {Bazzan},
  {Behnke}, {Bejger}, {Bell}, {Bell}, {Berger}, {Bergman}, {Bergmann}, {Berry},
  {Bersanetti}, {Bertolini}, {Betzwieser}, {Bhagwat}, {Bhand are}, {Bilenko},
  {Billingsley}, {Birch}, {Birney}, {Biscans}, {Bisht}, {Bitossi}, {Biwer},
  {Bizouard}, {Blackburn}, {Blair}, {Blair}, {Blair}, {Bloemen}, {Bock},
  {Bodiya}, {Boer}, {Bogaert}, {Bogan}, {Bohe}, {Bojtos}, {Bond}, {Bondu},
  {Bonnand}, {Boom}, {Bork}, {Boschi}, {Bose}, {Bouffanais}, {Bozzi},
  {Bradaschia}, {Brady}, {Braginsky}, {Branchesi}, {Brau}, {Briant}, {Brillet},
  {Brinkmann}, {Brisson}, {Brockill}, {Brooks}, {Brown}, {Brown}, {Brown},
  {Buchanan}, {Buikema}, {Bulik}, {Bulten}, {Buonanno}, {Buskulic}, {Buy},
  {Byer}, {Cadonati}, {Cagnoli}, {Cahillane}, {Calder{\'o}n Bustillo},
  {Callister}, {Calloni}, {Camp}, {Cannon}, {Cao}, {Capano}, {Capocasa},
  {Carbognani}, {Caride}, {Casanueva Diaz}, {Casentini}, {Caudill},
  {Cavagli{\`a}}, {Cavalier}, {Cavalieri}, {Cella}, {Cepeda}, {Cerboni
  Baiardi}, {Cerretani}, {Cesarini}, {Chakraborty}, {Chalermsongsak},
  {Chamberlin}, {Chan}, {Chao}, {Charlton}, {Chassande-Mottin}, {Chen}, {Chen},
  {Cheng}, {Chincarini}, {Chiummo}, {Cho}, {Cho}, {Chow}, {Christensen}, {Chu},
  {Chua}, {Chung}, {Ciani}, {Clara}, {Clark}, {Cleva}, {Coccia}, {Cohadon},
  {Colla}, {Collette}, {Cominsky}, {Constancio}, {Conte}, {Conti}, {Cook},
  {Corbitt}, {Cornish}, {Corsi}, {Cortese}, {Costa}, {Coughlin}, {Coughlin},
  {Coulon}, {Countryman}, {Couvares}, {Cowan}, {Coward}, {Cowart}, {Coyne},
  {Coyne}, {Craig}, {Creighton}, {Cripe}, {Crowder}, {Cumming}, {Cunningham},
  {Cuoco}, {Dal Canton}, {Danilishin}, {D'Antonio}, {Danzmann}, {Darman},
  {Dattilo}, {Dave}, {Daveloza}, {Davier}, {Davies}, {Daw}, {Day}, {DeBra},
  {Debreczeni}, {Degallaix}, {De Laurentis}, {Del{\'e}glise}, {Del Pozzo},
  {Denker}, {Dent}, {Dereli}, {Dergachev}, {DeRosa}, {De Rosa}, {DeSalvo},
  {Dhurandhar}, {D{\'\i}az}, {Di Fiore}, {Di Giovanni}, {Di Lieto}, {Di Pace},
  {Di Palma}, {Di Virgilio}, {Dojcinoski}, {Dolique}, {Donovan}, {Dooley},
  {Doravari}, {Douglas}, {Downes}, {Drago}, {Drever}, {Driggers}, {Du},
  {Ducrot}, {Dwyer}, {Edo}, {Edwards}, {Effler}, {Eggenstein}, {Ehrens},
  {Eichholz}, {Eikenberry}, {Engels}, {Essick}, {Etzel}, {Evans}, {Evans},
  {Everett}, {Factourovich}, {Fafone}, {Fair}, {Fairhurst}, {Fan}, {Fang},
  {Farinon}, {Farr}, {Farr}, {Favata}, {Fays}, {Fehrmann}, {Fejer}, {Ferrante},
  {Ferreira}, {Ferrini}, {Fidecaro}, {Fiori}, {Fiorucci}, {Fisher}, {Flaminio},
  {Fletcher}, {Fournier}, {Franco}, {Frasca}, {Frasconi}, {Frei}, {Freise},
  {Frey}, {Frey}, {Fricke}, {Fritschel}, {Frolov}, {Fulda}, {Fyffe}, {Gabbard},
  {Gair}, {Gammaitoni}, {Gaonkar}, {Garufi}, {Gatto}, {Gaur}, {Gehrels},
  {Gemme}, {Gendre}, {Genin}, {Gennai}, {George}, {Gergely}, {Germain},
  {Ghosh}, {Ghosh}, {Giaime}, {Giardina}, {Giazotto}, {Gill}, {Glaefke},
  {Goetz}, {Goetz}, {Gondan}, {Gonz{\'a}lez}, {Gonzalez Castro}, {Gopakumar},
  {Gordon}, {Gorodetsky}, {Gossan}, {Gosselin}, {Gouaty}, {Graef}, {Graff},
  {Granata}, {Grant}, {Gras}, {Gray}, {Greco}, {Green}, {Groot}, {Grote},
  {Grunewald}, {Guidi}, {Guo}, {Gupta}, {Gupta}, {Gushwa}, {Gustafson},
  {Gustafson}, {Hacker}, {Hall}, {Hall}, {Hammond}, {Haney}, {Hanke}, {Hanks},
  {Hanna}, {Hannam}, {Hanson}, {Hardwick}, {Haris}, {Harms}, {Harry}, {Harry},
  {Hart}, {Hartman}, {Haster}, {Haughian}, {Heidmann}, {Heintze}, {Heitmann},
  {Hello}, {Hemming}, {Hendry}, {Heng}, {Hennig}, {Heptonstall}, {Heurs},
  {Hild}, {Hoak}, {Hodge}, {Hofman}, {Hollitt}, {Holt}, {Holz}, {Hopkins},
  {Hosken}, {Hough}, {Houston}, {Howell}, {Hu}, {Huang}, {Huerta}, {Huet},
  {Hughey}, {Husa}, {Huttner}, {Huynh-Dinh}, {Idrisy}, {Indik}, {Ingram},
  {Inta}, {Isa}, {Isac}, {Isi}, {Islas}, {Isogai}, {Iyer}, {Izumi}, {Jacqmin},
  {Jang}, {Jani}, {Jaranowski}, {Jawahar}, {Jim{\'e}nez-Forteza}, {Johnson},
  {Jones}, {Jones}, {Jonker}, {Ju}, {Kalaghatgi}, {Kalogera}, {Kandhasamy},
  {Kang}, {Kanner}, {Karki}, {Kasprzack}, {Katsavounidis}, {Katzman}, {Kaufer},
  {Kaur}, {Kawabe}, {Kawazoe}, {K{\'e}f{\'e}lian}, {Kehl}, {Keitel}, {Kelley},
  {Kells}, {Kennedy}, {Key}, {Khalaidovski}, {Khalili}, {Khan}, {Khan}, {Khan},
  {Khazanov}, {Kijbunchoo}, {Kim}, {Kim}, {Kim}, {Kim}, {Kim}, {Kim}, {King},
  {King}, {Kinzel}, {Kissel}, {Kleybolte}, {Klimenko}, {Koehlenbeck},
  {Kokeyama}, {Koley}, {Kondrashov}, {Kontos}, {Korobko}, {Korth}, {Kowalska},
  {Kozak}, {Kringel}, {Kr{\'o}lak}, {Krueger}, {Kuehn}, {Kumar}, {Kuo},
  {Kutynia}, {Lackey}, {Land ry}, {Lange}, {Lantz}, {Lasky}, {Lazzarini},
  {Lazzaro}, {Leaci}, {Leavey}, {Lebigot}, {Lee}, {Lee}, {Lee}, {Lee}, {Lenon},
  {Leonardi}, {Leong}, {Leroy}, {Letendre}, {Levin}, {Levine}, {Li}, {Libson},
  {Littenberg}, {Lockerbie}, {Logue}, {Lombardi}, {Lord}, {Lorenzini},
  {Loriette}, {Lormand}, {Losurdo}, {Lough}, {L{\"u}ck}, {Lundgren}, {Luo},
  {Lynch}, {Ma}, {MacDonald}, {Machenschalk}, {MacInnis}, {Macleod},
  {Maga{\~n}a-Sandoval}, {Magee}, {Mageswaran}, {Majorana}, {Maksimovic},
  {Malvezzi}, {Man}, {Mandel}, {Mandic}, {Mangano}, {Mansell}, {Manske},
  {Mantovani}, {Marchesoni}, {Marion}, {M{\'a}rka}, {M{\'a}rka}, {Markosyan},
  {Maros}, {Martelli}, {Martellini}, {Martin}, {Martin}, {Martynov}, {Marx},
  {Mason}, {Masserot}, {Massinger}, {Masso-Reid}, {Matichard}, {Matone},
  {Mavalvala}, {Mazumder}, {Mazzolo}, {McCarthy}, {McClelland}, {McCormick},
  {McGuire}, {McIntyre}, {McIver}, {McManus}, {McWilliams}, {Meacher},
  {Meadors}, {Meidam}, {Melatos}, {Mendell}, {Mendoza-Gandara}, {Mercer},
  {Merilh}, {Merzougui}, {Meshkov}, {Messenger}, {Messick}, {Meyers},
  {Mezzani}, {Miao}, {Michel}, {Middleton}, {Mikhailov}, {Milano}, {Miller},
  {Millhouse}, {Minenkov}, {Ming}, {Mirshekari}, {Mishra}, {Mitra},
  {Mitrofanov}, {Mitselmakher}, {Mittleman}, {Moggi}, {Mohan}, {Mohapatra},
  {Montani}, {Moore}, {Moore}, {Moraru}, {Moreno}, {Morriss}, {Mossavi},
  {Mours}, {Mow-Lowry}, {Mueller}, {Mueller}, {Muir}, {Mukherjee}, {Mukherjee},
  {Mukherjee}, {Mukund}, {Mullavey}, {Munch}, {Murphy}, {Murray}, {Mytidis},
  {Nardecchia}, {Naticchioni}, {Nayak}, {Necula}, {Nedkova}, {Nelemans},
  {Neri}, {Neunzert}, {Newton}, {Nguyen}, {Nielsen}, {Nissanke}, {Nitz},
  {Nocera}, {Nolting}, {Normandin}, {Nuttall}, {Oberling}, {Ochsner}, {O'Dell},
  {Oelker}, {Ogin}, {Oh}, {Oh}, {Ohme}, {Oliver}, {Oppermann}, {Oram},
  {O'Reilly}, {O'Shaughnessy}, {Ottaway}, {Ottens}, {Overmier}, {Owen}, {Pai},
  {Pai}, {Palamos}, {Palashov}, {Palomba}, {Pal-Singh}, {Pan}, {Pankow},
  {Pannarale}, {Pant}, {Paoletti}, {Paoli}, {Papa}, {Paris}, {Parker},
  {Pascucci}, {Pasqualetti}, {Passaquieti}, {Passuello}, {Patricelli},
  {Patrick}, {Pearlstone}, {Pedraza}, {Pedurand}, {Pekowsky}, {Pele}, {Penn},
  {Perreca}, {Phelps}, {Piccinni}, {Pichot}, {Piergiovanni}, {Pierro},
  {Pillant}, {Pinard}, {Pinto}, {Pitkin}, {Poggiani}, {Popolizio}, {Post},
  {Powell}, {Prasad}, {Predoi}, {Premachandra}, {Prestegard}, {Price},
  {Prijatelj}, {Principe}, {Privitera}, {Prodi}, {Prokhorov}, {Puncken},
  {Punturo}, {Puppo}, {P{\"u}rrer}, {Qi}, {Qin}, {Quetschke}, {Quintero},
  {Quitzow-James}, {Raab}, {Rabeling}, {Radkins}, {Raffai}, {Raja},
  {Rakhmanov}, {Rapagnani}, {Raymond}, {Razzano}, {Re}, {Read}, {Reed},
  {Regimbau}, {Rei}, {Reid}, {Reitze}, {Rew}, {Reyes}, {Ricci}, {Riles},
  {Robertson}, {Robie}, {Robinet}, {Rocchi}, {Rolland}, {Rollins}, {Roma},
  {Romano}, {Romanov}, {Romie}, {Rosi{\'n}ska}, {Rowan}, {R{\"u}diger},
  {Ruggi}, {Ryan}, {Sachdev}, {Sadecki}, {Sadeghian}, {Salconi}, {Saleem},
  {Salemi}, {Samajdar}, {Sammut}, {Sanchez}, {Sand berg}, {Sandeen}, {Sanders},
  {Sassolas}, {Sathyaprakash}, {Saulson}, {Sauter}, {Savage}, {Sawadsky},
  {Schale}, {Schilling}, {Schmidt}, {Schmidt}, {Schnabel}, {Schofield},
  {Sch{\"o}nbeck}, {Schreiber}, {Schuette}, {Schutz}, {Scott}, {Scott},
  {Sellers}, {Sengupta}, {Sentenac}, {Sequino}, {Sergeev}, {Serna},
  {Setyawati}, {Sevigny}, {Shaddock}, {Shah}, {Shahriar}, {Shaltev}, {Shao},
  {Shapiro}, {Shawhan}, {Sheperd}, {Shoemaker}, {Shoemaker}, {Siellez},
  {Siemens}, {Sigg}, {Silva}, {Simakov}, {Singer}, {Singer}, {Singh}, {Singh},
  {Singhal}, {Sintes}, {Slagmolen}, {Smith}, {Smith}, {Smith}, {Son}, {Sorazu},
  {Sorrentino}, {Souradeep}, {Srivastava}, {Staley}, {Steinke}, {Steinlechner},
  {Steinlechner}, {Steinmeyer}, {Stephens}, {Stone}, {Strain}, {Straniero},
  {Stratta}, {Strauss}, {Strigin}, {Sturani}, {Stuver}, {Summerscales}, {Sun},
  {Sutton}, {Swinkels}, {Szczepa{\'n}czyk}, {Tacca}, {Talukder}, {Tanner},
  {T{\'a}pai}, {Tarabrin}, {Taracchini}, {Taylor}, {Theeg},
  {Thirugnanasambandam}, {Thomas}, {Thomas}, {Thomas}, {Thorne}, {Thorne},
  {Thrane}, {Tiwari}, {Tiwari}, {Tokmakov}, {Tomlinson}, {Tonelli}, {Torres},
  {Torrie}, {T{\"o}yr{\"a}}, {Travasso}, {Traylor}, {Trifir{\`o}}, {Tringali},
  {Trozzo}, {Tse}, {Turconi}, {Tuyenbayev}, {Ugolini}, {Unnikrishnan}, {Urban},
  {Usman}, {Vahlbruch}, {Vajente}, {Valdes}, {van Bakel}, {van Beuzekom}, {van
  den Brand}, {Van Den Broeck}, {Vander-Hyde}, {van der Schaaf}, {van
  Heijningen}, {van Veggel}, {Vardaro}, {Vass}, {Vas{\'u}th}, {Vaulin},
  {Vecchio}, {Vedovato}, {Veitch}, {Veitch}, {Venkateswara}, {Verkindt},
  {Vetrano}, {Vicer{\'e}}, {Vinciguerra}, {Vine}, {Vinet}, {Vitale}, {Vo},
  {Vocca}, {Vorvick}, {Voss}, {Vousden}, {Vyatchanin}, {Wade}, {Wade}, {Wade},
  {Walker}, {Wallace}, {Walsh}, {Wang}, {Wang}, {Wang}, {Wang}, {Wang}, {Ward},
  {Warner}, {Was}, {Weaver}, {Wei}, {Weinert}, {Weinstein}, {Weiss}, {Welborn},
  {Wen}, {We{\ss}els}, {Westphal}, {Wette}, {Whelan}, {Whitcomb}, {White},
  {Whiting}, {Williams}, {Williamson}, {Willis}, {Willke}, {Wimmer}, {Winkler},
  {Wipf}, {Wittel}, {Woan}, {Worden}, {Wright}, {Wu}, {Yablon}, {Yam},
  {Yamamoto}, {Yancey}, {Yap}, {Yu}, {Yvert}, {Zadro{\.Z}ny}, {Zangrando},
  {Zanolin}, {Zendri}, {Zevin}, {Zhang}, {Zhang}, {Zhang}, {Zhang}, {Zhao},
  {Zhou}, {Zhou}, {Zhu}, {Zucker}, {Zuraw}, {Zweizig}, {LIGO Scientific
  Collaboration}, \& {Virgo Collaboration}}]{2016PhRvL.116m1103A}
---. 2016{\natexlab{b}},
  \href{http://dx.doi.org/10.1103/PhysRevLett.116.131103}{\JournalTitle{\prl},
  116, 131103}

\bibitem[{{Abbott} {et~al.}(2016{\natexlab{c}}){Abbott}, {Abbott}, {Abbott},
  {Abernathy}, {Acernese}, {Ackley}, {Adams}, {Adams}, {Addesso}, {Adhikari},
  {Adya}, {Affeldt}, {Agathos}, {Agatsuma}, {Aggarwal}, {Aguiar}, {Ain},
  {Ajith}, {Allen}, {Allocca}, {Altin}, {Amariutei}, {Anderson}, {Anderson},
  {Arai}, {Araya}, {Arceneaux}, {Areeda}, {Arnaud}, {Arun}, {Ashton}, {Ast},
  {Aston}, {Astone}, {Aufmuth}, {Aulbert}, {Babak}, {Baker}, {Baldaccini},
  {Ballardin}, {Ballmer}, {Barayoga}, {Barclay}, {Barish}, {Barker}, {Barone},
  {Barr}, {Barsotti}, {Barsuglia}, {Barta}, {Bartlett}, {Bartos}, {Bassiri},
  {Basti}, {Batch}, {Baune}, {Bavigadda}, {Bazzan}, {Behnke}, {Bejger},
  {Belczynski}, {Bell}, {Bell}, {Berger}, {Bergman}, {Bergmann}, {Berry},
  {Bersanetti}, {Bertolini}, {Betzwieser}, {Bhagwat}, {Bhandare}, {Bilenko},
  {Billingsley}, {Birch}, {Birney}, {Biscans}, {Bisht}, {Bitossi}, {Biwer},
  {Bizouard}, {Blackburn}, {Blair}, {Blair}, {Blair}, {Bloemen}, {Bock},
  {Bodiya}, {Boer}, {Bogaert}, {Bogan}, {Bohe}, {Bojtos}, {Bond}, {Bondu},
  {Bonnand}, {Bork}, {Boschi}, {Bose}, {Bozzi}, {Bradaschia}, {Brady},
  {Braginsky}, {Branchesi}, {Brau}, {Briant}, {Brillet}, {Brinkmann},
  {Brisson}, {Brockill}, {Brooks}, {Brown}, {Brown}, {Brown}, {Buchanan},
  {Buikema}, {Bulik}, {Bulten}, {Buonanno}, {Buskulic}, {Buy}, {Byer},
  {Cadonati}, {Cagnoli}, {Cahillane}, {Calder{\'o}n Bustillo}, {Callister},
  {Calloni}, {Camp}, {Cannon}, {Cao}, {Capano}, {Capocasa}, {Carbognani},
  {Caride}, {Casanueva Diaz}, {Casentini}, {Caudill}, {Cavagli{\`a}},
  {Cavalier}, {Cavalieri}, {Cella}, {Cepeda}, {Cerboni Baiardi}, {Cerretani},
  {Cesarini}, {Chakraborty}, {Chalermsongsak}, {Chamberlin}, {Chan}, {Chao},
  {Charlton}, {Chassand e-Mottin}, {Chen}, {Chen}, {Cheng}, {Chincarini},
  {Chiummo}, {Cho}, {Cho}, {Chow}, {Christensen}, {Chu}, {Chua}, {Chung},
  {Ciani}, {Clara}, {Clark}, {Cleva}, {Coccia}, {Cohadon}, {Colla}, {Collette},
  {Constancio}, {Conte}, {Conti}, {Cook}, {Corbitt}, {Cornish}, {Corsi},
  {Cortese}, {Costa}, {Coughlin}, {Coughlin}, {Coulon}, {Countryman},
  {Couvares}, {Coward}, {Cowart}, {Coyne}, {Coyne}, {Craig}, {Creighton},
  {Cripe}, {Crowder}, {Cumming}, {Cunningham}, {Cuoco}, {Dal Canton},
  {Danilishin}, {D'Antonio}, {Danzmann}, {Darman}, {Dattilo}, {Dave},
  {Daveloza}, {Davier}, {Davies}, {Daw}, {Day}, {DeBra}, {Debreczeni},
  {Degallaix}, {De Laurentis}, {Del{\'e}glise}, {Del Pozzo}, {Denker}, {Dent},
  {Dereli}, {Dergachev}, {DeRosa}, {De Rosa}, {DeSalvo}, {Dhurandhar},
  {D{\'\i}az}, {Di Fiore}, {Di Giovanni}, {Di Lieto}, {Di Palma}, {Di
  Virgilio}, {Dojcinoski}, {Dolique}, {Donovan}, {Dooley}, {Doravari},
  {Douglas}, {Downes}, {Drago}, {Drever}, {Driggers}, {Du}, {Ducrot}, {Dwyer},
  {Edo}, {Edwards}, {Effler}, {Eggenstein}, {Ehrens}, {Eichholz}, {Eikenberry},
  {Engels}, {Essick}, {Etzel}, {Evans}, {Evans}, {Everett}, {Factourovich},
  {Fafone}, {Fair}, {Fairhurst}, {Fan}, {Fang}, {Farinon}, {Farr}, {Farr},
  {Favata}, {Fays}, {Fehrmann}, {Fejer}, {Ferrante}, {Ferreira}, {Ferrini},
  {Fidecaro}, {Fiori}, {Fisher}, {Flaminio}, {Fletcher}, {Fournier}, {Franco},
  {Frasca}, {Frasconi}, {Frei}, {Freise}, {Frey}, {Fricke}, {Fritschel},
  {Frolov}, {Fulda}, {Fyffe}, {Gabbard}, {Gair}, {Gammaitoni}, {Gaonkar},
  {Garufi}, {Gatto}, {Gaur}, {Gehrels}, {Gemme}, {Gendre}, {Genin}, {Gennai},
  {George}, {Gergely}, {Germain}, {Ghosh}, {Ghosh}, {Giaime}, {Giardina},
  {Giazotto}, {Gill}, {Glaefke}, {Goetz}, {Goetz}, {Gondan}, {Gonz{\'a}lez},
  {Gonzalez Castro}, {Gopakumar}, {Gordon}, {Gorodetsky}, {Gossan}, {Gosselin},
  {Gouaty}, {Graef}, {Graff}, {Granata}, {Grant}, {Gras}, {Gray}, {Greco},
  {Green}, {Groot}, {Grote}, {Grunewald}, {Guidi}, {Guo}, {Gupta}, {Gupta},
  {Gushwa}, {Gustafson}, {Gustafson}, {Hacker}, {Hall}, {Hall}, {Hammond},
  {Haney}, {Hanke}, {Hanks}, {Hanna}, {Hannam}, {Hanson}, {Hardwick}, {Harms},
  {Harry}, {Harry}, {Hart}, {Hartman}, {Haster}, {Haughian}, {Heidmann},
  {Heintze}, {Heitmann}, {Hello}, {Hemming}, {Hendry}, {Heng}, {Hennig},
  {Heptonstall}, {Heurs}, {Hild}, {Hoak}, {Hodge}, {Hofman}, {Hollitt}, {Holt},
  {Holz}, {Hopkins}, {Hosken}, {Hough}, {Houston}, {Howell}, {Hu}, {Huang},
  {Huerta}, {Huet}, {Hughey}, {Husa}, {Huttner}, {Huynh-Dinh}, {Idrisy},
  {Indik}, {Ingram}, {Inta}, {Isa}, {Isac}, {Isi}, {Islas}, {Isogai}, {Iyer},
  {Izumi}, {Jacqmin}, {Jang}, {Jani}, {Jaranowski}, {Jawahar},
  {Jim{\'e}nez-Forteza}, {Johnson}, {Jones}, {Jones}, {Jonker}, {Ju}, {K},
  {Kalaghatgi}, {Kalogera}, {Kandhasamy}, {Kang}, {Kanner}, {Karki},
  {Kasprzack}, {Katsavounidis}, {Katzman}, {Kaufer}, {Kaur}, {Kawabe},
  {Kawazoe}, {K{\'e}f{\'e}lian}, {Kehl}, {Keitel}, {Kelley}, {Kells},
  {Kennedy}, {Key}, {Khalaidovski}, {Khalili}, {Khan}, {Khan}, {Khazanov},
  {Kijbunchoo}, {Kim}, {Kim}, {Kim}, {Kim}, {Kim}, {Kim}, {King}, {King},
  {Kinzel}, {Kissel}, {Kleybolte}, {Klimenko}, {Koehlenbeck}, {Kokeyama},
  {Koley}, {Kondrashov}, {Kontos}, {Korobko}, {Korth}, {Kowalska}, {Kozak},
  {Kringel}, {Krishnan}, {Kr{\'o}lak}, {Krueger}, {Kuehn}, {Kumar}, {Kuo},
  {Kutynia}, {Lackey}, {Landry}, {Lange}, {Lantz}, {Lasky}, {Lazzarini},
  {Lazzaro}, {Leaci}, {Leavey}, {Lebigot}, {Lee}, {Lee}, {Lee}, {Lee}, {Lenon},
  {Leonardi}, {Leong}, {Leroy}, {Letendre}, {Levin}, {Levine}, {Li}, {Libson},
  {Littenberg}, {Lockerbie}, {Logue}, {Lombardi}, {Lord}, {Lorenzini},
  {Loriette}, {Lormand}, {Losurdo}, {Lough}, {L{\"u}ck}, {Lundgren}, {Luo},
  {Lynch}, {Ma}, {MacDonald}, {Machenschalk}, {MacInnis}, {Macleod},
  {Magana-Sandoval}, {Magee}, {Mageswaran}, {Majorana}, {Maksimovic},
  {Malvezzi}, {Man}, {Mandel}, {Mandic}, {Mangano}, {Mansell}, {Manske},
  {Mantovani}, {Marchesoni}, {Marion}, {M{\'a}rka}, {M{\'a}rka}, {Markosyan},
  {Maros}, {Martelli}, {Martellini}, {Martin}, {Martin}, {Martynov}, {Marx},
  {Mason}, {Masserot}, {Massinger}, {Masso-Reid}, {Matichard}, {Matone},
  {Mavalvala}, {Mazumder}, {Mazzolo}, {McCarthy}, {McClelland}, {McCormick},
  {McGuire}, {McIntyre}, {McIver}, {McManus}, {McWilliams}, {Meacher},
  {Meadors}, {Meidam}, {Melatos}, {Mendell}, {Mendoza-Gandara}, {Mercer},
  {Merilh}, {Merzougui}, {Meshkov}, {Messenger}, {Messick}, {Meyers},
  {Mezzani}, {Miao}, {Michel}, {Middleton}, {Mikhailov}, {Milano}, {Miller},
  {Millhouse}, {Minenkov}, {Ming}, {Mirshekari}, {Mishra}, {Mitra},
  {Mitrofanov}, {Mitselmakher}, {Mittleman}, {Moggi}, {Mohan}, {Mohapatra},
  {Montani}, {Moore}, {Moore}, {Moraru}, {Moreno}, {Morriss}, {Mossavi},
  {Mours}, {Mow-Lowry}, {Mueller}, {Mueller}, {Muir}, {Mukherjee}, {Mukherjee},
  {Mukherjee}, {Mullavey}, {Munch}, {Murphy}, {Murray}, {Mytidis},
  {Nardecchia}, {Naticchioni}, {Nayak}, {Necula}, {Nedkova}, {Nelemans},
  {Neri}, {Neunzert}, {Newton}, {Nguyen}, {Nielsen}, {Nissanke}, {Nitz},
  {Nocera}, {Nolting}, {Normandin}, {Nuttall}, {Oberling}, {Ochsner}, {O'Dell},
  {Oelker}, {Ogin}, {Oh}, {Oh}, {Ohme}, {Oliver}, {Oppermann}, {Oram},
  {O'Reilly}, {O'Shaughnessy}, {Ott}, {Ottaway}, {Ottens}, {Overmier}, {LIGO
  Scientific Collaboration}, \& {Virgo Collaboration}}]{2016LRR....19....1A}
---. 2016{\natexlab{c}},
  \href{http://dx.doi.org/10.1007/lrr-2016-1}{\JournalTitle{Living Reviews in
  Relativity}, 19, 1}

\bibitem[{{Abbott} {et~al.}(2017{\natexlab{a}}){Abbott}, {Abbott}, {Abbott},
  {Acernese}, {Ackley}, {Adams}, {Adams}, {Addesso}, {Adhikari}, {Adya},
  {Affeldt}, {Afrough}, {Agarwal}, {Agathos}, {Agatsuma}, {Aggarwal}, {Aguiar},
  {Aiello}, {Ain}, {Ajith}, {Allen}, {Allen}, {Allocca}, {Altin}, {Amato},
  {Ananyeva}, {Anderson}, {Anderson}, {Angelova}, {Antier}, {Appert}, {Arai},
  {Araya}, {Areeda}, {Arnaud}, {Arun}, {Ascenzi}, {Ashton}, {Ast}, {Aston},
  {Astone}, {Atallah}, {Aufmuth}, {Aulbert}, {Aultoneal}, {Austin},
  {Avila-Alvarez}, {Babak}, {Bacon}, {Bader}, {Bae}, {Baker}, {Baldaccini},
  {Ballardin}, {Ballmer}, {Banagiri}, {Barayoga}, {Barclay}, {Barish},
  {Barker}, {Barkett}, {Barone}, {Barr}, {Barsotti}, {Barsuglia}, {Barta},
  {Bartlett}, {Bartos}, {Bassiri}, {Basti}, {Batch}, {Bawaj}, {Bayley},
  {Bazzan}, {B{\'e}csy}, {Beer}, {Bejger}, {Belahcene}, {Bell}, {Berger},
  {Bergmann}, {Bero}, {Berry}, {Bersanetti}, {Bertolini}, {Betzwieser},
  {Bhagwat}, {Bhandare}, {Bilenko}, {Billingsley}, {Billman}, {Birch},
  {Birney}, {Birnholtz}, {Biscans}, {Biscoveanu}, {Bisht}, {Bitossi}, {Biwer},
  {Bizouard}, {Blackburn}, {Blackman}, {Blair}, {Blair}, {Blair}, {Bloemen},
  {Bock}, {Bode}, {Boer}, {Bogaert}, {Bohe}, {Bondu}, {Bonilla}, {Bonnand},
  {Boom}, {Bork}, {Boschi}, {Bose}, {Bossie}, {Bouffanais}, {Bozzi},
  {Bradaschia}, {Brady}, {Branchesi}, {Brau}, {Briant}, {Brillet}, {Brinkmann},
  {Brisson}, {Brockill}, {Broida}, {Brooks}, {Brown}, {Brown}, {Brunett},
  {Buchanan}, {Buikema}, {Bulik}, {Bulten}, {Buonanno}, {Buskulic}, {Buy},
  {Byer}, {Cabero}, {Cadonati}, {Cagnoli}, {Cahillane}, {Bustillo},
  {Callister}, {Calloni}, {Camp}, {Canepa}, {Canizares}, {Cannon}, {Cao},
  {Cao}, {Capano}, {Capocasa}, {Carbognani}, {Caride}, {Carney}, {Diaz},
  {Casentini}, {Caudill}, {Cavagli{\`a}}, {Cavalier}, {Cavalieri}, {Cella},
  {Cepeda}, {Cerd{\'a}-Dur{\'a}n}, {Cerretani}, {Cesarini}, {Chamberlin},
  {Chan}, {Chao}, {Charlton}, {Chase}, {Chassand e-Mottin}, {Chatterjee},
  {Chatziioannou}, {Cheeseboro}, {Chen}, {Chen}, {Chen}, {Cheng}, {Chia},
  {Chincarini}, {Chiummo}, {Chmiel}, {Cho}, {Cho}, {Chow}, {Christensen},
  {Chu}, {Chua}, {Chua}, {Chung}, {Chung}, {Ciani}, {Ciolfi}, {Cirelli},
  {Cirone}, {Clara}, {Clark}, {Clearwater}, {Cleva}, {Cocchieri}, {Coccia},
  {Cohadon}, {Cohen}, {Colla}, {Collette}, {Cominsky}, {Constancio}, {Conti},
  {Cooper}, {Corban}, {Corbitt}, {Cordero-Carri{\'o}n}, {Corley}, {Cornish},
  {Corsi}, {Cortese}, {Costa}, {Coughlin}, {Coughlin}, {Coulon}, {Countryman},
  {Couvares}, {Covas}, {Cowan}, {Coward}, {Cowart}, {Coyne}, {Coyne},
  {Creighton}, {Creighton}, {Cripe}, {Crowder}, {Cullen}, {Cumming},
  {Cunningham}, {Cuoco}, {Dal Canton}, {D{\'a}lya}, {Danilishin}, {D'Antonio},
  {Danzmann}, {Dasgupta}, {da Silva Costa}, {Datrier}, {Dattilo}, {Dave},
  {Davier}, {Davis}, {Daw}, {Day}, {de}, {Debra}, {Degallaix}, {de Laurentis},
  {Del{\'e}glise}, {Del Pozzo}, {Demos}, {Denker}, {Dent}, {de Pietri},
  {Dergachev}, {De Rosa}, {Derosa}, {de Rossi}, {Desalvo}, {de Varona},
  {Devenson}, {Dhurandhar}, {D{\'\i}az}, {di Fiore}, {di Giovanni}, {di
  Girolamo}, {di Lieto}, {di Pace}, {di Palma}, {di Renzo}, {Doctor},
  {Dolique}, {Donovan}, {Dooley}, {Doravari}, {Dorrington}, {Douglas}, {Dovale
  {\'A}lvarez}, {Downes}, {Drago}, {Dreissigacker}, {Driggers}, {Du}, {Ducrot},
  {Dupej}, {Dwyer}, {Edo}, {Edwards}, {Effler}, {Eggenstein}, {Ehrens},
  {Eichholz}, {Eikenberry}, {Eisenstein}, {Essick}, {Estevez}, {Etienne},
  {Etzel}, {Evans}, {Evans}, {Factourovich}, {Fafone}, {Fair}, {Fairhurst},
  {Fan}, {Farinon}, {Farr}, {Farr}, {Fauchon-Jones}, {Favata}, {Fays}, {Fee},
  {Fehrmann}, {Feicht}, {Fejer}, {Fernandez-Galiana}, {Ferrante}, {Ferreira},
  {Ferrini}, {Fidecaro}, {Finstad}, {Fiori}, {Fiorucci}, {Fishbach}, {Fisher},
  {Fitz-Axen}, {Flaminio}, {Fletcher}, {Fong}, {Font}, {Forsyth}, {Forsyth},
  {Fournier}, {Frasca}, {Frasconi}, {Frei}, {Freise}, {Frey}, {Frey}, {Fries},
  {Fritschel}, {Frolov}, {Fulda}, {Fyffe}, {Gabbard}, {Gadre}, {Gaebel},
  {Gair}, {Gammaitoni}, {Ganija}, {Gaonkar}, {Garcia-Quiros}, {Garufi},
  {Gateley}, {Gaudio}, {Gaur}, {Gayathri}, {Gehrels}, {Gemme}, {Genin},
  {Gennai}, {George}, {George}, {Gergely}, {Germain}, {Ghonge}, {Ghosh},
  {Ghosh}, {Ghosh}, {Giaime}, {Giardina}, {Giazotto}, {Gill}, {Glover},
  {Goetz}, {Goetz}, {Gomes}, {Goncharov}, {Gonz{\'a}lez}, {Castro},
  {Gopakumar}, {Gorodetsky}, {Gossan}, {Gosselin}, {Gouaty}, {Grado}, {Graef},
  {Granata}, {Grant}, {Gras}, {Gray}, {Greco}, {Green}, {Gretarsson}, {Groot},
  {Grote}, {Grunewald}, {Gruning}, {Guidi}, {Guo}, {Gupta}, {Gupta}, {Gushwa},
  {Gustafson}, {Gustafson}, {Halim}, {Hall}, {Hall}, {Hamilton}, {Hammond},
  {Haney}, {Hanke}, {Hanks}, {Hanna}, {Hannam}, {Hannuksela}, {Hanson},
  {Hardwick}, {Harms}, {Harry}, {Harry}, {Hart}, {Haster}, {Haughian}, {Healy},
  {Heidmann}, {Heintze}, {Heitmann}, {Hello}, {Hemming}, {Hendry}, {Heng},
  {Hennig}, {Heptonstall}, {Heurs}, {Hild}, {Hinderer}, {Hoak}, {Hofman},
  {Holt}, {Holz}, {Hopkins}, {Horst}, {Hough}, {Houston}, {Howell}, {Hreibi},
  {Hu}, {Huerta}, {Huet}, {Hughey}, {Husa}, {Huttner}, {Huynh-Dinh}, {Indik},
  {Inta}, {Intini}, {Isa}, {Isac}, {Isi}, {Iyer}, {Izumi}, {Jacqmin}, {Jani},
  {Jaranowski}, {Jawahar}, {Jim{\'e}nez-Forteza}, {Johnson}, {Jones}, {Jones},
  {Jonker}, {Ju}, {Junker}, {Kalaghatgi}, {Kalogera}, {Kamai}, {Kand hasamy},
  {Kang}, {Kanner}, {Kapadia}, {Karki}, {Karvinen}, {Kasprzack}, {Katolik},
  {Katsavounidis}, {Katzman}, {Kaufer}, {Kawabe}, {K{\'e}f{\'e}lian}, {Keitel},
  {Kemball}, {Kennedy}, {Kent}, {Key}, {Khalili}, {Khan}, {Khan}, {Khan},
  {Khazanov}, {Kijbunchoo}, {Kim}, {Kim}, {Kim}, {Kim}, {Kim}, {Kim},
  {Kimbrell}, {King}, {King}, {Kinley-Hanlon}, {Kirchhoff}, {Kissel},
  {Kleybolte}, {Klimenko}, {Knowles}, {Koch}, {Koehlenbeck}, {Koley},
  {Kondrashov}, {Kontos}, {Korobko}, {Korth}, {Kowalska}, {Kozak},
  {Kr{\"a}mer}, {Kringel}, {Krishnan}, {Kr{\'o}lak}, {Kuehn}, {Kumar}, {Kumar},
  {Kumar}, {Kuo}, {Kutynia}, {Kwang}, {Lackey}, {Lai}, {Landry}, {Lang},
  {Lange}, {Lantz}, {Lanza}, {Lartaux-Vollard}, {Lasky}, {Laxen}, {Lazzarini},
  {Lazzaro}, {Leaci}, {Leavey}, {Lee}, {Lee}, {Lee}, {Lee}, {Lee}, {Lehmann},
  {Lenon}, {Leonardi}, {Leroy}, {Letendre}, {Levin}, {Li}, {Linker},
  {Littenberg}, {Liu}, {Liu}, {Lo}, {Lockerbie}, {London}, {Lord}, {Lorenzini},
  {Loriette}, {Lormand}, {Losurdo}, {Lough}, {Lousto}, {Lovelace}, {L{\"u}ck},
  {Lumaca}, {Lundgren}, {Lynch}, {Ma}, {Macas}, {Macfoy}, {Machenschalk},
  {Macinnis}, {MacLeod}, {Hernandez}, {Maga{\~n}a-Sand oval}, {Zertuche},
  {Magee}, {Majorana}, {Maksimovic}, {Man}, {Mandic}, {Mangano}, {Mansell},
  {Manske}, {Mantovani}, {Marchesoni}, {Marion}, {M{\'a}rka}, {M{\'a}rka},
  {Markakis}, {Markosyan}, {Markowitz}, {Maros}, {Marquina}, {Martelli},
  {Martellini}, {Martin}, {Martin}, {Martynov}, {Mason}, {Massera}, {Masserot},
  {Massinger}, {Masso-Reid}, {Mastrogiovanni}, {Matas}, {Matichard}, {Matone},
  {Mavalvala}, {Mazumder}, {McCarthy}, {McClelland}, {McCormick}, {McCuller},
  {McGuire}, {McIntyre}, {McIver}, {McManus}, {McNeill}, {McRae}, {McWilliams},
  {Meacher}, {Meadors}, {Mehmet}, {Meidam}, {Mejuto-Villa}, {Melatos},
  {Mendell}, {Mercer}, {Merilh}, {Merzougui}, {Meshkov}, {Messenger},
  {Messick}, {Metzdorff}, {Meyers}, {Miao}, {Michel}, {Middleton}, {Mikhailov},
  {Milano}, {Miller}, {Miller}, {Miller}, {Millhouse}, {Milovich-Goff},
  {Minazzoli}, {Minenkov}, {Ming}, {Mishra}, {Mitra}, {Mitrofanov},
  {Mitselmakher}, {Mittleman}, {Moffa}, {Moggi}, {Mogushi}, {Mohan},
  {Mohapatra}, {Montani}, {Moore}, {Moraru}, {Moreno}, {Morriss}, {Mours},
  {Mow-Lowry}, {Mueller}, {Muir}, {Mukherjee}, {Mukherjee}, {Mukherjee},
  {Mukund}, {Mullavey}, {Munch}, {Mu{\~n}iz}, {Muratore}, {Murray}, {Napier},
  {Nardecchia}, {Naticchioni}, {Nayak}, {Neilson}, {Nelemans}, {Nelson},
  {Nery}, {Neunzert}, {Nevin}, {Newport}, {Newton}, {Ng}, {Nguyen}, {Nichols},
  {Nielsen}, {Nissanke}, {Nitz}, {Noack}, {Nocera}, {Nolting}, {North},
  {Nuttall}, {Oberling}, {O'Dea}, {Ogin}, {Oh}, {Oh}, {Ohme}, {Okada},
  {Oliver}, {Oppermann}, {Oram}, {O'Reilly}, {Ormiston}, {Ortega},
  {O'Shaughnessy}, {Ossokine}, {Ottaway}, {Overmier}, {Owen}, {Pace}, {Page},
  {Page}, {Pai}, {Pai}, {Palamos}, {Palashov}, {Palomba}, {Pal-Singh}, {Pan},
  {Pan}, {Pang}, {Pang}, {Pankow}, {Pannarale}, {Pant}, {Paoletti}, {Paoli},
  {Papa}, {Parida}, {Parker}, {Pascucci}, {Pasqualetti}, {Passaquieti},
  {Passuello}, {Patil}, {Patricelli}, {Pearlstone}, {Pedraza}, {Pedurand},
  {Pekowsky}, {Pele}, {Penn}, {Perez}, {Perreca}, {Perri}, {Pfeiffer},
  {Phelps}, {Piccinni}, {Pichot}, {Piergiovanni}, {Pierro}, {Pillant},
  {Pinard}, {Pinto}, {Pirello}, {Pitkin}, {Poe}, {Poggiani}, {Popolizio},
  {Porter}, {Post}, {Powell}, {Prasad}, {Pratt}, {Pratten}, {Predoi},
  {Prestegard}, {Prijatelj}, {Principe}, {Privitera}, {Prodi}, {Prokhorov},
  {Puncken}, {Punturo}, {Puppo}, {P{\"u}rrer}, {Qi}, {Quetschke}, {Quintero},
  {Quitzow-James}, {Raab}, {Rabeling}, {Radkins}, {Raffai}, {Raja}, {Rajan},
  {Rajbhandari}, {Rakhmanov}, {Ramirez}, {Ramos-Buades}, {Rapagnani},
  {Raymond}, {Razzano}, {Read}, {Regimbau}, {Rei}, {Reid}, {Reitze}, {Ren},
  {Reyes}, {Ricci}, {Ricker}, {Rieger}, {Riles}, {Rizzo}, {Robertson}, {Robie},
  {Robinet}, {Rocchi}, {Rolland}, {Rollins}, {Roma}, {Romano}, {Romano},
  {Romel}, {Romie}, {Rosi{\'n}ska}, {Ross}, {Rowan}, {R{\"u}diger}, {Ruggi},
  {Rutins}, {Ryan}, {Sachdev}, {Sadecki}, {Sadeghian}, {Sakellariadou},
  {Salconi}, {Saleem}, {Salemi}, {Samajdar}, {Sammut}, {Sampson}, {Sanchez},
  {Sanchez}, {Sanchis-Gual}, {Sand berg}, {Sanders}, {Sassolas},
  {Sathyaprakash}, {Saulson}, {Sauter}, {Savage}, {Sawadsky}, {Schale},
  {Scheel}, {Scheuer}, {Schmidt}, {Schmidt}, {Schnabel}, {Schofield},
  {Sch{\"o}nbeck}, {Schreiber}, {Schuette}, {Schulte}, {Schutz}, {Schwalbe},
  {Scott}, {Scott}, {Seidel}, {Sellers}, {Sengupta}, {Sentenac}, {Sequino},
  {Sergeev}, {Shaddock}, {Shaffer}, {Shah}, {Shahriar}, {Shaner}, {Shao},
  {Shapiro}, {Shawhan}, {Sheperd}, {Shoemaker}, {Shoemaker}, {Siellez},
  {Siemens}, {Sieniawska}, {Sigg}, {Silva}, {Singer}, {Singh}, {Singhal},
  {Sintes}, {Slagmolen}, {Smith}, {Smith}, {Smith}, {Somala}, {Son},
  {Sonnenberg}, {Sorazu}, {Sorrentino}, {Souradeep}, {Spencer}, {Srivastava},
  {Staats}, {Staley}, {Steer}, {Steinke}, {Steinlechner}, {Steinlechner},
  {Steinmeyer}, {Stevenson}, {Stone}, {Stops}, {Strain}, {Stratta}, {Strigin},
  {Strunk}, {Sturani}, {Stuver}, {Summerscales}, {Sun}, {Sunil}, {Suresh},
  {Sutton}, {Swinkels}, {Szczepa{\'n}czyk}, {Tacca}, {Tait}, {Talbot},
  {Talukder}, {Tanner}, {T{\'a}pai}, {Taracchini}, {Tasson}, {Taylor},
  {Taylor}, {Tewari}, {Theeg}, {Thies}, {Thomas}, {Thomas}, {Thomas}, {Thorne},
  {Thrane}, {Tiwari}, {Tiwari}, {Tokmakov}, {Toland}, {Tonelli}, {Tornasi},
  {Torres-Forn{\'e}}, {Torrie}, {T{\"o}yr{\"a}}, {Travasso}, {Traylor},
  {Trinastic}, {Tringali}, {Trozzo}, {Tsang}, {Tse}, {Tso}, {Tsukada}, {Tsuna},
  {Tuyenbayev}, {Ueno}, {Ugolini}, {Unnikrishnan}, {Urban}, {Usman},
  {Vahlbruch}, {Vajente}, {Valdes}, {van Bakel}, {van Beuzekom}, {van den
  Brand}, {van den Broeck}, {Vander-Hyde}, {van der Schaaf}, {van Heijningen},
  {van Veggel}, {Vardaro}, {Varma}, {Vass}, {Vas{\'u}th}, {Vecchio},
  {Vedovato}, {Veitch}, {Veitch}, {Venkateswara}, {Venugopalan}, {Verkindt},
  {Vetrano}, {Vicer{\'e}}, {Viets}, {Vinciguerra}, {Vine}, {Vinet}, {Vitale},
  {Vo}, {Vocca}, {Vorvick}, {Vyatchanin}, {Wade}, {Wade}, {Wade}, {Walet},
  {Walker}, {Wallace}, {Walsh}, {Wang}, {Wang}, {Wang}, {Wang}, {Wang}, {Ward},
  {Warner}, {Was}, {Watchi}, {Weaver}, {Wei}, {Weinert}, {Weinstein}, {Weiss},
  {Wen}, \& {Wessel}}]{2017Natur.551...85A}
---. 2017{\natexlab{a}},
  \href{http://dx.doi.org/10.1038/nature24471}{\JournalTitle{\nat}, 551, 85}

\bibitem[{{Abbott} {et~al.}(2017{\natexlab{b}}){Abbott}, {Abbott}, {Abbott},
  {Abernathy}, {Ackley}, {Adams}, {Addesso}, {Adhikari}, {Adya}, {Affeldt},
  {Aggarwal}, {Aguiar}, {Ain}, {Ajith}, {Allen}, {Altin}, {Anderson},
  {Anderson}, {Arai}, {Araya}, {Arceneaux}, {Areeda}, {Arun}, {Ashton}, {Ast},
  {Aston}, {Aufmuth}, {Aulbert}, {Babak}, {Baker}, {Ballmer}, {Barayoga},
  {Barclay}, {Barish}, {Barker}, {Barr}, {Barsotti}, {Bartlett}, {Bartos},
  {Bassiri}, {Batch}, {Baune}, {Bell}, {Berger}, {Bergmann}, {Berry},
  {Betzwieser}, {Bhagwat}, {Bhand are}, {Bilenko}, {Billingsley}, {Birch},
  {Birney}, {Biscans}, {Bisht}, {Biwer}, {Blackburn}, {Blair}, {Blair},
  {Blair}, {Bock}, {Bogan}, {Bohe}, {Bond}, {Bork}, {Bose}, {Brady},
  {Braginsky}, {Brau}, {Brinkmann}, {Brockill}, {Broida}, {Brooks}, {Brown},
  {Brown}, {Brown}, {Brunett}, {Buchanan}, {Buikema}, {Buonanno}, {Byer},
  {Cabero}, {Cadonati}, {Cahillane}, {Calder{\'o}n Bustillo}, {Callister},
  {Camp}, {Cannon}, {Cao}, {Capano}, {Caride}, {Caudill}, {Cavagli{\`a}},
  {Cepeda}, {Chamberlin}, {Chan}, {Chao}, {Charlton}, {Cheeseboro}, {Chen},
  {Chen}, {Cheng}, {Cho}, {Cho}, {Chow}, {Christensen}, {Chu}, {Chung},
  {Ciani}, {Clara}, {Clark}, {Collette}, {Cominsky}, {Constancio}, {Cook},
  {Corbitt}, {Cornish}, {Corsi}, {Costa}, {Coughlin}, {Coughlin}, {Countryman},
  {Couvares}, {Cowan}, {Coward}, {Cowart}, {Coyne}, {Coyne}, {Craig},
  {Creighton}, {Cripe}, {Crowder}, {Cumming}, {Cunningham}, {Dal Canton},
  {Danilishin}, {Danzmann}, {Darman}, {Dasgupta}, {Da Silva Costa}, {Dave},
  {Davies}, {Daw}, {De}, {DeBra}, {Del Pozzo}, {Denker}, {Dent}, {Dergachev},
  {DeRosa}, {DeSalvo}, {Devine}, {Dhurand har}, {D{\'\i}az}, {Di Palma},
  {Donovan}, {Dooley}, {Doravari}, {Douglas}, {Downes}, {Drago}, {Drever},
  {Driggers}, {Dwyer}, {Edo}, {Edwards}, {Effler}, {Eggenstein}, {Ehrens},
  {Eichholz}, {Eikenberry}, {Engels}, {Essick}, {Etzel}, {Evans}, {Evans},
  {Everett}, {Factourovich}, {Fair}, {Fairhurst}, {Fan}, {Fang}, {Farr},
  {Farr}, {Favata}, {Fays}, {Fehrmann}, {Fejer}, {Fenyvesi}, {Ferreira},
  {Fisher}, {Fletcher}, {Frei}, {Freise}, {Frey}, {Fritschel}, {Frolov},
  {Fulda}, {Fyffe}, {Gabbard}, {Gair}, {Gaonkar}, {Gaur}, {Gehrels}, {Geng},
  {George}, {Gergely}, {Ghosh}, {Ghosh}, {Giaime}, {Giardina}, {Gill},
  {Glaefke}, {Goetz}, {Goetz}, {Gondan}, {Gonz{\'a}lez}, {Gopakumar}, {Gordon},
  {Gorodetsky}, {Gossan}, {Graef}, {Graff}, {Grant}, {Gras}, {Gray}, {Green},
  {Grote}, {Grunewald}, {Guo}, {Gupta}, {Gupta}, {Gushwa}, {Gustafson},
  {Gustafson}, {Hacker}, {Hall}, {Hall}, {Hammond}, {Haney}, {Hanke}, {Hanks},
  {Hanna}, {Hannam}, {Hanson}, {Hardwick}, {Harry}, {Harry}, {Hart}, {Hartman},
  {Haster}, {Haughian}, {Heintze}, {Hendry}, {Heng}, {Hennig}, {Henry},
  {Heptonstall}, {Heurs}, {Hild}, {Hoak}, {Holt}, {Holz}, {Hopkins}, {Hough},
  {Houston}, {Howell}, {Hu}, {Huang}, {Huerta}, {Hughey}, {Husa}, {Huttner},
  {Huynh-Dinh}, {Indik}, {Ingram}, {Inta}, {Isa}, {Isi}, {Isogai}, {Iyer},
  {Izumi}, {Jang}, {Jani}, {Jawahar}, {Jian}, {Jim{\'e}nez-Forteza}, {Johnson},
  {Jones}, {Jones}, {Ju}, {Haris}, {Kalaghatgi}, {Kalogera}, {Kandhasamy},
  {Kang}, {Kanner}, {Kapadia}, {Karki}, {Karvinen}, {Kasprzack},
  {Katsavounidis}, {Katzman}, {Kaufer}, {Kaur}, {Kawabe}, {Kehl}, {Keitel},
  {Kelley}, {Kells}, {Kennedy}, {Key}, {Khalili}, {Khan}, {Khan}, {Khazanov},
  {Kijbunchoo}, {Kim}, {Kim}, {Kim}, {Kim}, {Kim}, {Kim}, {Kim}, {Kimbrell},
  {King}, {King}, {Kissel}, {Klein}, {Kleybolte}, {Klimenko}, {Koehlenbeck},
  {Kondrashov}, {Kontos}, {Korobko}, {Korth}, {Kozak}, {Kringel}, {Krueger},
  {Kuehn}, {Kumar}, {Kumar}, {Kuo}, {Lackey}, {Landry}, {Lange}, {Lantz},
  {Lasky}, {Laxen}, {Lazzarini}, {Leavey}, {Lebigot}, {Lee}, {Lee}, {Lee},
  {Lee}, {Lenon}, {Leong}, {Levin}, {Lewis}, {Li}, {Libson}, {Littenberg},
  {Lockerbie}, {Lombardi}, {London}, {Lord}, {Lormand }, {Lough}, {L{\"u}ck},
  {Lundgren}, {Lynch}, {Ma}, {Machenschalk}, {MacInnis}, {Macleod},
  {Maga{\~n}a-Sandoval}, {Maga{\~n}a Zertuche}, {Magee}, {Mandic}, {Mangano},
  {Mansell}, {Manske}, {M{\'a}rka}, {M{\'a}rka}, {Markosyan}, {Maros},
  {Martin}, {Martynov}, {Mason}, {Massinger}, {Masso-Reid}, {Matichard},
  {Matone}, {Mavalvala}, {Mazumder}, {McCarthy}, {McClelland}, {McCormick},
  {McGuire}, {McIntyre}, {McIver}, {McManus}, {McRae}, {McWilliams}, {Meacher},
  {Meadors}, {Melatos}, {Mendell}, {Mercer}, {Merilh}, {Meshkov}, {Messenger},
  {Messick}, {Meyers}, {Miao}, {Middleton}, {Mikhailov}, {Miller}, {Miller},
  {Miller}, {Miller}, {Millhouse}, {Ming}, {Mirshekari}, {Mishra}, {Mitra},
  {Mitrofanov}, {Mitselmakher}, {Mittleman}, {Mohapatra}, {Moore}, {Moore},
  {Moraru}, {Moreno}, {Morriss}, {Mossavi}, {Mow-Lowry}, {Mueller}, {Muir},
  {Mukherjee}, {Mukherjee}, {Mukherjee}, {Mukund}, {Mullavey}, {Munch},
  {Murphy}, {Murray}, {Mytidis}, {Nayak}, {Nedkova}, {Nelson}, {Neunzert},
  {Newton}, {Nguyen}, {Nielsen}, {Nitz}, {Nolting}, {Normandin}, {Nuttall},
  {Oberling}, {Ochsner}, {O'Dell}, {Oelker}, {Ogin}, {Oh}, {Oh}, {Ohme},
  {Oliver}, {Oppermann}, {Oram}, {O'Reilly}, {O'Shaughnessy}, {Ottaway},
  {Overmier}, {Owen}, {Pai}, {Pai}, {Palamos}, {Palashov}, {Pal-Singh}, {Pan},
  {Pankow}, {Pannarale}, {Pant}, {Papa}, {Paris}, {Parker}, {Pascucci},
  {Patrick}, {Pearlstone}, {Pedraza}, {Pekowsky}, {Pele}, {Penn}, {Perreca},
  {Perri}, {Phelps}, {Pierro}, {Pinto}, {Pitkin}, {Poe}, {Post}, {Powell},
  {Prasad}, {Predoi}, {Prestegard}, {Price}, {Prijatelj}, {Principe},
  {Privitera}, {Prokhorov}, {Puncken}, {P{\"u}rrer}, {Qi}, {Qin}, {Qiu},
  {Quetschke}, {Quintero}, {Quitzow-James}, {Raab}, {Rabeling}, {Radkins},
  {Raffai}, {Raja}, {Rajan}, {Rakhmanov}, {Raymond}, {Read}, {Reed}, {Reid},
  {Reitze}, {Rew}, {Reyes}, {Riles}, {Rizzo}, {Robertson}, {Robie}, {Rollins},
  {Roma}, {Romanov}, {Romie}, {Rowan}, {R{\"u}diger}, {Ryan}, {Sachdev},
  {Sadecki}, {Sadeghian}, {Sakellariadou}, {Saleem}, {Salemi}, {Samajdar},
  {Sammut}, {Sanchez}, {Sand berg}, {Sandeen}, {Sanders}, {Sathyaprakash},
  {Saulson}, {Sauter}, {Savage}, {Sawadsky}, {Schale}, {Schilling}, {Schmidt},
  {Schmidt}, {Schnabel}, {Schofield}, {Sch{\"o}nbeck}, {Schreiber}, {Schuette},
  {Schutz}, {Scott}, {Scott}, {Sellers}, {Sengupta}, {Sergeev}, {Shaddock},
  {Shaffer}, {Shahriar}, {Shaltev}, {Shapiro}, {Shawhan}, {Sheperd},
  {Shoemaker}, {Shoemaker}, {Siellez}, {Siemens}, {Sigg}, {Silva}, {Singer},
  {Singer}, {Singh}, {Singh}, {Sintes}, {Slagmolen}, {Smith}, {Smith}, {Smith},
  {Son}, {Sorazu}, {Souradeep}, {Srivastava}, {Staley}, {Steinke},
  {Steinlechner}, {Steinlechner}, {Steinmeyer}, {Stephens}, {Stone}, {Strain},
  {Strauss}, {Strigin}, {Sturani}, {Stuver}, {Summerscales}, {Sun}, {Sunil},
  {Sutton}, {Szczepa{\'n}czyk}, {Talukder}, {Tanner}, {T{\'a}pai}, {Tarabrin},
  {Taracchini}, {Taylor}, {Theeg}, {Thirugnanasambandam}, {Thomas}, {Thomas},
  {Thomas}, {Thorne}, {Thrane}, {Tiwari}, {Tokmakov}, {Toland}, {Tomlinson},
  {Tornasi}, {Torres}, {Torrie}, {T{\"o}yr{\"a}}, {Traylor}, {Trifir{\`o}},
  {Tse}, {Tuyenbayev}, {Ugolini}, {Unnikrishnan}, {Urban}, {Usman},
  {Vahlbruch}, {Vajente}, {Valdes}, {Vand er-Hyde}, {van Veggel}, {Vass},
  {Vaulin}, {Vecchio}, {Veitch}, {Veitch}, {Venkateswara}, {Vinciguerra},
  {Vine}, {Vitale}, {Vo}, {Vorvick}, {Voss}, {Vousden}, {Vyatchanin}, {Wade},
  {Wade}, {Wade}, {Walker}, {Wallace}, {Walsh}, {Wang}, {Wang}, {Wang}, {Wang},
  {Ward}, {Warner}, {Weaver}, {Weinert}, {Weinstein}, {Weiss}, {Wen},
  {We{\ss}els}, {Westphal}, {Wette}, {Whelan}, {Whiting}, {Williams},
  {Williamson}, {Willis}, {Willke}, {Wimmer}, {Winkler}, {Wipf}, {Wittel},
  {Woan}, {Woehler}, {Worden}, {Wright}, {Wu}, {Wu}, {Yablon}, {Yam},
  {Yamamoto}, {Yancey}, {Yu}, {Zanolin}, {Zevin}, {Zhang}, {Zhang}, {Zhang},
  {Zhao}, {Zhou}, {Zhou}, {Zhu}, {Zucker}, {Zuraw}, {Zweizig}, {(LIGO
  Scientific Collaboration}, \& {Harms}}]{2017CQGra..34d4001A}
---. 2017{\natexlab{b}},
  \href{http://dx.doi.org/10.1088/1361-6382/aa51f4}{\JournalTitle{Classical and
  Quantum Gravity}, 34, 044001}

\bibitem[{{Abbott} {et~al.}(2017{\natexlab{c}}){Abbott}, {Abbott}, {Abbott},
  {Acernese}, {Ackley}, {Adams}, {Adams}, {Addesso}, {Adhikari}, {Adya},
  {Affeldt}, {Afrough}, {Agarwal}, {Agathos}, {Agatsuma}, {Aggarwal}, {Aguiar},
  {Aiello}, {Ain}, {Ajith}, {Allen}, {Allen}, {Allocca}, {Altin}, {Amato},
  {Ananyeva}, {Anderson}, {Anderson}, {Angelova}, {Antier}, {Appert}, {Arai},
  {Araya}, {Areeda}, {Arnaud}, {Arun}, {Ascenzi}, {Ashton}, {Ast}, {Aston},
  {Astone}, {Atallah}, {Aufmuth}, {Aulbert}, {AultONeal}, {Austin},
  {Avila-Alvarez}, {Babak}, {Bacon}, {Bader}, {Bae}, {Bailes}, {Baker},
  {Baldaccini}, {Ballardin}, {Ballmer}, {Banagiri}, {Barayoga}, {Barclay},
  {Barish}, {Barker}, {Barkett}, {Barone}, {Barr}, {Barsotti}, {Barsuglia},
  {Barta}, {Barthelmy}, {Bartlett}, {Bartos}, {Bassiri}, {Basti}, {Batch},
  {Bawaj}, {Bayley}, {Bazzan}, {B{\'e}csy}, {Beer}, {Bejger}, {Belahcene},
  {Bell}, {Berger}, {Bergmann}, {Bernuzzi}, {Bero}, {Berry}, {Bersanetti},
  {Bertolini}, {Betzwieser}, {Bhagwat}, {Bhandare}, {Bilenko}, {Billingsley},
  {Billman}, {Birch}, {Birney}, {Birnholtz}, {Biscans}, {Biscoveanu}, {Bisht},
  {Bitossi}, {Biwer}, {Bizouard}, {Blackburn}, {Blackman}, {Blair}, {Blair},
  {Blair}, {Bloemen}, {Bock}, {Bode}, {Boer}, {Bogaert}, {Bohe}, {Bondu},
  {Bonilla}, {Bonnand}, {Boom}, {Bork}, {Boschi}, {Bose}, {Bossie},
  {Bouffanais}, {Bozzi}, {Bradaschia}, {Brady}, {Branchesi}, {Brau}, {Briant},
  {Brillet}, {Brinkmann}, {Brisson}, {Brockill}, {Broida}, {Brooks}, {Brown},
  {Brown}, {Brunett}, {Buchanan}, {Buikema}, {Bulik}, {Bulten}, {Buonanno},
  {Buskulic}, {Buy}, {Byer}, {Cabero}, {Cadonati}, {Cagnoli}, {Cahillane},
  {Calder{\'o}n Bustillo}, {Callister}, {Calloni}, {Camp}, {Canepa},
  {Canizares}, {Cannon}, {Cao}, {Cao}, {Capano}, {Capocasa}, {Carbognani},
  {Caride}, {Carney}, {Carullo}, {Casanueva Diaz}, {Casentini}, {Caudill},
  {Cavagli{\`a}}, {Cavalier}, {Cavalieri}, {Cella}, {Cepeda},
  {Cerd{\'a}-Dur{\'a}n}, {Cerretani}, {Cesarini}, {Chamberlin}, {Chan}, {Chao},
  {Charlton}, {Chase}, {Chassande-Mottin}, {Chatterjee}, {Chatziioannou},
  {Cheeseboro}, {Chen}, {Chen}, {Chen}, {Cheng}, {Chia}, {Chincarini},
  {Chiummo}, {Chmiel}, {Cho}, {Cho}, {Chow}, {Christensen}, {Chu}, {Chua},
  {Chua}, {Chung}, {Chung}, {Ciani}, {Ciolfi}, {Cirelli}, {Cirone}, {Clara},
  {Clark}, {Clearwater}, {Cleva}, {Cocchieri}, {Coccia}, {Cohadon}, {Cohen},
  {Colla}, {Collette}, {Cominsky}, {Constancio}, {Conti}, {Cooper}, {Corban},
  {Corbitt}, {Cordero-Carri{\'o}n}, {Corley}, {Cornish}, {Corsi}, {Cortese},
  {Costa}, {Coughlin}, {Coughlin}, {Coulon}, {Countryman}, {Couvares}, {Covas},
  {Cowan}, {Coward}, {Cowart}, {Coyne}, {Coyne}, {Creighton}, {Creighton},
  {Cripe}, {Crowder}, {Cullen}, {Cumming}, {Cunningham}, {Cuoco}, {Dal Canton},
  {D{\'a}lya}, {Danilishin}, {D'Antonio}, {Danzmann}, {Dasgupta}, {Da Silva
  Costa}, {Dattilo}, {Dave}, {Davier}, {Davis}, {Daw}, {Day}, {De}, {DeBra},
  {Degallaix}, {De Laurentis}, {Del{\'e}glise}, {Del Pozzo}, {Demos}, {Denker},
  {Dent}, {De Pietri}, {Dergachev}, {De Rosa}, {DeRosa}, {De Rossi}, {DeSalvo},
  {de Varona}, {Devenson}, {Dhurandhar}, {D{\'\i}az}, {Dietrich}, {Di Fiore},
  {Di Giovanni}, {Di Girolamo}, {Di Lieto}, {Di Pace}, {Di Palma}, {Di Renzo},
  {Doctor}, {Dolique}, {Donovan}, {Dooley}, {Doravari}, {Dorrington},
  {Douglas}, {Dovale {\'A}lvarez}, {Downes}, {Drago}, {Dreissigacker},
  {Driggers}, {Du}, {Ducrot}, {Dudi}, {Dupej}, {Dwyer}, {Edo}, {Edwards},
  {Effler}, {Eggenstein}, {Ehrens}, {Eichholz}, {Eikenberry}, {Eisenstein},
  {Essick}, {Estevez}, {Etienne}, {Etzel}, {Evans}, {Evans}, {Factourovich},
  {Fafone}, {Fair}, {Fairhurst}, {Fan}, {Farinon}, {Farr}, {Farr},
  {Fauchon-Jones}, {Favata}, {Fays}, {Fee}, {Fehrmann}, {Feicht}, {Fejer},
  {Fernandez-Galiana}, {Ferrante}, {Ferreira}, {Ferrini}, {Fidecaro},
  {Finstad}, {Fiori}, {Fiorucci}, {Fishbach}, {Fisher}, {Fitz-Axen},
  {Flaminio}, {Fletcher}, {Fong}, {Font}, {Forsyth}, {Forsyth}, {Fournier},
  {Frasca}, {Frasconi}, {Frei}, {Freise}, {Frey}, {Frey}, {Fries}, {Fritschel},
  {Frolov}, {Fulda}, {Fyffe}, {Gabbard}, {Gadre}, {Gaebel}, {Gair},
  {Gammaitoni}, {Ganija}, {Gaonkar}, {Garcia-Quiros}, {Garufi}, {Gateley},
  {Gaudio}, {Gaur}, {Gayathri}, {Gehrels}, {Gemme}, {Genin}, {Gennai},
  {George}, {George}, {Gergely}, {Germain}, {Ghonge}, {Ghosh}, {Ghosh},
  {Ghosh}, {Giaime}, {Giardina}, {Giazotto}, {Gill}, {Glover}, {Goetz},
  {Goetz}, {Gomes}, {Goncharov}, {Gonz{\'a}lez}, {Gonzalez Castro},
  {Gopakumar}, {Gorodetsky}, {Gossan}, {Gosselin}, {Gouaty}, {Grado}, {Graef},
  {Granata}, {Grant}, {Gras}, {Gray}, {Greco}, {Green}, {Gretarsson}, {Groot},
  {Grote}, {Grunewald}, {Gruning}, {Guidi}, {Guo}, {Gupta}, {Gupta}, {Gushwa},
  {Gustafson}, {Gustafson}, {Halim}, {Hall}, {Hall}, {Hamilton}, {Hammond},
  {Haney}, {Hanke}, {Hanks}, {Hanna}, {Hannam}, {Hannuksela}, {Hanson},
  {Hardwick}, {Harms}, {Harry}, {Harry}, {Hart}, {Haster}, {Haughian}, {Healy},
  {Heidmann}, {Heintze}, {Heitmann}, {Hello}, {Hemming}, {Hendry}, {Heng},
  {Hennig}, {Heptonstall}, {Heurs}, {Hild}, {Hinderer}, {Ho}, {Hoak}, {Hofman},
  {Holt}, {Holz}, {Hopkins}, {Horst}, {Hough}, {Houston}, {Howell}, {Hreibi},
  {Hu}, {Huerta}, {Huet}, {Hughey}, {Husa}, {Huttner}, {Huynh-Dinh}, {Indik},
  {Inta}, {Intini}, {Isa}, {Isac}, {Isi}, {Iyer}, {Izumi}, {Jacqmin}, {Jani},
  {Jaranowski}, {Jawahar}, {Jim{\'e}nez-Forteza}, {Johnson},
  {Johnson-McDaniel}, {Jones}, {Jones}, {Jonker}, {Ju}, {Junker}, {Kalaghatgi},
  {Kalogera}, {Kamai}, {Kand hasamy}, {Kang}, {Kanner}, {Kapadia}, {Karki},
  {Karvinen}, {Kasprzack}, {Kastaun}, {Katolik}, {Katsavounidis}, {Katzman},
  {Kaufer}, {Kawabe}, {K{\'e}f{\'e}lian}, {Keitel}, {Kemball}, {Kennedy},
  {Kent}, {Key}, {Khalili}, {Khan}, {Khan}, {Khan}, {Khazanov}, {Kijbunchoo},
  {Kim}, {Kim}, {Kim}, {Kim}, {Kim}, {Kim}, {Kimbrell}, {King}, {King},
  {Kinley-Hanlon}, {Kirchhoff}, {Kissel}, {Kleybolte}, {Klimenko}, {Knowles},
  {Koch}, {Koehlenbeck}, {Koley}, {Kondrashov}, {Kontos}, {Korobko}, {Korth},
  {Kowalska}, {Kozak}, {Kr{\"a}mer}, {Kringel}, {Krishnan}, {Kr{\'o}lak},
  {Kuehn}, {Kumar}, {Kumar}, {Kumar}, {Kuo}, {Kutynia}, {Kwang}, {Lackey},
  {Lai}, {Landry}, {Lang}, {Lange}, {Lantz}, {Lanza}, {Larson},
  {Lartaux-Vollard}, {Lasky}, {Laxen}, {Lazzarini}, {Lazzaro}, {Leaci},
  {Leavey}, {Lee}, {Lee}, {Lee}, {Lee}, {Lee}, {Lehmann}, {Lenon}, {Leon},
  {Leonardi}, {Leroy}, {Letendre}, {Levin}, {Li}, {Linker}, {Littenberg},
  {Liu}, {Liu}, {Lo}, {Lockerbie}, {London}, {Lord}, {Lorenzini}, {Loriette},
  {Lormand}, {Losurdo}, {Lough}, {Lousto}, {Lovelace}, {L{\"u}ck}, {Lumaca},
  {Lundgren}, {Lynch}, {Ma}, {Macas}, {Macfoy}, {Machenschalk}, {MacInnis},
  {Macleod}, {Maga{\~n}a Hernandez}, {Maga{\~n}a-Sandoval}, {Maga{\~n}a
  Zertuche}, {Magee}, {Majorana}, {Maksimovic}, {Man}, {Mandic}, {Mangano},
  {Mansell}, {Manske}, {Mantovani}, {Marchesoni}, {Marion}, {M{\'a}rka},
  {M{\'a}rka}, {Markakis}, {Markosyan}, {Markowitz}, {Maros}, {Marquina},
  {Marsh}, {Martelli}, {Martellini}, {Martin}, {Martin}, {Martynov}, {Marx},
  {Mason}, {Massera}, {Masserot}, {Massinger}, {Masso-Reid}, {Mastrogiovanni},
  {Matas}, {Matichard}, {Matone}, {Mavalvala}, {Mazumder}, {McCarthy},
  {McClelland}, {McCormick}, {McCuller}, {McGuire}, {McIntyre}, {McIver},
  {McManus}, {McNeill}, {McRae}, {McWilliams}, {Meacher}, {Meadors}, {Mehmet},
  {Meidam}, {Mejuto-Villa}, {Melatos}, {Mendell}, {Mercer}, {Merilh},
  {Merzougui}, {Meshkov}, {Messenger}, {Messick}, {Metzdorff}, {Meyers},
  {Miao}, {Michel}, {Middleton}, {Mikhailov}, {Milano}, {Miller}, {Miller},
  {Miller}, {Millhouse}, {Milovich-Goff}, {Minazzoli}, {Minenkov}, {Ming},
  {Mishra}, {Mitra}, {Mitrofanov}, {Mitselmakher}, {Mittleman}, {Moffa},
  {Moggi}, {Mogushi}, {Mohan}, {Mohapatra}, {Molina}, {Montani}, {Moore},
  {Moraru}, {Moreno}, {Morisaki}, {Morriss}, {Mours}, {Mow-Lowry}, {Mueller},
  {Muir}, {Mukherjee}, {Mukherjee}, {Mukherjee}, {Mukund}, {Mullavey}, {Munch},
  {Mu{\~n}iz}, {Muratore}, {Murray}, {Nagar}, {Napier}, {Nardecchia},
  {Naticchioni}, {Nayak}, {Neilson}, {Nelemans}, {Nelson}, {Nery}, {Neunzert},
  {Nevin}, {Newport}, {Newton}, {Ng}, {Nguyen}, {Nguyen}, {Nichols}, {Nielsen},
  {Nissanke}, {Nitz}, {Noack}, {Nocera}, {Nolting}, {North}, {Nuttall},
  {Oberling}, {O'Dea}, {Ogin}, {Oh}, {Oh}, {Ohme}, {Okada}, {Oliver},
  {Oppermann}, {Oram}, {O'Reilly}, {Ormiston}, {Ortega}, {O'Shaughnessy},
  {Ossokine}, {Ottaway}, {Overmier}, {Owen}, {Pace}, {Page}, {Page}, {Pai},
  {Pai}, {Palamos}, {Palashov}, {Palomba}, {Pal-Singh}, {Pan}, {Pan}, {Pang},
  {Pang}, {Pankow}, {Pannarale}, {Pant}, {Paoletti}, {Paoli}, {Papa}, {Parida},
  {Parker}, {Pascucci}, {Pasqualetti}, {Passaquieti}, {Passuello}, {Patil},
  {Patricelli}, {Pearlstone}, {Pedraza}, {Pedurand}, {Pekowsky}, {Pele},
  {Penn}, {Perez}, {Perreca}, {Perri}, {Pfeiffer}, {Phelps}, {Piccinni},
  {Pichot}, {Piergiovanni}, {Pierro}, {Pillant}, {Pinard}, {Pinto}, {Pirello},
  {Pitkin}, {Poe}, {Poggiani}, {Popolizio}, {Porter}, {Post}, {Powell},
  {Prasad}, {Pratt}, {Pratten}, {Predoi}, {Prestegard}, {Prijatelj},
  {Principe}, {Privitera}, {Prix}, {Prodi}, {Prokhorov}, {Puncken}, {Punturo},
  {Puppo}, {P{\"u}rrer}, {Qi}, {Quetschke}, {Quintero}, {Quitzow-James},
  {Raab}, {Rabeling}, {Radkins}, {Raffai}, {Raja}, {Rajan}, {Rajbhandari},
  {Rakhmanov}, {Ramirez}, {Ramos-Buades}, {Rapagnani}, {Raymond}, {Razzano},
  {Read}, {Regimbau}, {Rei}, {Reid}, {Reitze}, {Ren}, {Reyes}, {Ricci},
  {Ricker}, {Rieger}, {Riles}, {Rizzo}, {Robertson}, {Robie}, {Robinet},
  {Rocchi}, {Rolland}, {Rollins}, {Roma}, {Romano}, {Romano}, {Romel}, {Romie},
  {Rosi{\'n}ska}, {Ross}, {Rowan}, {R{\"u}diger}, {Ruggi}, {Rutins}, {Ryan},
  {Sachdev}, {Sadecki}, {Sadeghian}, {Sakellariadou}, {Salconi}, {Saleem},
  {Salemi}, {Samajdar}, {Sammut}, {Sampson}, {Sanchez}, {Sanchez},
  {Sanchis-Gual}, {Sand berg}, {Sanders}, {Sassolas}, {Sathyaprakash},
  {Saulson}, {Sauter}, {Savage}, {Sawadsky}, {Schale}, {Scheel}, {Scheuer},
  {Schmidt}, {Schmidt}, {Schnabel}, {Schofield}, {Sch{\"o}nbeck}, {Schreiber},
  {Schuette}, {Schulte}, {Schutz}, {Schwalbe}, {Scott}, {Scott}, {Seidel},
  {Sellers}, {Sengupta}, {Sentenac}, {Sequino}, {Sergeev}, {Shaddock},
  {Shaffer}, {Shah}, {Shahriar}, {Shaner}, {Shao}, {Shapiro}, {Shawhan},
  {Sheperd}, {Shoemaker}, {Shoemaker}, {Siellez}, {Siemens}, {Sieniawska},
  {Sigg}, {Silva}, {Singer}, {Singh}, {Singhal}, {Sintes}, {Slagmolen},
  {Smith}, {Smith}, {Smith}, {Somala}, {Son}, {Sonnenberg}, {Sorazu},
  {Sorrentino}, {Souradeep}, {Spencer}, {Srivastava}, {Staats}, {Staley},
  {Steinke}, {Steinlechner}, {Steinlechner}, {Steinmeyer}, {Stevenson},
  {Stone}, {Stops}, {Strain}, {Stratta}, {Strigin}, {Strunk}, {Sturani},
  {Stuver}, {Summerscales}, {Sun}, {Sunil}, {Suresh}, {Sutton}, {Swinkels},
  {Szczepa{\'n}czyk}, {Tacca}, {Tait}, {Talbot}, {Talukder}, {Tanner},
  {T{\'a}pai}, {Taracchini}, {Tasson}, {Taylor}, {Taylor}, {Tewari}, {Theeg},
  {Thies}, {Thomas}, {Thomas}, {Thomas}, {Thorne}, {Thorne}, {Thrane},
  {Tiwari}, {Tiwari}, {Tokmakov}, {Toland}, {Tonelli}, {Tornasi},
  {Torres-Forn{\'e}}, {Torrie}, {T{\"o}yr{\"a}}, {Travasso}, {Traylor},
  {Trinastic}, {Tringali}, {Trozzo}, {Tsang}, {Tse}, {Tso}, {Tsukada}, {Tsuna},
  {Tuyenbayev}, {Ueno}, {Ugolini}, {Unnikrishnan}, {Urban}, {Usman},
  {Vahlbruch}, {Vajente}, {Valdes}, {LIGO Scientific Collaboration}, \& {Virgo
  Collaboration}}]{2017PhRvL.119p1101A}
---. 2017{\natexlab{c}},
  \href{http://dx.doi.org/10.1103/PhysRevLett.119.161101}{\JournalTitle{\prl},
  119, 161101}

\bibitem[{{Abbott} {et~al.}(2017{\natexlab{d}}){Abbott}, {Abbott}, {Abbott},
  {Acernese}, {Ackley}, {Adams}, {Adams}, {Addesso}, {Adhikari}, {Adya},
  {Affeldt}, {Afrough}, {Agarwal}, {Agathos}, {Agatsuma}, {Aggarwal}, {Aguiar},
  {Aiello}, {Ain}, {Ajith}, {Allen}, {Allen}, {Allocca}, {Altin}, {Amato},
  {Ananyeva}, {Anderson}, {Anderson}, {Angelova}, {Antier}, {Appert}, {Arai},
  {Araya}, {Areeda}, {Arnaud}, {Arun}, {Ascenzi}, {Ashton}, {Ast}, {Aston},
  {Astone}, {Atallah}, {Aufmuth}, {Aulbert}, {AultONeal}, {Austin},
  {Avila-Alvarez}, {Babak}, {Bacon}, {Bader}, {Bae}, {Baker}, {Baldaccini},
  {Ballardin}, {Ballmer}, {Banagiri}, {Barayoga}, {Barclay}, {Barish},
  {Barker}, {Barkett}, {Barone}, {Barr}, {Barsotti}, {Barsuglia}, {Barta},
  {Barthelmy}, {Bartlett}, {Bartos}, {Bassiri}, {Basti}, {Batch}, {Bawaj},
  {Bayley}, {Bazzan}, {B{\'e}csy}, {Beer}, {Bejger}, {Belahcene}, {Bell},
  {Berger}, {Bergmann}, {Bero}, {Berry}, {Bersanetti}, {Bertolini},
  {Betzwieser}, {Bhagwat}, {Bhandare}, {Bilenko}, {Billingsley}, {Billman},
  {Birch}, {Birney}, {Birnholtz}, {Biscans}, {Biscoveanu}, {Bisht}, {Bitossi},
  {Biwer}, {Bizouard}, {Blackburn}, {Blackman}, {Blair}, {Blair}, {Blair},
  {Bloemen}, {Bock}, {Bode}, {Boer}, {Bogaert}, {Bohe}, {Bondu}, {Bonilla},
  {Bonnand}, {Boom}, {Bork}, {Boschi}, {Bose}, {Bossie}, {Bouffanais}, {Bozzi},
  {Bradaschia}, {Brady}, {Branchesi}, {Brau}, {Briant}, {Brillet}, {Brinkmann},
  {Brisson}, {Brockill}, {Broida}, {Brooks}, {Brown}, {Brown}, {Brunett},
  {Buchanan}, {Buikema}, {Bulik}, {Bulten}, {Buonanno}, {Buskulic}, {Buy},
  {Byer}, {Cabero}, {Cadonati}, {Cagnoli}, {Cahillane}, {Calder{\'o}n
  Bustillo}, {Callister}, {Calloni}, {Camp}, {Canepa}, {Canizares}, {Cannon},
  {Cao}, {Cao}, {Capano}, {Capocasa}, {Carbognani}, {Caride}, {Carney},
  {Casanueva Diaz}, {Casentini}, {Caudill}, {Cavagli{\`a}}, {Cavalier},
  {Cavalieri}, {Cella}, {Cepeda}, {Cerd{\'a}-Dur{\'a}n}, {Cerretani},
  {Cesarini}, {Chamberlin}, {Chan}, {Chao}, {Charlton}, {Chase}, {Chassand
  e-Mottin}, {Chatterjee}, {Chatziioannou}, {Cheeseboro}, {Chen}, {Chen},
  {Chen}, {Cheng}, {Chia}, {Chincarini}, {Chiummo}, {Chmiel}, {Cho}, {Cho},
  {Chow}, {Christensen}, {Chu}, {Chua}, {Chua}, {Chung}, {Chung}, {Ciani},
  {Ciolfi}, {Cirelli}, {Cirone}, {Clara}, {Clark}, {Clearwater}, {Cleva},
  {Cocchieri}, {Coccia}, {Cohadon}, {Cohen}, {Colla}, {Collette}, {Cominsky},
  {Constancio}, {Conti}, {Cooper}, {Corban}, {Corbitt}, {Cordero-Carri{\'o}n},
  {Corley}, {Cornish}, {Corsi}, {Cortese}, {Costa}, {Coughlin}, {Coughlin},
  {Coulon}, {Countryman}, {Couvares}, {Covas}, {Cowan}, {Coward}, {Cowart},
  {Coyne}, {Coyne}, {Creighton}, {Creighton}, {Cripe}, {Crowder}, {Cullen},
  {Cumming}, {Cunningham}, {Cuoco}, {Dal Canton}, {D{\'a}lya}, {Danilishin},
  {D'Antonio}, {Danzmann}, {Dasgupta}, {Da Silva Costa}, {Dattilo}, {Dave},
  {Davier}, {Davis}, {Daw}, {Day}, {De}, {DeBra}, {Degallaix}, {De Laurentis},
  {Del{\'e}glise}, {Del Pozzo}, {Demos}, {Denker}, {Dent}, {De Pietri},
  {Dergachev}, {De Rosa}, {DeRosa}, {De Rossi}, {DeSalvo}, {de Varona},
  {Devenson}, {Dhurandhar}, {D{\'\i}az}, {Di Fiore}, {Di Giovanni}, {Di
  Girolamo}, {Di Lieto}, {Di Pace}, {Di Palma}, {Di Renzo}, {Doctor},
  {Dolique}, {Donovan}, {Dooley}, {Doravari}, {Dorrington}, {Douglas}, {Dovale
  {\'A}lvarez}, {Downes}, {Drago}, {Dreissigacker}, {Driggers}, {Du}, {Ducrot},
  {Dupej}, {Dwyer}, {Edo}, {Edwards}, {Effler}, {Ehrens}, {Eichholz},
  {Eikenberry}, {Eisenstein}, {Essick}, {Estevez}, {Etienne}, {Etzel}, {Evans},
  {Evans}, {Factourovich}, {Fafone}, {Fair}, {Fairhurst}, {Fan}, {Farinon},
  {Farr}, {Farr}, {Fauchon-Jones}, {Favata}, {Fays}, {Fee}, {Fehrmann},
  {Feicht}, {Fejer}, {Fernand ez-Galiana}, {Ferrante}, {Ferreira}, {Ferrini},
  {Fidecaro}, {Finstad}, {Fiori}, {Fiorucci}, {Fishbach}, {Fisher},
  {Fitz-Axen}, {Flaminio}, {Fletcher}, {Fong}, {Font}, {Forsyth}, {Forsyth},
  {Fournier}, {Frasca}, {Frasconi}, {Frei}, {Freise}, {Frey}, {Frey}, {Fries},
  {Fritschel}, {Frolov}, {Fulda}, {Fyffe}, {Gabbard}, {Gadre}, {Gaebel},
  {Gair}, {Gammaitoni}, {Ganija}, {Gaonkar}, {Garcia-Quiros}, {Garufi},
  {Gateley}, {Gaudio}, {Gaur}, {Gayathri}, {Gehrels}, {Gemme}, {Genin},
  {Gennai}, {George}, {George}, {Gergely}, {Germain}, {Ghonge}, {Ghosh},
  {Ghosh}, {Ghosh}, {Giaime}, {Giardina}, {Giazotto}, {Gill}, {Glover},
  {Goetz}, {Goetz}, {Gomes}, {Goncharov}, {Gonz{\'a}lez}, {Gonzalez Castro},
  {Gopakumar}, {Gorodetsky}, {Gossan}, {Gosselin}, {Gouaty}, {Grado}, {Graef},
  {Granata}, {Grant}, {Gras}, {Gray}, {Greco}, {Green}, {Gretarsson},
  {Griswold}, {Groot}, {Grote}, {Grunewald}, {Gruning}, {Guidi}, {Guo},
  {Gupta}, {Gupta}, {Gushwa}, {Gustafson}, {Gustafson}, {Halim}, {Hall},
  {Hall}, {Hamilton}, {Hammond}, {Haney}, {Hanke}, {Hanks}, {Hanna}, {Hannam},
  {Hannuksela}, {Hanson}, {Hardwick}, {Harms}, {Harry}, {Harry}, {Hart},
  {Haster}, {Haughian}, {Healy}, {Heidmann}, {Heintze}, {Heitmann}, {Hello},
  {Hemming}, {Hendry}, {Heng}, {Hennig}, {Heptonstall}, {Heurs}, {Hild},
  {Hinderer}, {Hoak}, {Hofman}, {Holt}, {Holz}, {Hopkins}, {Horst}, {Hough},
  {Houston}, {Howell}, {Hreibi}, {Hu}, {Huerta}, {Huet}, {Hughey}, {Husa},
  {Huttner}, {Huynh-Dinh}, {Indik}, {Inta}, {Intini}, {Isa}, {Isac}, {Isi},
  {Iyer}, {Izumi}, {Jacqmin}, {Jani}, {Jaranowski}, {Jawahar},
  {Jim{\'e}nez-Forteza}, {Johnson}, {Jones}, {Jones}, {Jonker}, {Ju}, {Junker},
  {Kalaghatgi}, {Kalogera}, {Kamai}, {Kandhasamy}, {Kang}, {Kanner}, {Kapadia},
  {Karki}, {Karvinen}, {Kasprzack}, {Katolik}, {Katsavounidis}, {Katzman},
  {Kaufer}, {Kawabe}, {K{\'e}f{\'e}lian}, {Keitel}, {Kemball}, {Kennedy},
  {Kent}, {Key}, {Khalili}, {Khan}, {Khan}, {Khan}, {Khazanov}, {Kijbunchoo},
  {Kim}, {Kim}, {Kim}, {Kim}, {Kim}, {Kim}, {Kimbrell}, {King}, {King},
  {Kinley-Hanlon}, {Kirchhoff}, {Kissel}, {Kleybolte}, {Klimenko}, {Knowles},
  {Koch}, {Koehlenbeck}, {Koley}, {Kondrashov}, {Kontos}, {Korobko}, {Korth},
  {Kowalska}, {Kozak}, {Kr{\"a}mer}, {Kringel}, {Krishnan}, {Kr{\'o}lak},
  {Kuehn}, {Kumar}, {Kumar}, {Kumar}, {Kuo}, {Kutynia}, {Kwang}, {Lackey},
  {Lai}, {Landry}, {Lang}, {Lange}, {Lantz}, {Lanza}, {Larson},
  {Lartaux-Vollard}, {Lasky}, {Laxen}, {Lazzarini}, {Lazzaro}, {Leaci},
  {Leavey}, {Lee}, {Lee}, {Lee}, {Lee}, {Lee}, {Lehmann}, {Lenon}, {Leonardi},
  {Leroy}, {Letendre}, {Levin}, {Li}, {Linker}, {Littenberg}, {Liu}, {Lo},
  {Lockerbie}, {London}, {Lord}, {Lorenzini}, {Loriette}, {Lormand}, {Losurdo},
  {Lough}, {Lousto}, {Lovelace}, {L{\"u}ck}, {Lumaca}, {Lundgren}, {Lynch},
  {Ma}, {Macas}, {Macfoy}, {Machenschalk}, {MacInnis}, {Macleod}, {Maga{\~n}a
  Hernandez}, {Maga{\~n}a-Sand oval}, {Maga{\~n}a Zertuche}, {Magee},
  {Majorana}, {Maksimovic}, {Man}, {Mandic}, {Mangano}, {Mansell}, {Manske},
  {Mantovani}, {Marchesoni}, {Marion}, {M{\'a}rka}, {M{\'a}rka}, {Markakis},
  {Markosyan}, {Markowitz}, {Maros}, {Marquina}, {Marsh}, {Martelli},
  {Martellini}, {Martin}, {Martin}, {Martynov}, {Mason}, {Massera}, {Masserot},
  {Massinger}, {Masso-Reid}, {Mastrogiovanni}, {Matas}, {Matichard}, {Matone},
  {Mavalvala}, {Mazumder}, {McCarthy}, {McClelland}, {McCormick}, {McCuller},
  {McGuire}, {McIntyre}, {McIver}, {McManus}, {McNeill}, {McRae}, {McWilliams},
  {Meacher}, {Meadors}, {Mehmet}, {Meidam}, {Mejuto-Villa}, {Melatos},
  {Mendell}, {Mercer}, {Merilh}, {Merzougui}, {Meshkov}, {Messenger},
  {Messick}, {Metzdorff}, {Meyers}, {Miao}, {Michel}, {Middleton}, {Mikhailov},
  {Milano}, {Miller}, {Miller}, {Miller}, {Millhouse}, {Milovich-Goff},
  {Minazzoli}, {Minenkov}, {Ming}, {Mishra}, {Mitra}, {Mitrofanov},
  {Mitselmakher}, {Mittleman}, {Moffa}, {Moggi}, {Mogushi}, {Mohan},
  {Mohapatra}, {Montani}, {Moore}, {Moraru}, {Moreno}, {Morriss}, {Mours},
  {Mow-Lowry}, {Mueller}, {Muir}, {Mukherjee}, {Mukherjee}, {Mukherjee},
  {Mukund}, {Mullavey}, {Munch}, {Mu{\~n}iz}, {Muratore}, {Murray}, {Napier},
  {Nardecchia}, {Naticchioni}, {Nayak}, {Neilson}, {Nelemans}, {Nelson},
  {Nery}, {Neunzert}, {Nevin}, {Newport}, {Newton}, {Ng}, {Nguyen}, {Nguyen},
  {Nichols}, {Nielsen}, {Nissanke}, {Nitz}, {Noack}, {Nocera}, {Nolting},
  {North}, {Nuttall}, {Oberling}, {O'Dea}, {Ogin}, {Oh}, {Oh}, {Ohme}, {Okada},
  {Oliver}, {Oppermann}, {Oram}, {O'Reilly}, {Ormiston}, {Ortega},
  {O'Shaughnessy}, {Ossokine}, {Ottaway}, {Overmier}, {Owen}, {Pace}, {Page},
  {Page}, {Pai}, {Pai}, {Palamos}, {Palashov}, {Palomba}, {Pal-Singh}, {Pan},
  {Pan}, {Pang}, {Pang}, {Pankow}, {Pannarale}, {Pant}, {Paoletti}, {Paoli},
  {Papa}, {Parida}, {Parker}, {Pascucci}, {Pasqualetti}, {Passaquieti},
  {Passuello}, {Patil}, {Patricelli}, {Pearlstone}, {Pedraza}, {Pedurand},
  {Pekowsky}, {Pele}, {Penn}, {Perez}, {Perreca}, {Perri}, {Pfeiffer},
  {Phelps}, {Piccinni}, {Pichot}, {Piergiovanni}, {Pierro}, {Pillant},
  {Pinard}, {Pinto}, {Pirello}, {Pitkin}, {Poe}, {Poggiani}, {Popolizio},
  {Porter}, {Post}, {Powell}, {Prasad}, {Pratt}, {Pratten}, {Predoi},
  {Prestegard}, {Price}, {Prijatelj}, {Principe}, {Privitera}, {Prodi},
  {Prokhorov}, {Puncken}, {Punturo}, {Puppo}, {P{\"u}rrer}, {Qi}, {Quetschke},
  {Quintero}, {Quitzow-James}, {Raab}, {Rabeling}, {Radkins}, {Raffai}, {Raja},
  {Rajan}, {Rajbhandari}, {Rakhmanov}, {Ramirez}, {Ramos-Buades}, {Rapagnani},
  {Raymond}, {Razzano}, {Read}, {Regimbau}, {Rei}, {Reid}, {Reitze}, {Ren},
  {Reyes}, {Ricci}, {Ricker}, {Rieger}, {Riles}, {Rizzo}, {Robertson}, {Robie},
  {Robinet}, {Rocchi}, {Rolland}, {Rollins}, {Roma}, {Romano}, {Romel},
  {Romie}, {Rosi{\'n}ska}, {Ross}, {Rowan}, {R{\"u}diger}, {Ruggi}, {Rutins},
  {Ryan}, {Sachdev}, {Sadecki}, {Sadeghian}, {Sakellariadou}, {Salconi},
  {Saleem}, {Salemi}, {Samajdar}, {Sammut}, {Sampson}, {Sanchez}, {Sanchez},
  {Sanchis-Gual}, {Sand berg}, {Sanders}, {Sassolas}, {Sathyaprakash},
  {Saulson}, {Sauter}, {Savage}, {Sawadsky}, {Schale}, {Scheel}, {Scheuer},
  {Schmidt}, {Schmidt}, {Schnabel}, {Schofield}, {Sch{\"o}nbeck}, {Schreiber},
  {Schuette}, {Schulte}, {Schutz}, {Schwalbe}, {Scott}, {Scott}, {Seidel},
  {Sellers}, {Sengupta}, {Sentenac}, {Sequino}, {Sergeev}, {Shaddock},
  {Shaffer}, {Shah}, {Shahriar}, {Shaner}, {Shao}, {Shapiro}, {Shawhan},
  {Sheperd}, {Shoemaker}, {Shoemaker}, {Siellez}, {Siemens}, {Sieniawska},
  {Sigg}, {Silva}, {Singer}, {Singh}, {Singhal}, {Sintes}, {Slagmolen},
  {Smith}, {Smith}, {Smith}, {Somala}, {Son}, {Sonnenberg}, {Sorazu},
  {Sorrentino}, {Souradeep}, {Spencer}, {Srivastava}, {Staats}, {Staley},
  {Steinke}, {Steinlechner}, {Steinlechner}, {Steinmeyer}, {Stevenson},
  {Stone}, {Stops}, {Strain}, {Stratta}, {Strigin}, {Strunk}, {Sturani},
  {Stuver}, {Summerscales}, {Sun}, {Sunil}, {Suresh}, {Sutton}, {Swinkels},
  {Szczepa{\'n}czyk}, {Tacca}, {Tait}, {Talbot}, {Talukder}, {Tanner},
  {T{\'a}pai}, {Taracchini}, {Tasson}, {Taylor}, {Taylor}, {Tewari}, {Theeg},
  {Thies}, {Thomas}, {Thomas}, {Thomas}, {Thorne}, {Thorne}, {Thrane},
  {Tiwari}, {Tiwari}, {Tokmakov}, {Toland}, {Tonelli}, {Tornasi},
  {Torres-Forn{\'e}}, {Torrie}, {T{\"o}yr{\"a}}, {Travasso}, {Traylor},
  {Trinastic}, {Tringali}, {Trozzo}, {Tsang}, {Tse}, {Tso}, {Tsukada}, {Tsuna},
  {Tuyenbayev}, {Ueno}, {Ugolini}, {Unnikrishnan}, {Urban}, {Usman},
  {Vahlbruch}, {Vajente}, {Valdes}, {van Bakel}, {van Beuzekom}, {van den
  Brand}, {Van Den Broeck}, {Vander-Hyde}, {van der Schaaf}, {van Heijningen},
  {van Veggel}, {Vardaro}, {Varma}, {Vass}, {Vas{\'u}th}, {Vecchio},
  {Vedovato}, {Veitch}, {Veitch}, {Venkateswara}, {Venugopalan}, {Verkindt},
  {Vetrano}, {Vicer{\'e}}, {Viets}, {Vinciguerra}, {Vine}, {Vinet}, {Vitale},
  {Vo}, {Vocca}, {Vorvick}, {Vyatchanin}, {Wade}, {Wade}, {Wade}, {Walet},
  {Walker}, {Wallace}, {Walsh}, {Wang}, {Wang}, {Wang}, {Wang}, {Wang}, {Ward},
  {Warner}, {Was}, {Watchi}, {Weaver}, {Wei}, {Weinert}, {Weinstein}, {Weiss},
  {Wen}, \& {South Africa/MeerKAT}}]{2017ApJ...848L..12A}
---. 2017{\natexlab{d}},
  \href{http://dx.doi.org/10.3847/2041-8213/aa91c9}{\JournalTitle{\apj}, 848,
  L12}

\bibitem[{{Abbott} {et~al.}(2017{\natexlab{e}}){Abbott}, {Abbott}, {Abbott},
  {Acernese}, {Ackley}, {Adams}, {Adams}, {Addesso}, {Adhikari}, {Adya},
  {Affeldt}, {Afrough}, {Agarwal}, {Agathos}, {Agatsuma}, {Aggarwal}, {Aguiar},
  {Aiello}, {Ain}, {Ajith}, {Allen}, {Allen}, {Allocca}, {Altin}, {Amato},
  {Ananyeva}, {Anderson}, {Anderson}, {Angelova}, {Antier}, {Appert}, {Arai},
  {Araya}, {Areeda}, {Arnaud}, {Arun}, {Ascenzi}, {Ashton}, {Ast}, {Aston},
  {Astone}, {Atallah}, {Aufmuth}, {Aulbert}, {AultONeal}, {Austin},
  {Avila-Alvarez}, {Babak}, {Bacon}, {Bader}, {Bae}, {Baker}, {Baldaccini},
  {Ballardin}, {Ballmer}, {Banagiri}, {Barayoga}, {Barclay}, {Barish},
  {Barker}, {Barkett}, {Barone}, {Barr}, {Barsotti}, {Barsuglia}, {Barta},
  {Bartlett}, {Bartos}, {Bassiri}, {Basti}, {Batch}, {Bawaj}, {Bayley},
  {Bazzan}, {B{\'e}csy}, {Beer}, {Bejger}, {Belahcene}, {Bell}, {Berger},
  {Bergmann}, {Bero}, {Berry}, {Bersanetti}, {Bertolini}, {Betzwieser},
  {Bhagwat}, {Bhandare}, {Bilenko}, {Billingsley}, {Billman}, {Birch},
  {Birney}, {Birnholtz}, {Biscans}, {Biscoveanu}, {Bisht}, {Bitossi}, {Biwer},
  {Bizouard}, {Blackburn}, {Blackman}, {Blair}, {Blair}, {Blair}, {Bloemen},
  {Bock}, {Bode}, {Boer}, {Bogaert}, {Bohe}, {Bondu}, {Bonilla}, {Bonnand},
  {Boom}, {Bork}, {Boschi}, {Bose}, {Bossie}, {Bouffanais}, {Bozzi},
  {Bradaschia}, {Brady}, {Branchesi}, {Brau}, {Briant}, {Brillet}, {Brinkmann},
  {Brisson}, {Brockill}, {Broida}, {Brooks}, {Brown}, {Brunett}, {Buchanan},
  {Buikema}, {Bulik}, {Bulten}, {Buonanno}, {Buskulic}, {Buy}, {Byer},
  {Cabero}, {Cadonati}, {Cagnoli}, {Cahillane}, {Calder{\'o}n Bustillo},
  {Callister}, {Calloni}, {Camp}, {Canepa}, {Canizares}, {Cannon}, {Cao},
  {Cao}, {Capano}, {Capocasa}, {Carbognani}, {Caride}, {Carney}, {Casanueva
  Diaz}, {Casentini}, {Caudill}, {Cavagli{\`a}}, {Cavalier}, {Cavalieri},
  {Cella}, {Cepeda}, {Cerd{\'a}-Dur{\'a}n}, {Cerretani}, {Cesarini},
  {Chamberlin}, {Chan}, {Chao}, {Charlton}, {Chase}, {Chassande-Mottin},
  {Chatterjee}, {Cheeseboro}, {Chen}, {Chen}, {Chen}, {Cheng}, {Chia},
  {Chincarini}, {Chiummo}, {Chmiel}, {Cho}, {Cho}, {Chow}, {Christensen},
  {Chu}, {Chua}, {Chua}, {Chung}, {Chung}, {Ciani}, {Ciolfi}, {Cirelli},
  {Cirone}, {Clara}, {Clark}, {Clearwater}, {Cleva}, {Cocchieri}, {Coccia},
  {Cohadon}, {Cohen}, {Colla}, {Collette}, {Cominsky}, {Constancio}, {Conti},
  {Cooper}, {Corban}, {Corbitt}, {Cordero-Carri{\'o}n}, {Corley}, {Corsi},
  {Cortese}, {Costa}, {Coughlin}, {Coughlin}, {Coulon}, {Countryman},
  {Couvares}, {Covas}, {Cowan}, {Coward}, {Cowart}, {Coyne}, {Coyne},
  {Creighton}, {Creighton}, {Cripe}, {Crowder}, {Cullen}, {Cumming},
  {Cunningham}, {Cuoco}, {Dal Canton}, {D{\'a}lya}, {Danilishin}, {D'Antonio},
  {Danzmann}, {Dasgupta}, {Da Silva Costa}, {Dattilo}, {Dave}, {Davier},
  {Davis}, {Daw}, {Day}, {De}, {DeBra}, {Degallaix}, {De Laurentis},
  {Del{\'e}glise}, {Del Pozzo}, {Demos}, {Denker}, {Dent}, {De Pietri},
  {Dergachev}, {De Rosa}, {DeRosa}, {De Rossi}, {DeSalvo}, {de Varona},
  {Devenson}, {Dhurandhar}, {D{\'\i}az}, {Di Fiore}, {Di Giovanni}, {Di
  Girolamo}, {Di Lieto}, {Di Pace}, {Di Palma}, {Di Renzo}, {Doctor},
  {Dolique}, {Donovan}, {Dooley}, {Doravari}, {Dorrington}, {Douglas}, {Dovale
  {\'A}lvarez}, {Downes}, {Drago}, {Dreissigacker}, {Driggers}, {Du}, {Ducrot},
  {Dupej}, {Dwyer}, {Edo}, {Edwards}, {Effler}, {Eggenstein}, {Ehrens},
  {Eichholz}, {Eikenberry}, {Eisenstein}, {Essick}, {Estevez}, {Etienne},
  {Etzel}, {Evans}, {Evans}, {Factourovich}, {Fafone}, {Fair}, {Fairhurst},
  {Fan}, {Farinon}, {Farr}, {Farr}, {Fauchon-Jones}, {Favata}, {Fays}, {Fee},
  {Fehrmann}, {Feicht}, {Fejer}, {Fernandez-Galiana}, {Ferrante}, {Ferreira},
  {Ferrini}, {Fidecaro}, {Finstad}, {Fiori}, {Fiorucci}, {Fishbach}, {Fisher},
  {Fitz-Axen}, {Flaminio}, {Fletcher}, {Fong}, {Font}, {Forsyth}, {Forsyth},
  {Fournier}, {Frasca}, {Frasconi}, {Frei}, {Freise}, {Frey}, {Frey}, {Fries},
  {Fritschel}, {Frolov}, {Fulda}, {Fyffe}, {Gabbard}, {Gadre}, {Gaebel},
  {Gair}, {Gammaitoni}, {Ganija}, {Gaonkar}, {Garcia-Quiros}, {Garufi},
  {Gateley}, {Gaudio}, {Gaur}, {Gayathri}, {Gehrels}, {Gemme}, {Genin},
  {Gennai}, {George}, {George}, {Gergely}, {Germain}, {Ghonge}, {Ghosh},
  {Ghosh}, {Ghosh}, {Giaime}, {Giardina}, {Giazotto}, {Gill}, {Glover},
  {Goetz}, {Goetz}, {Gomes}, {Goncharov}, {Gonzalez Castro}, {Gopakumar},
  {Gorodetsky}, {Gossan}, {Gosselin}, {Gouaty}, {Grado}, {Graef}, {Granata},
  {Grant}, {Gras}, {Gray}, {Greco}, {Green}, {Gretarsson}, {Groot}, {Grote},
  {Grunewald}, {Gruning}, {Guidi}, {Guo}, {Gupta}, {Gupta}, {Gushwa},
  {Gustafson}, {Gustafson}, {Halim}, {Hall}, {Hall}, {Hamilton}, {Hammond},
  {Haney}, {Hanke}, {Hanks}, {Hanna}, {Hannam}, {Hannuksela}, {Hanson},
  {Hardwick}, {Harms}, {Harry}, {Harry}, {Hart}, {Haster}, {Haughian}, {Healy},
  {Heidmann}, {Heintze}, {Heitmann}, {Hello}, {Hemming}, {Hendry}, {Heng},
  {Hennig}, {Heptonstall}, {Heurs}, {Hild}, {Hinderer}, {Hoak}, {Hofman},
  {Holgado}, {Holt}, {Holz}, {Hopkins}, {Horst}, {Hough}, {Houston}, {Howell},
  {Hreibi}, {Hu}, {Huerta}, {Huet}, {Hughey}, {Husa}, {Huttner}, {Huynh-Dinh},
  {Indik}, {Inta}, {Intini}, {Isa}, {Isac}, {Isi}, {Iyer}, {Izumi}, {Jacqmin},
  {Jani}, {Jaranowski}, {Jawahar}, {Jim{\'e}nez-Forteza}, {Johnson}, {Jones},
  {Jones}, {Jonker}, {Ju}, {Junker}, {Kalaghatgi}, {Kalogera}, {Kamai},
  {Kandhasamy}, {Kang}, {Kanner}, {Kapadia}, {Karki}, {Karvinen}, {Kasprzack},
  {Katolik}, {Katsavounidis}, {Katzman}, {Kaufer}, {Kawabe},
  {K{\'e}f{\'e}lian}, {Keitel}, {Kemball}, {Kennedy}, {Kent}, {Key}, {Khalili},
  {Khan}, {Khan}, {Khan}, {Khazanov}, {Kijbunchoo}, {Kim}, {Kim}, {Kim}, {Kim},
  {Kim}, {Kim}, {Kimball}, {Kimbrell}, {King}, {King}, {Kinley-Hanlon},
  {Kirchhoff}, {Kissel}, {Kleybolte}, {Klimenko}, {Knowles}, {Koch},
  {Koehlenbeck}, {Koley}, {Kondrashov}, {Kontos}, {Korobko}, {Korth},
  {Kowalska}, {Kozak}, {Kr{\"a}mer}, {Kringel}, {Kr{\'o}lak}, {Kuehn}, {Kumar},
  {Kumar}, {Kumar}, {Kuo}, {Kutynia}, {Kwang}, {Lackey}, {Lai}, {Landry},
  {Lang}, {Lange}, {Lantz}, {Lanza}, {Larson}, {Lartaux-Vollard}, {Lasky},
  {Laxen}, {Lazzarini}, {Lazzaro}, {Leaci}, {Leavey}, {Lee}, {Lee}, {Lee},
  {Lee}, {Lee}, {Lehmann}, {Lenon}, {Leonardi}, {Leroy}, {Letendre}, {Levin},
  {Li}, {Linker}, {Littenberg}, {Liu}, {Lo}, {Lockerbie}, {London}, {Lord},
  {Lorenzini}, {Loriette}, {Lormand}, {Losurdo}, {Lough}, {Lousto}, {Lovelace},
  {L{\"u}ck}, {Lumaca}, {Lundgren}, {Lynch}, {Ma}, {Macas}, {Macfoy},
  {Machenschalk}, {MacInnis}, {Macleod}, {Maga{\~n}a Hernandez},
  {Maga{\~n}a-Sand oval}, {Maga{\~n}a Zertuche}, {Magee}, {Majorana},
  {Maksimovic}, {Man}, {Mandic}, {Mangano}, {Mansell}, {Manske}, {Mantovani},
  {Marchesoni}, {Marion}, {M{\'a}rka}, {M{\'a}rka}, {Markakis}, {Markosyan},
  {Markowitz}, {Maros}, {Marquina}, {Martelli}, {Martellini}, {Martin},
  {Martin}, {Martynov}, {Mason}, {Massera}, {Masserot}, {Massinger},
  {Masso-Reid}, {Mastrogiovanni}, {Matas}, {Matichard}, {Matone}, {Mavalvala},
  {Mazumder}, {McCarthy}, {McClelland}, {McCormick}, {McCuller}, {McGuire},
  {McIntyre}, {McIver}, {McManus}, {McNeill}, {McRae}, {McWilliams}, {Meacher},
  {Meadors}, {Mehmet}, {Meidam}, {Mejuto-Villa}, {Melatos}, {Mendell},
  {Mercer}, {Merilh}, {Merzougui}, {Meshkov}, {Messenger}, {Messick},
  {Metzdorff}, {Meyers}, {Miao}, {Michel}, {Middleton}, {Mikhailov}, {Milano},
  {Miller}, {Miller}, {Miller}, {Millhouse}, {Milovich-Goff}, {Minazzoli},
  {Minenkov}, {Ming}, {Mishra}, {Mitra}, {Mitrofanov}, {Mitselmakher},
  {Mittleman}, {Moffa}, {Moggi}, {Mogushi}, {Mohan}, {Mohapatra}, {Montani},
  {Moore}, {Moraru}, {Moreno}, {Morriss}, {Mours}, {Mow-Lowry}, {Mueller},
  {Muir}, {Mukherjee}, {Mukherjee}, {Mukherjee}, {Mukund}, {Mullavey}, {Munch},
  {Mu{\~n}iz}, {Muratore}, {Murray}, {Napier}, {Nardecchia}, {Naticchioni},
  {Nayak}, {Neilson}, {Nelemans}, {Nelson}, {Nery}, {Neunzert}, {Nevin},
  {Newport}, {Newton}, {Ng}, {Nguyen}, {Nichols}, {Nielsen}, {Nissanke},
  {Nitz}, {Noack}, {Nocera}, {Nolting}, {North}, {Nuttall}, {Oberling},
  {O'Dea}, {Ogin}, {Oh}, {Oh}, {Ohme}, {Okada}, {Oliver}, {Oppermann}, {Oram},
  {O'Reilly}, {Ormiston}, {Ortega}, {O'Shaughnessy}, {Ossokine}, {Ottaway},
  {Overmier}, {Owen}, {Pace}, {Page}, {Page}, {Pai}, {Pai}, {Palamos},
  {Palashov}, {Palomba}, {Pal-Singh}, {Pan}, {Pan}, {Pang}, {Pang}, {Pankow},
  {Pannarale}, {Pant}, {Paoletti}, {Paoli}, {Papa}, {Parida}, {Parker},
  {Pascucci}, {Pasqualetti}, {Passaquieti}, {Passuello}, {Patil}, {Patricelli},
  {Pearlstone}, {Pedraza}, {Pedurand}, {Pekowsky}, {Pele}, {Penn}, {Perez},
  {Perreca}, {Perri}, {Pfeiffer}, {Phelps}, {Piccinni}, {Pichot},
  {Piergiovanni}, {Pierro}, {Pillant}, {Pinard}, {Pinto}, {Pirello}, {Pitkin},
  {Poe}, {Poggiani}, {Popolizio}, {Porter}, {Post}, {Powell}, {Prasad},
  {Pratt}, {Pratten}, {Predoi}, {Prestegard}, {Prijatelj}, {Principe},
  {Privitera}, {Prodi}, {Prokhorov}, {Puncken}, {Punturo}, {Puppo},
  {P{\"u}rrer}, {Qi}, {Quetschke}, {Quintero}, {Quitzow-James}, {Rabeling},
  {Radkins}, {Raffai}, {Raja}, {Rajan}, {Rajbhandari}, {Rakhmanov}, {Ramirez},
  {Ramos-Buades}, {Rapagnani}, {Raymond}, {Razzano}, {Read}, {Regimbau}, {Rei},
  {Reid}, {Reitze}, {Ren}, {Reyes}, {Ricci}, {Ricker}, {Rieger}, {Riles},
  {Rizzo}, {Robertson}, {Robie}, {Robinet}, {Rocchi}, {Rolland}, {Rollins},
  {Roma}, {Romano}, {Romel}, {Romie}, {Rosi{\'n}ska}, {Ross}, {Rowan},
  {R{\"u}diger}, {Ruggi}, {Rutins}, {Ryan}, {Sachdev}, {Sadecki}, {Sadeghian},
  {Sakellariadou}, {Salconi}, {Saleem}, {Salemi}, {Samajdar}, {Sammut},
  {Sampson}, {Sanchez}, {Sanchez}, {Sanchis-Gual}, {Sand berg}, {Sanders},
  {Sassolas}, {Sathyaprakash}, {Sauter}, {Savage}, {Sawadsky}, {Schale},
  {Scheel}, {Scheuer}, {Schmidt}, {Schmidt}, {Schnabel}, {Schofield},
  {Sch{\"o}nbeck}, {Schreiber}, {Schuette}, {Schulte}, {Schutz}, {Schwalbe},
  {Scott}, {Scott}, {Seidel}, {Sellers}, {Sengupta}, {Sentenac}, {Sequino},
  {Sergeev}, {Shaddock}, {Shaffer}, {Shah}, {Shahriar}, {Shaner}, {Shao},
  {Shapiro}, {Shawhan}, {Sheperd}, {Shoemaker}, {Shoemaker}, {Siellez},
  {Siemens}, {Sieniawska}, {Sigg}, {Silva}, {Singer}, {Singh}, {Singhal},
  {Sintes}, {Slagmolen}, {Smith}, {Smith}, {Smith}, {Somala}, {Son},
  {Sonnenberg}, {Sorazu}, {Sorrentino}, {Souradeep}, {Spencer}, {Srivastava},
  {Staats}, {Staley}, {Steinke}, {Steinlechner}, {Steinlechner}, {Steinmeyer},
  {Stevenson}, {Stone}, {Stops}, {Strain}, {Stratta}, {Strigin}, {Strunk},
  {Sturani}, {Stuver}, {Summerscales}, {Sun}, {Sunil}, {Suresh}, {Sutton},
  {Swinkels}, {Szczepa{\'n}czyk}, {Tacca}, {Tait}, {Talbot}, {Talukder},
  {Tanner}, {T{\'a}pai}, {Taracchini}, {Tasson}, {Taylor}, {Taylor}, {Tewari},
  {Theeg}, {Thies}, {Thomas}, {Thomas}, {Thomas}, {Thorne}, {Thrane}, {Tiwari},
  {Tiwari}, {Tokmakov}, {Toland}, {Tonelli}, {Tornasi}, {Torres-Forn{\'e}},
  {Torrie}, {T{\"o}yr{\"a}}, {Travasso}, {Traylor}, {Trinastic}, {Tringali},
  {Trozzo}, {Tsang}, {Tse}, {Tso}, {Tsukada}, {Tsuna}, {Tuyenbayev}, {Ueno},
  {Ugolini}, {Unnikrishnan}, {Urban}, {Usman}, {Vahlbruch}, {Vajente},
  {Valdes}, {van Bakel}, {van Beuzekom}, {van den Brand}, {Van Den Broeck},
  {Vander-Hyde}, {van der Schaaf}, {van Heijningen}, {van Veggel}, {Vardaro},
  {Varma}, {Vass}, {Vas{\'u}th}, {Vecchio}, {Vedovato}, {Veitch}, {Veitch},
  {Venkateswara}, {Venugopalan}, {Verkindt}, {Vetrano}, {Vicer{\'e}}, {Viets},
  {Vinciguerra}, {Vine}, {Vinet}, {Vitale}, {Vo}, {Vocca}, {Vorvick},
  {Vyatchanin}, {Wade}, {Wade}, {Wade}, {Walet}, {Walker}, {Wallace}, {Walsh},
  {Wang}, {Wang}, {Wang}, {Wang}, {Wang}, {Ward}, {Warner}, {Was}, {Watchi},
  {Weaver}, {Wei}, {Weinert}, {(LIGO Scientific Collaboration}, \& {Virgo
  Collaboration}}]{2017ApJ...850L..40A}
---. 2017{\natexlab{e}},
  \href{http://dx.doi.org/10.3847/2041-8213/aa93fc}{\JournalTitle{\apj}, 850,
  L40}

\bibitem[{{Abbott} {et~al.}(2018{\natexlab{a}}){Abbott}, {Abbott}, {Abbott},
  {Acernese}, {Ackley}, {Adams}, {Adams}, {Addesso}, {Adhikari}, {Adya},
  {Affeldt}, {Agarwal}, {Agathos}, {Agatsuma}, {Aggarwal}, {Aguiar}, {Aiello},
  {Ain}, {Ajith}, {Allen}, {Allen}, {Allocca}, {Aloy}, {Altin}, {Amato},
  {Ananyeva}, {Anderson}, {Anderson}, {Angelova}, {Antier}, {Appert}, {Arai},
  {Araya}, {Areeda}, {Ar{\`e}ne}, {Arnaud}, {Arun}, {Ascenzi}, {Ashton}, {Ast},
  {Aston}, {Astone}, {Atallah}, {Aubin}, {Aufmuth}, {Aulbert}, {AultONeal},
  {Austin}, {Avila-Alvarez}, {Babak}, {Bacon}, {Badaracco}, {Bader}, {Bae},
  {Baker}, {Baldaccini}, {Ballardin}, {Ballmer}, {Banagiri}, {Barayoga},
  {Barclay}, {Barish}, {Barker}, {Barkett}, {Barnum}, {Barone}, {Barr},
  {Barsotti}, {Barsuglia}, {Barta}, {Bartlett}, {Bartos}, {Bassiri}, {Basti},
  {Batch}, {Bawaj}, {Bayley}, {Bazzan}, {B{\'e}csy}, {Beer}, {Bejger},
  {Belahcene}, {Bell}, {Beniwal}, {Bensch}, {Berger}, {Bergmann}, {Bernuzzi},
  {Bero}, {Berry}, {Bersanetti}, {Bertolini}, {Betzwieser}, {Bhandare},
  {Bilenko}, {Bilgili}, {Billingsley}, {Billman}, {Birch}, {Birney},
  {Birnholtz}, {Biscans}, {Biscoveanu}, {Bisht}, {Bitossi}, {Bizouard},
  {Blackburn}, {Blackman}, {Blair}, {Blair}, {Blair}, {Bloemen}, {Bock},
  {Bode}, {Boer}, {Boetzel}, {Bogaert}, {Bohe}, {Bondu}, {Bonilla}, {Bonnand},
  {Booker}, {Boom}, {Booth}, {Bork}, {Boschi}, {Bose}, {Bossie}, {Bossilkov},
  {Bosveld}, {Bouffanais}, {Bozzi}, {Bradaschia}, {Brady}, {Bramley},
  {Branchesi}, {Brau}, {Briant}, {Brighenti}, {Brillet}, {Brinkmann},
  {Brisson}, {Brockill}, {Brooks}, {Brown}, {Brunett}, {Buchanan}, {Buikema},
  {Bulik}, {Bulten}, {Buonanno}, {Buskulic}, {Buy}, {Byer}, {Cabero},
  {Cadonati}, {Cagnoli}, {Cahillane}, {Calder{\'o}n Bustillo}, {Callister},
  {Calloni}, {Camp}, {Canepa}, {Canizares}, {Cannon}, {Cao}, {Cao}, {Capano},
  {Capocasa}, {Carbognani}, {Caride}, {Carney}, {Carullo}, {Casanueva Diaz},
  {Casentini}, {Caudill}, {Cavagli{\`a}}, {Cavalier}, {Cavalieri}, {Cella},
  {Cepeda}, {Cerd{\'a}-Dur{\'a}n}, {Cerretani}, {Cesarini}, {Chaibi},
  {Chamberlin}, {Chan}, {Chao}, {Charlton}, {Chase}, {Chassande-Mottin},
  {Chatterjee}, {Chatziioannou}, {Cheeseboro}, {Chen}, {Chen}, {Chen}, {Cheng},
  {Chia}, {Chincarini}, {Chiummo}, {Chmiel}, {Cho}, {Cho}, {Chow},
  {Christensen}, {Chu}, {Chua}, {Chua}, {Chung}, {Chung}, {Ciani}, {Ciobanu},
  {Ciolfi}, {Cipriano}, {Cirelli}, {Cirone}, {Clara}, {Clark}, {Clearwater},
  {Cleva}, {Cocchieri}, {Coccia}, {Cohadon}, {Cohen}, {Colla}, {Collette},
  {Collins}, {Cominsky}, {Constancio}, {Conti}, {Cooper}, {Corban}, {Corbitt},
  {Cordero-Carri{\'o}n}, {Corley}, {Cornish}, {Corsi}, {Cortese}, {Costa},
  {Cotesta}, {Coughlin}, {Coughlin}, {Coulon}, {Countryman}, {Couvares},
  {Covas}, {Cowan}, {Coward}, {Cowart}, {Coyne}, {Coyne}, {Creighton},
  {Creighton}, {Cripe}, {Crowder}, {Cullen}, {Cumming}, {Cunningham}, {Cuoco},
  {Canton}, {D{\'a}lya}, {Danilishin}, {D'Antonio}, {Danzmann}, {Dasgupta}, {Da
  Silva Costa}, {Dattilo}, {Dave}, {Davier}, {Davis}, {Daw}, {Day}, {DeBra},
  {Deenadayalan}, {Degallaix}, {De Laurentis}, {Del{\'e}glise}, {Del Pozzo},
  {Demos}, {Denker}, {Dent}, {De Pietri}, {Derby}, {Dergachev}, {De Rosa}, {De
  Rossi}, {DeSalvo}, {de Varona}, {Dhurandhar}, {D{\'\i}az}, {Dietrich}, {Di
  Fiore}, {Di Giovanni}, {Di Girolamo}, {Di Lieto}, {Ding}, {Di Pace}, {Di
  Palma}, {Di Renzo}, {Dmitriev}, {Doctor}, {Dolique}, {Donovan}, {Dooley},
  {Doravari}, {Dorrington}, {Dovale {\'A}lvarez}, {Downes}, {Drago},
  {Dreissigacker}, {Driggers}, {Du}, {Dupej}, {Dwyer}, {Easter}, {Edo},
  {Edwards}, {Effler}, {Eggenstein}, {Ehrens}, {Eichholz}, {Eikenberry},
  {Eisenmann}, {Eisenstein}, {Essick}, {Estelles}, {Estevez}, {Etienne},
  {Etzel}, {Evans}, {Evans}, {Fafone}, {Fair}, {Fairhurst}, {Fan}, {Farinon},
  {Farr}, {Farr}, {Fauchon-Jones}, {Favata}, {Fays}, {Fee}, {Fehrmann},
  {Feicht}, {Fejer}, {Feng}, {Fernandez-Galiana}, {Ferrante}, {Ferreira},
  {Ferrini}, {Fidecaro}, {Fiori}, {Fiorucci}, {Fishbach}, {Fisher}, {Fishner},
  {Fitz-Axen}, {Flaminio}, {Fletcher}, {Fong}, {Font}, {Forsyth}, {Forsyth},
  {Fournier}, {Frasca}, {Frasconi}, {Frei}, {Freise}, {Frey}, {Frey},
  {Fritschel}, {Frolov}, {Fulda}, {Fyffe}, {Gabbard}, {Gadre}, {Gaebel},
  {Gair}, {Gammaitoni}, {Ganija}, {Gaonkar}, {Garcia},
  {Garc{\'\i}a-Quir{\'o}s}, {Garufi}, {Gateley}, {Gaudio}, {Gaur}, {Gayathri},
  {Gemme}, {Genin}, {Gennai}, {George}, {George}, {Gergely}, {Germain},
  {Ghonge}, {Ghosh}, {Ghosh}, {Ghosh}, {Giacomazzo}, {Giaime}, {Giardina},
  {Giazotto}, {Gill}, {Giordano}, {Glover}, {Goetz}, {Goetz}, {Goncharov},
  {Gonz{\'a}lez}, {Gonzalez Castro}, {Gopakumar}, {Gorodetsky}, {Gossan},
  {Gosselin}, {Gouaty}, {Grado}, {Graef}, {Granata}, {Grant}, {Gras}, {Gray},
  {Greco}, {Green}, {Green}, {Gretarsson}, {Groot}, {Grote}, {Grunewald},
  {Gruning}, {Guidi}, {Gulati}, {Guo}, {Gupta}, {Gupta}, {Gushwa}, {Gustafson},
  {Gustafson}, {Halim}, {Hall}, {Hall}, {Hamilton}, {Hamilton}, {Hammond},
  {Haney}, {Hanke}, {Hanks}, {Hanna}, {Hannam}, {Hannuksela}, {Hanson},
  {Hardwick}, {Harms}, {Harry}, {Harry}, {Hart}, {Haster}, {Haughian}, {Healy},
  {Heidmann}, {Heintze}, {Heitmann}, {Hello}, {Hemming}, {Hendry}, {Heng},
  {Hennig}, {Heptonstall}, {Hernandez}, {Heurs}, {Hild}, {Hinderer}, {Ho},
  {Hoak}, {Hochheim}, {Hofman}, {Holland}, {Holt}, {Holz}, {Hopkins}, {Horst},
  {Hough}, {Houston}, {Howell}, {Hreibi}, {Huerta}, {Huet}, {Hughey}, {Hulko},
  {Husa}, {Huttner}, {Huynh-Dinh}, {Iess}, {Indik}, {Ingram}, {Inta}, {Intini},
  {Irwin}, {Isa}, {Isac}, {Isi}, {Iyer}, {Izumi}, {Jacqmin}, {Jani},
  {Jaranowski}, {Johnson}, {Johnson}, {Jones}, {Jones}, {Jonker}, {Ju},
  {Junker}, {Kalaghatgi}, {Kalogera}, {Kamai}, {Kandhasamy}, {Kang}, {Kanner},
  {Kapadia}, {Karki}, {Karvinen}, {Kasprzack}, {Katolik}, {Katsanevas},
  {Katsavounidis}, {Katzman}, {Kaufer}, {Kawabe}, {Keerthana},
  {K{\'e}f{\'e}lian}, {Keitel}, {Kemball}, {Kennedy}, {Key}, {Khalili},
  {Khamesra}, {Khan}, {Khan}, {Khan}, {Khan}, {Khazanov}, {Kijbunchoo}, {Kim},
  {Kim}, {Kim}, {Kim}, {Kim}, {Kim}, {King}, {King}, {Kinley-Hanlon},
  {Kirchhoff}, {Kissel}, {Kleybolte}, {Klimenko}, {Knowles}, {Koch},
  {Koehlenbeck}, {Koley}, {Kondrashov}, {Kontos}, {Korobko}, {Korth},
  {Kowalska}, {Kozak}, {Kr{\"a}mer}, {Kringel}, {Krishnan}, {Kr{\'o}lak},
  {Kuehn}, {Kumar}, {Kumar}, {Kumar}, {Kuo}, {Kutynia}, {Kwang}, {Lackey},
  {Lai}, {Landry}, {Landry}, {Lang}, {Lange}, {Lantz}, {Lanza},
  {Lartaux-Vollard}, {Lasky}, {Laxen}, {Lazzarini}, {Lazzaro}, {Leaci},
  {Leavey}, {Lee}, {Lee}, {Lee}, {Lee}, {Lee}, {Lehmann}, {Lenon}, {Leonardi},
  {Leroy}, {Letendre}, {Levin}, {Li}, {Li}, {Li}, {Linker}, {Littenberg},
  {Liu}, {Liu}, {Lo}, {Lockerbie}, {London}, {Longo}, {Lorenzini}, {Loriette},
  {Lormand}, {Losurdo}, {Lough}, {Lousto}, {Lovelace}, {L{\"u}ck}, {Lumaca},
  {Lundgren}, {Lynch}, {Ma}, {Macas}, {Macfoy}, {Machenschalk}, {MacInnis},
  {Macleod}, {Maga{\~n}a Hernandez}, {Maga{\~n}a-Sandoval}, {Maga{\~n}a
  Zertuche}, {Magee}, {Majorana}, {Maksimovic}, {Man}, {Mandic}, {Mangano},
  {Mansell}, {Manske}, {Mantovani}, {Marchesoni}, {Marion}, {M{\'a}rka},
  {M{\'a}rka}, {Markakis}, {Markosyan}, {Markowitz}, {Maros}, {Marquina},
  {Martelli}, {Martellini}, {Martin}, {Martin}, {Martynov}, {Mason}, {Massera},
  {Masserot}, {Massinger}, {Masso-Reid}, {Mastrogiovanni}, {Matas},
  {Matichard}, {Matone}, {Mavalvala}, {Mazumder}, {McCann}, {McCarthy},
  {McClelland }, {McCormick}, {McCuller}, {McGuire}, {McIver}, {McManus},
  {McRae}, {McWilliams}, {Meacher}, {Meadors}, {Mehmet}, {Meidam},
  {Mejuto-Villa}, {Melatos}, {Mendell}, {Mendoza-Gandara}, {Mercer}, {Mereni},
  {Merilh}, {Merzougui}, {Meshkov}, {Messenger}, {Messick}, {Metzdorff},
  {Meyers}, {Miao}, {Michel}, {Middleton}, {Mikhailov}, {Milano}, {Miller},
  {Miller}, {Miller}, {Miller}, {Millhouse}, {Mills}, {Milovich-Goff},
  {Minazzoli}, {Minenkov}, {Ming}, {Mishra}, {Mitra}, {Mitrofanov},
  {Mitselmakher}, {Mittleman}, {Moffa}, {Mogushi}, {Mohan}, {Mohapatra},
  {Montani}, {Moore}, {Moraru}, {Moreno}, {Morisaki}, {Mours}, {Mow-Lowry},
  {Mueller}, {Muir}, {Mukherjee}, {Mukherjee}, {Mukherjee}, {Mukund},
  {Mullavey}, {Munch}, {Mu{\~n}iz}, {Muratore}, {Murray}, {Nagar}, {Napier},
  {Nardecchia}, {Naticchioni}, {Nayak}, {Neilson}, {Nelemans}, {Nelson},
  {Nery}, {Neunzert}, {Nevin}, {Newport}, {Ng}, {Ng}, {Nguyen}, {Nguyen},
  {Nichols}, {Nielsen}, {Nissanke}, {Nitz}, {Nocera}, {Nolting}, {North},
  {Nuttall}, {Obergaulinger}, {Oberling}, {O'Brien}, {O'Dea}, {Ogin}, {Oh},
  {Oh}, {Ohme}, {Ohta}, {Okada}, {Oliver}, {Oppermann}, {Oram}, {O'Reilly},
  {Ormiston}, {Ortega}, {O'Shaughnessy}, {Ossokine}, {Ottaway}, {Overmier},
  {Owen}, {Pace}, {Pagano}, {Page}, {Page}, {Pai}, {Pai}, {Palamos},
  {Palashov}, {LIGO Scientific Collaboration}, \& {Virgo
  Collaboration}}]{2018PhRvL.121p1101A}
---. 2018{\natexlab{a}},
  \href{http://dx.doi.org/10.1103/PhysRevLett.121.161101}{\JournalTitle{\prl},
  121, 161101}

\bibitem[{{Abbott} {et~al.}(2018{\natexlab{b}}){Abbott}, {Abbott}, {Abbott},
  {Abernathy}, {Acernese}, {Ackley}, {Adams}, {Adams}, {Addesso}, {Adhikari},
  {Adya}, {Affeldt}, {Agathos}, {Agatsuma}, {Aggarwal}, {Aguiar}, {Aiello},
  {Ain}, {Ajith}, {Akutsu}, {Allen}, {Allocca}, {Altin}, {Ananyeva},
  {Anderson}, {Anderson}, {Ando}, {Appert}, {Arai}, {Araya}, {Araya}, {Areeda},
  {Arnaud}, {Arun}, {Asada}, {Ascenzi}, {Ashton}, {Aso}, {Ast}, {Aston},
  {Astone}, {Atsuta}, {Aufmuth}, {Aulbert}, {Avila-Alvarez}, {Awai}, {Babak},
  {Bacon}, {Bader}, {Baiotti}, {Baker}, {Baldaccini}, {Ballardin}, {Ballmer},
  {Barayoga}, {Barclay}, {Barish}, {Barker}, {Barone}, {Barr}, {Barsotti},
  {Barsuglia}, {Barta}, {Bartlett}, {Barton}, {Bartos}, {Bassiri}, {Basti},
  {Batch}, {Baune}, {Bavigadda}, {Bazzan}, {B{\'e}csy}, {Beer}, {Bejger},
  {Belahcene}, {Belgin}, {Bell}, {Berger}, {Bergmann}, {Berry}, {Bersanetti},
  {Bertolini}, {Betzwieser}, {Bhagwat}, {Bhandare}, {Bilenko}, {Billingsley},
  {Billman}, {Birch}, {Birney}, {Birnholtz}, {Biscans}, {Bisht}, {Bitossi},
  {Biwer}, {Bizouard}, {Blackburn}, {Blackman}, {Blair}, {Blair}, {Blair},
  {Bloemen}, {Bock}, {Boer}, {Bogaert}, {Bohe}, {Bondu}, {Bonnand }, {Boom},
  {Bork}, {Boschi}, {Bose}, {Bouffanais}, {Bozzi}, {Bradaschia}, {Brady},
  {Braginsky}, {Branchesi}, {Brau}, {Briant}, {Brillet}, {Brinkmann},
  {Brisson}, {Brockill}, {Broida}, {Brooks}, {Brown}, {Brown}, {Brown},
  {Brunett}, {Buchanan}, {Buikema}, {Bulik}, {Bulten}, {Buonanno}, {Buskulic},
  {Buy}, {Byer}, {Cabero}, {Cadonati}, {Cagnoli}, {Cahillane}, {Calder{\'o}n
  Bustillo}, {Callister}, {Calloni}, {Camp}, {Cannon}, {Cao}, {Cao}, {Capano},
  {Capocasa}, {Carbognani}, {Caride}, {Casanueva Diaz}, {Casentini}, {Caudill},
  {Cavagli{\`a}}, {Cavalier}, {Cavalieri}, {Cella}, {Cepeda}, {Cerboni
  Baiardi}, {Cerretani}, {Cesarini}, {Chamberlin}, {Chan}, {Chao}, {Charlton},
  {Chassande-Mottin}, {Cheeseboro}, {Chen}, {Chen}, {Cheng}, {Chincarini},
  {Chiummo}, {Chmiel}, {Cho}, {Cho}, {Chow}, {Christensen}, {Chu}, {Chua},
  {Chua}, {Chung}, {Ciani}, {Clara}, {Clark}, {Cleva}, {Cocchieri}, {Coccia},
  {Cohadon}, {Colla}, {Collette}, {Cominsky}, {Constancio}, {Conti}, {Cooper},
  {Corbitt}, {Cornish}, {Corsi}, {Cortese}, {Costa}, {Coughlin}, {Coughlin},
  {Coulon}, {Countryman}, {Couvares}, {Covas}, {Cowan}, {Coward}, {Cowart},
  {Coyne}, {Coyne}, {Creighton}, {Creighton}, {Cripe}, {Crowder}, {Cullen},
  {Cumming}, {Cunningham}, {Cuoco}, {Dal Canton}, {Danilishin}, {D'Antonio},
  {Danzmann}, {Dasgupta}, {Da Silva Costa}, {Dattilo}, {Dave}, {Davier},
  {Davies}, {Davis}, {Daw}, {Day}, {Day}, {De}, {DeBra}, {Debreczeni},
  {Degallaix}, {De Laurentis}, {Del{\'e}glise}, {Del Pozzo}, {Denker}, {Dent},
  {Dergachev}, {De Rosa}, {DeRosa}, {DeSalvo}, {Devine}, {Dhurandhar},
  {D{\'\i}az}, {Fiore}, {Giovanni}, {Girolamo}, {Lieto}, {Pace}, {Palma},
  {Virgilio}, {Doctor}, {Doi}, {Dolique}, {Donovan}, {Dooley}, {Doravari},
  {Dorrington}, {Douglas}, {Dovale {\'A}lvarez}, {Downes}, {Drago}, {Drever},
  {Driggers}, {Du}, {Ducrot}, {Dwyer}, {Eda}, {Edo}, {Edwards}, {Effler},
  {Eggenstein}, {Ehrens}, {Eichholz}, {Eikenberry}, {Eisenstein}, {Essick},
  {Etienne}, {Etzel}, {Evans}, {Evans}, {Everett}, {Factourovich}, {Fafone},
  {Fair}, {Fairhurst}, {Fan}, {Farinon}, {Farr}, {Farr}, {Fauchon-Jones},
  {Favata}, {Fays}, {Fehrmann}, {Fejer}, {Fern{\'a}ndez Galiana}, {Ferrante},
  {Ferreira}, {Ferrini}, {Fidecaro}, {Fiori}, {Fiorucci}, {Fisher}, {Flaminio},
  {Fletcher}, {Fong}, {Forsyth}, {Fournier}, {Frasca}, {Frasconi}, {Frei},
  {Freise}, {Frey}, {Frey}, {Fries}, {Fritschel}, {Frolov}, {Fujii},
  {Fujimoto}, {Fulda}, {Fyffe}, {Gabbard}, {Gadre}, {Gaebel}, {Gair},
  {Gammaitoni}, {Gaonkar}, {Garufi}, {Gaur}, {Gayathri}, {Gehrels}, {Gemme},
  {Genin}, {Gennai}, {George}, {Gergely}, {Germain}, {Ghonge}, {Ghosh},
  {Ghosh}, {Ghosh}, {Giaime}, {Giardina}, {Giazotto}, {Gill}, {Glaefke},
  {Goetz}, {Goetz}, {Gondan}, {Gonz{\'a}lez}, {Gonzalez Castro}, {Gopakumar},
  {Gorodetsky}, {Gossan}, {Gosselin}, {Gouaty}, {Grado}, {Graef}, {Granata},
  {Grant}, {Gras}, {Gray}, {Greco}, {Green}, {Groot}, {Grote}, {Grunewald},
  {Guidi}, {Guo}, {Gupta}, {Gupta}, {Gushwa}, {Gustafson}, {Gustafson},
  {Hacker}, {Hagiwara}, {Hall}, {Hall}, {Hammond}, {Haney}, {Hanke}, {Hanks},
  {Hanna}, {Hannam}, {Hanson}, {Hardwick}, {Harms}, {Harry}, {Harry}, {Hart},
  {Hartman}, {Haster}, {Haughian}, {Hayama}, {Healy}, {Heidmann}, {Heintze},
  {Heitmann}, {Hello}, {Hemming}, {Hendry}, {Heng}, {Hennig}, {Henry},
  {Heptonstall}, {Heurs}, {Hild}, {Hirose}, {Hoak}, {Hofman}, {Holt}, {Holz},
  {Hopkins}, {Hough}, {Houston}, {Howell}, {Hu}, {Huerta}, {Huet}, {Hughey},
  {Husa}, {Huttner}, {Huynh-Dinh}, {Indik}, {Ingram}, {Inta}, {Ioka}, {Isa},
  {Isac}, {Isi}, {Isogai}, {Itoh}, {Iyer}, {Izumi}, {Jacqmin}, {Jani},
  {Jaranowski}, {Jawahar}, {Jim{\'e}nez-Forteza}, {Johnson}, {Jones}, {Jones},
  {Jonker}, {Ju}, {Junker}, {Kagawa}, {Kajita}, {Kakizaki}, {Kalaghatgi},
  {Kalogera}, {Kamiizumi}, {Kanda}, {Kand hasamy}, {Kanemura}, {Kaneyama},
  {Kang}, {Kanner}, {Karki}, {Karvinen}, {Kasprzack}, {Kataoka},
  {Katsavounidis}, {Katzman}, {Kaufer}, {Kaur}, {Kawabe}, {Kawai}, {Kawamura},
  {K{\'e}f{\'e}lian}, {Keitel}, {Kelley}, {Kennedy}, {Key}, {Khalili}, {Khan},
  {Khan}, {Khan}, {Khazanov}, {Kijbunchoo}, {Kim}, {Kim}, {Kim}, {Kim}, {Kim},
  {Kim}, {Kimbrell}, {Kimura}, {King}, {King}, {Kirchhoff}, {Kissel}, {Klein},
  {Kleybolte}, {Klimenko}, {Koch}, {Koehlenbeck}, {Kojima}, {Kokeyama},
  {Koley}, {Komori}, {Kondrashov}, {Kontos}, {Korobko}, {Korth}, {Kotake},
  {Kowalska}, {Kozak}, {Kr{\"a}mer}, {Kringel}, {Krishnan}, {Kr{\'o}lak},
  {Kuehn}, {Kumar}, {Kumar}, {Kumar}, {Kuo}, {Kuroda}, {Kutynia}, {Kuwahara},
  {Lackey}, {Landry}, {Lang}, {Lange}, {Lantz}, {Lanza}, {Lartaux-Vollard},
  {Lasky}, {Laxen}, {Lazzarini}, {Lazzaro}, {Leaci}, {Leavey}, {Lebigot},
  {Lee}, {Lee}, {Lee}, {Lee}, {Lee}, {Lehmann}, {Lenon}, {Leonardi}, {Leong},
  {Leroy}, {Letendre}, {Levin}, {Li}, {Libson}, {Littenberg}, {Liu},
  {Lockerbie}, {Lombardi}, {London}, {Lord}, {Lorenzini}, {Loriette},
  {Lormand}, {Losurdo}, {Lough}, {Lousto}, {Lovelace}, {L{\"u}ck}, {Lundgren},
  {Lynch}, {Ma}, {Macfoy}, {Machenschalk}, {MacInnis}, {Macleod},
  {Maga{\~n}a-Sandoval}, {Majorana}, {Maksimovic}, {Malvezzi}, {Man}, {Mandic},
  {Mangano}, {Mano}, {Mansell}, {Manske}, {Mantovani}, {Marchesoni}, {Marchio},
  {Marion}, {M{\'a}rka}, {M{\'a}rka}, {Markosyan}, {Maros}, {Martelli},
  {Martellini}, {Martin}, {Martynov}, {Mason}, {Masserot}, {Massinger},
  {Masso-Reid}, {Mastrogiovanni}, {Matichard}, {Matone}, {Matsumoto},
  {Matsushima}, {Mavalvala}, {Mazumder}, {McCarthy}, {McClelland },
  {McCormick}, {McGrath}, {McGuire}, {McIntyre}, {McIver}, {McManus}, {McRae},
  {McWilliams}, {Meacher}, {Meadors}, {Meidam}, {Melatos}, {Mendell},
  {Mendoza-Gand ara}, {Mercer}, {Merilh}, {Merzougui}, {Meshkov}, {Messenger},
  {Messick}, {Metzdorff}, {Meyers}, {Mezzani}, {Miao}, {Michel}, {Michimura},
  {Middleton}, {Mikhailov}, {Milano}, {Miller}, {Miller}, {Miller}, \&
  {Miller}}]{2018LRR....21....3A}
---. 2018{\natexlab{b}},
  \href{http://dx.doi.org/10.1007/s41114-018-0012-9}{\JournalTitle{Living
  Reviews in Relativity}, 21, 3}

\bibitem[{{Abbott} {et~al.}(2019){Abbott}, {Abbott}, {Abbott}, {Acernese},
  {Ackley}, {Adams}, {Adams}, {Addesso}, {Adhikari}, {Adya}, {Affeldt},
  {Agarwal}, {Agathos}, {Agatsuma}, {Aggarwal}, {Aguiar}, {Aiello}, {Ain},
  {Ajith}, {Allen}, {Allen}, {Allocca}, {Aloy}, {Altin}, {Amato}, {Ananyeva},
  {Anderson}, {Anderson}, {Angelova}, {Antier}, {Appert}, {Arai}, {Araya},
  {Areeda}, {Ar{\`e}ne}, {Arnaud}, {Arun}, {Ascenzi}, {Ashton}, {Ast}, {Aston},
  {Astone}, {Atallah}, {Aubin}, {Aufmuth}, {Aulbert}, {AultONeal}, {Austin},
  {Avila-Alvarez}, {Babak}, {Bacon}, {Badaracco}, {Bader}, {Bae}, {Baker},
  {Baldaccini}, {Ballardin}, {Ballmer}, {Banagiri}, {Barayoga}, {Barclay},
  {Barish}, {Barker}, {Barkett}, {Barnum}, {Barone}, {Barr}, {Barsotti},
  {Barsuglia}, {Barta}, {Bartlett}, {Bartos}, {Bassiri}, {Basti}, {Batch},
  {Bawaj}, {Bayley}, {Bazzan}, {B{\'e}csy}, {Beer}, {Bejger}, {Belahcene},
  {Bell}, {Beniwal}, {Bensch}, {Berger}, {Bergmann}, {Bernuzzi}, {Bero},
  {Berry}, {Bersanetti}, {Bertolini}, {Betzwieser}, {Bhandare}, {Bilenko},
  {Bilgili}, {Billingsley}, {Billman}, {Birch}, {Birney}, {Birnholtz},
  {Biscans}, {Biscoveanu}, {Bisht}, {Bitossi}, {Bizouard}, {Blackburn},
  {Blackman}, {Blair}, {Blair}, {Blair}, {Bloemen}, {Bock}, {Bode}, {Boer},
  {Boetzel}, {Bogaert}, {Bohe}, {Bondu}, {Bonilla}, {Bonnand}, {Booker},
  {Boom}, {Booth}, {Bork}, {Boschi}, {Bose}, {Bossie}, {Bossilkov}, {Bosveld},
  {Bouffanais}, {Bozzi}, {Bradaschia}, {Brady}, {Bramley}, {Branchesi}, {Brau},
  {Briant}, {Brighenti}, {Brillet}, {Brinkmann}, {Brisson}, {Brockill},
  {Brooks}, {Brown}, {Brunett}, {Buchanan}, {Buikema}, {Bulik}, {Bulten},
  {Buonanno}, {Buskulic}, {Buy}, {Byer}, {Cabero}, {Cadonati}, {Cagnoli},
  {Cahillane}, {Calder{\'o}n Bustillo}, {Callister}, {Calloni}, {Camp},
  {Canepa}, {Canizares}, {Cannon}, {Cao}, {Cao}, {Capano}, {Capocasa},
  {Carbognani}, {Caride}, {Carney}, {Casanueva Diaz}, {Casentini}, {Caudill},
  {Cavagli{\`a}}, {Cavalier}, {Cavalieri}, {Cella}, {Cepeda},
  {Cerd{\'a}-Dur{\'a}n}, {Cerretani}, {Cesarini}, {Chaibi}, {Chamberlin},
  {Chan}, {Chao}, {Charlton}, {Chase}, {Chassande-Mottin}, {Chatterjee},
  {Chatziioannou}, {Cheeseboro}, {Chen}, {Chen}, {Chen}, {Cheng}, {Chia},
  {Chincarini}, {Chiummo}, {Chmiel}, {Cho}, {Cho}, {Chow}, {Christensen},
  {Chu}, {Chua}, {Chua}, {Chung}, {Chung}, {Ciani}, {Ciobanu}, {Ciolfi},
  {Cipriano}, {Cirelli}, {Cirone}, {Clara}, {Clark}, {Clearwater}, {Cleva},
  {Cocchieri}, {Coccia}, {Cohadon}, {Cohen}, {Colla}, {Collette}, {Collins},
  {Cominsky}, {Constancio}, {Conti}, {Cooper}, {Corban}, {Corbitt},
  {Cordero-Carri{\'o}n}, {Corley}, {Cornish}, {Corsi}, {Cortese}, {Costa},
  {Cotesta}, {Coughlin}, {Coughlin}, {Coulon}, {Countryman}, {Couvares},
  {Covas}, {Cowan}, {Coward}, {Cowart}, {Coyne}, {Coyne}, {Creighton},
  {Creighton}, {Cripe}, {Crowder}, {Cullen}, {Cumming}, {Cunningham}, {Cuoco},
  {Canton}, {D{\'a}lya}, {Danilishin}, {D'Antonio}, {Danzmann}, {Dasgupta},
  {Costa}, {Dattilo}, {Dave}, {Davier}, {Davis}, {Daw}, {Day}, {DeBra},
  {Deenadayalan}, {Degallaix}, {De Laurentis}, {Del{\'e}glise}, {Del Pozzo},
  {Demos}, {Denker}, {Dent}, {De Pietri}, {Derby}, {Dergachev}, {De Rosa}, {De
  Rossi}, {DeSalvo}, {de Varona}, {Dhurandhar}, {D{\'\i}az}, {Di Fiore}, {Di
  Giovanni}, {Di Girolamo}, {Di Lieto}, {Ding}, {Di Pace}, {Di Palma}, {Di
  Renzo}, {Dmitriev}, {Doctor}, {Dolique}, {Donovan}, {Dooley}, {Doravari},
  {Dorrington}, {Dovale {\'A}lvarez}, {Downes}, {Drago}, {Dreissigacker},
  {Driggers}, {Du}, {Dupej}, {Dwyer}, {Easter}, {Edo}, {Edwards}, {Effler},
  {Eggenstein}, {Ehrens}, {Eichholz}, {Eikenberry}, {Eisenmann}, {Eisenstein},
  {Essick}, {Estelles}, {Estevez}, {Etienne}, {Etzel}, {Evans}, {Evans},
  {Fafone}, {Fair}, {Fairhurst}, {Fan}, {Farinon}, {Farr}, {Farr},
  {Fauchon-Jones}, {Favata}, {Fays}, {Fee}, {Fehrmann}, {Feicht}, {Fejer},
  {Feng}, {Fernandez-Galiana}, {Ferrante}, {Ferreira}, {Ferrini}, {Fidecaro},
  {Fiori}, {Fiorucci}, {Fishbach}, {Fisher}, {Fishner}, {Fitz-Axen},
  {Flaminio}, {Fletcher}, {Fong}, {Font}, {Forsyth}, {Forsyth}, {Fournier},
  {Frasca}, {Frasconi}, {Frei}, {Freise}, {Frey}, {Frey}, {Fritschel},
  {Frolov}, {Fulda}, {Fyffe}, {Gabbard}, {Gadre}, {Gaebel}, {Gair},
  {Gammaitoni}, {Ganija}, {Gaonkar}, {Garcia}, {Garc{\'\i}a-Quir{\'o}s},
  {Garufi}, {Gateley}, {Gaudio}, {Gaur}, {Gayathri}, {Gemme}, {Genin},
  {Gennai}, {George}, {George}, {Gergely}, {Germain}, {Ghonge}, {Ghosh},
  {Ghosh}, {Ghosh}, {Giacomazzo}, {Giaime}, {Giardina}, {Giazotto}, {Gill},
  {Giordano}, {Glover}, {Goetz}, {Goetz}, {Goncharov}, {Gonz{\'a}lez},
  {Gonzalez Castro}, {Gopakumar}, {Gorodetsky}, {Gossan}, {Gosselin}, {Gouaty},
  {Grado}, {Graef}, {Granata}, {Grant}, {Gras}, {Gray}, {Greco}, {Green},
  {Green}, {Gretarsson}, {Groot}, {Grote}, {Grunewald}, {Gruning}, {Guidi},
  {Gulati}, {Guo}, {Gupta}, {Gupta}, {Gushwa}, {Gustafson}, {Gustafson},
  {Halim}, {Hall}, {Hall}, {Hamilton}, {Hamilton}, {Hammond}, {Haney}, {Hanke},
  {Hanks}, {Hanna}, {Hannam}, {Hannuksela}, {Hanson}, {Hardwick}, {Harms},
  {Harry}, {Harry}, {Hart}, {Haster}, {Haughian}, {Healy}, {Heidmann},
  {Heintze}, {Heitmann}, {Hello}, {Hemming}, {Hendry}, {Heng}, {Hennig},
  {Heptonstall}, {Hernandez}, {Heurs}, {Hild}, {Hinderer}, {Hoak}, {Hochheim},
  {Hofman}, {Holland}, {Holt}, {Holz}, {Hopkins}, {Horst}, {Hough}, {Houston},
  {Howell}, {Hreibi}, {Huerta}, {Huet}, {Hughey}, {Hulko}, {Husa}, {Huttner},
  {Huynh-Dinh}, {Iess}, {Indik}, {Ingram}, {Inta}, {Intini}, {Isa}, {Isac},
  {Isi}, {Iyer}, {Izumi}, {Jacqmin}, {Jani}, {Jaranowski}, {Johnson},
  {Johnson}, {Jones}, {Jones}, {Jonker}, {Ju}, {Junker}, {Kalaghatgi},
  {Kalogera}, {Kamai}, {Kand hasamy}, {Kang}, {Kanner}, {Kapadia}, {Karki},
  {Karvinen}, {Kasprzack}, {Katolik}, {Katsanevas}, {Katsavounidis}, {Katzman},
  {Kaufer}, {Kawabe}, {Keerthana}, {K{\'e}f{\'e}lian}, {Keitel}, {Kemball},
  {Kennedy}, {Key}, {Khalili}, {Khamesra}, {Khan}, {Khan}, {Khan}, {Khan},
  {Khazanov}, {Kijbunchoo}, {Kim}, {Kim}, {Kim}, {Kim}, {Kim}, {Kim}, {King},
  {King}, {Kinley-Hanlon}, {Kirchhoff}, {Kissel}, {Kleybolte}, {Klimenko},
  {Knowles}, {Koch}, {Koehlenbeck}, {Koley}, {Kondrashov}, {Kontos}, {Korobko},
  {Korth}, {Kowalska}, {Kozak}, {Kr{\"a}mer}, {Kringel}, {Krishnan},
  {Kr{\'o}lak}, {Kuehn}, {Kumar}, {Kumar}, {Kumar}, {Kuo}, {Kutynia}, {Kwang},
  {Lackey}, {Lai}, {Landry}, {Lang}, {Lange}, {Lantz}, {Lanza},
  {Lartaux-Vollard}, {Lasky}, {Laxen}, {Lazzarini}, {Lazzaro}, {Leaci},
  {Leavey}, {Lee}, {Lee}, {Lee}, {Lee}, {Lee}, {Lehmann}, {Lenon}, {Leonardi},
  {Leroy}, {Letendre}, {Levin}, {Li}, {Li}, {Li}, {Linker}, {Littenberg},
  {Liu}, {Liu}, {Lo}, {Lockerbie}, {London}, {Longo}, {Lorenzini}, {Loriette},
  {Lormand}, {Losurdo}, {Lough}, {Lovelace}, {L{\"u}ck}, {Lumaca}, {Lundgren},
  {Lynch}, {Ma}, {Macas}, {Macfoy}, {Machenschalk}, {MacInnis}, {Macleod},
  {Maga{\~n}a Hernandez}, {Maga{\~n}a-Sand oval}, {Maga{\~n}a Zertuche},
  {Magee}, {Majorana}, {Maksimovic}, {Man}, {Mandic}, {Mangano}, {Mansell},
  {Manske}, {Mantovani}, {Marchesoni}, {Marion}, {M{\'a}rka}, {M{\'a}rka},
  {Markakis}, {Markosyan}, {Markowitz}, {Maros}, {Marquina}, {Martelli},
  {Martellini}, {Martin}, {Martin}, {Martynov}, {Mason}, {Massera}, {Masserot},
  {Massinger}, {Masso-Reid}, {Mastrogiovanni}, {Matas}, {Matichard}, {Matone},
  {Mavalvala}, {Mazumder}, {McCann}, {McCarthy}, {McClelland }, {McCormick},
  {McCuller}, {McGuire}, {McIver}, {McManus}, {McRae}, {McWilliams}, {Meacher},
  {Meadors}, {Mehmet}, {Meidam}, {Mejuto-Villa}, {Melatos}, {Mendell},
  {Mendoza-Gandara}, {Mercer}, {Mereni}, {Merilh}, {Merzougui}, {Meshkov},
  {Messenger}, {Messick}, {Metzdorff}, {Meyers}, {Miao}, {Michel}, {Middleton},
  {Mikhailov}, {Milano}, {Miller}, {Miller}, {Miller}, {Miller}, {Millhouse},
  {Mills}, {Milovich-Goff}, {Minazzoli}, {Minenkov}, {Ming}, {Mishra}, {Mitra},
  {Mitrofanov}, {Mitselmakher}, {Mittleman}, {Moffa}, {Mogushi}, {Mohan},
  {Mohapatra}, {Montani}, {Moore}, {Moraru}, {Moreno}, {Morisaki}, {Mours},
  {Mow-Lowry}, {Mueller}, {Muir}, {Mukherjee}, {Mukherjee}, {Mukherjee},
  {Mukund}, {Mullavey}, {Munch}, {Mu{\~n}iz}, {Muratore}, {Murray}, {Nagar},
  {Napier}, {Nardecchia}, {Naticchioni}, {Nayak}, {Neilson}, {Nelemans},
  {Nelson}, {Nery}, {Neunzert}, {Nevin}, {Newport}, {Ng}, {Ng}, {Nguyen},
  {Nguyen}, {Nichols}, {Nielsen}, {Nissanke}, {Nitz}, {Nocera}, {Nolting},
  {North}, {Nuttall}, {Obergaulinger}, {Oberling}, {O'Brien}, {O'Dea}, {Ogin},
  {Oh}, {Oh}, {Ohme}, {Ohta}, {Okada}, {Oliver}, {Oppermann}, {Oram},
  {O'Reilly}, {Ormiston}, {Ortega}, {O'Shaughnessy}, {Ossokine}, {Ottaway},
  {Overmier}, {Owen}, {Pace}, {Pagano}, {Page}, {Page}, {Pai}, {Pai},
  {Palamos}, {Palashov}, {Palomba}, {Pal-Singh}, {Pan}, {Pan}, {Pang}, {Pang},
  {Pankow}, {Pannarale}, {Pant}, {Paoletti}, {Paoli}, {Papa}, {Parida},
  {Parker}, {Pascucci}, {Pasqualetti}, {Passaquieti}, {Passuello}, {Patil},
  {Patricelli}, {Pearlstone}, {Pedersen}, {Pedraza}, {Pedurand}, {Pekowsky},
  {Pele}, {Penn}, {Perez}, {Perreca}, {Perri}, {Pfeiffer}, {Phelps}, {Phukon},
  {Piccinni}, {Pichot}, {Piergiovanni}, {Pierro}, {Pillant}, {Pinard}, {Pinto},
  {Pirello}, {Pitkin}, {Poggiani}, {Popolizio}, {Porter}, {Possenti}, {Post},
  {Powell}, {Prasad}, {Pratt}, {Pratten}, {Predoi}, {Prestegard}, {Principe},
  {Privitera}, {Prodi}, {Prokhorov}, {Puncken}, {Punturo}, {Puppo},
  {P{\"u}rrer}, {Qi}, {Quetschke}, {Quintero}, {Quitzow-James}, {Rabeling},
  {Radkins}, {Raffai}, {Raja}, {Rajan}, {Rajbhandari}, {Rakhmanov}, {Ramirez},
  {Ramos-Buades}, {Rana}, {Rapagnani}, {Raymond}, {Razzano}, {Read},
  {Regimbau}, {Rei}, {Reid}, {Reitze}, {Ren}, {Ricci}, {Ricker}, {Riles},
  {Rizzo}, {Robertson}, {Robie}, {Robinet}, {Robson}, {Rocchi}, {Rolland},
  {Rollins}, {Roma}, {Romano}, {Romel}, {Romie}, {Rosi{\'n}ska}, {Ross},
  {Rowan}, {R{\"u}diger}, {Ruggi}, {Rutins}, {Ryan}, {Sachdev}, {Sadecki},
  {Sakellariadou}, {Salconi}, {Saleem}, {Salemi}, {Samajdar}, {Sammut},
  {Sampson}, {Sanchez}, {Sanchez}, {Sanchis-Gual}, {Sandberg}, {Sand ers},
  {Sarin}, {Sassolas}, {Saulson}, {Sauter}, {Savage}, {Sawadsky}, {Schale},
  {Scheel}, {Scheuer}, {Schmidt}, {Schnabel}, {Schofield}, {Sch{\"o}nbeck},
  {Schreiber}, {Schuette}, {Schulte}, {Schutz}, {Schwalbe}, {Scott}, {Scott},
  {Seidel}, {Sellers}, {Sengupta}, {Sentenac}, {Sequino}, {Sergeev},
  {Setyawati}, {Shaddock}, {Shaffer}, {Shah}, {Shahriar}, {Shaner}, {Shao},
  {Shapiro}, {Shawhan}, {Shen}, {Shoemaker}, {Shoemaker}, {Siellez}, {Siemens},
  {Sieniawska}, {Sigg}, {Silva}, {Singer}, {Singh}, {Singhal}, {Sintes},
  {Slagmolen}, {Slaven-Blair}, {Smith}, {Smith}, {Smith}, {Somala}, {Son},
  {Sorazu}, {Sorrentino}, {Souradeep}, {Spencer}, {Srivastava}, {Staats},
  {Steinke}, {Steinlechner}, {Steinlechner}, {Steinmeyer}, {Steltner},
  {Stevenson}, {Stocks}, {Stone}, {Stops}, {Strain}, {Stratta}, {Strigin},
  {Strunk}, {Sturani}, {Stuver}, {Summerscales}, {Sun}, {Sunil}, {Suresh},
  {Sutton}, {Swinkels}, {Szczepa{\'n}czyk}, {Tacca}, {Tait}, {Talbot},
  {Talukder}, {Tanner}, {T{\'a}pai}, {Taracchini}, {Tasson}, {Taylor},
  {Taylor}, {Tewari}, {Theeg}, {Thies}, {Thomas}, {Thomas}, {Thomas}, {Thorne},
  {Thrane}, {Tiwari}, {Tiwari}, {Tokmakov}, {Toland}, {Tonelli}, {Tornasi},
  {Torres-Forn{\'e}}, {Torrie}, {T{\"o}yr{\"a}}, {Travasso}, {Traylor},
  {Trinastic}, {Tringali}, {Trozzo}, {Tsang}, {Tse}, {Tso}, {Tsuna}, {Tsukada},
  {Tuyenbayev}, {Ueno}, {Ugolini}, {Urban}, {Usman}, {Vahlbruch}, {Vajente},
  {Valdes}, {van Bakel}, {van Beuzekom}, {van den Brand}, {Van Den Broeck},
  {Vander-Hyde}, {van der Schaaf}, {van Heijningen}, {van Veggel}, {Vardaro},
  {Varma}, {Vass}, {Vas{\'u}th}, {Vecchio}, {Vedovato}, {Veitch}, {Veitch},
  {Venkateswara}, {Venugopalan}, {Verkindt}, {Vetrano}, {Vicer{\'e}}, {Viets},
  \& {Virgo Collaboration}}]{2019PhRvL.122f1104A}
---. 2019,
  \href{http://dx.doi.org/10.1103/PhysRevLett.122.061104}{\JournalTitle{\prl},
  122, 061104}

\bibitem[{{Accadia} {et~al.}(2012){Accadia}, {Acernese}, {Alshourbagy},
  {Amico}, {Antonucci}, {Aoudia}, {Arnaud}, {Arnault}, {Arun}, {Astone},
  {Avino}, {Babusci}, {Ballardin}, {Barone}, {Barrand}, {Barsotti},
  {Barsuglia}, {Basti}, {Bauer}, {Beauville}, {Bebronne}, {Bejger}, {Beker},
  {Bellachia}, {Belletoile}, {Beney}, {Bernardini}, {Bigotta}, {Bilhaut},
  {Birindelli}, {Bitossi}, {Bizouard}, {Blom}, {Boccara}, {Boget}, {Bondu},
  {Bonelli}, {Bonnand}, {Boschi}, {Bosi}, {Bouedo}, {Bouhou}, {Bozzi},
  {Bracci}, {Braccini}, {Bradaschia}, {Branchesi}, {Briant}, {Brillet},
  {Brisson}, {Brocco}, {Bulik}, {Bulten}, {Buskulic}, {Buy}, {Cagnoli},
  {Calamai}, {Calloni}, {Campagna}, {Canuel}, {Carbognani}, {Carbone},
  {Cavalier}, {Cavalieri}, {Cecchi}, {Cella}, {Cesarini}, {Chassande-Mottin},
  {Chatterji}, {Chiche}, {Chincarini}, {Chiummo}, {Christensen}, {Clapson},
  {Cleva}, {Coccia}, {Cohadon}, {Colacino}, {Colas}, {Colla}, {Colombini},
  {Conforto}, {Corsi}, {Cortese}, {Cottone}, {Coulon}, {Cuoco}, {D'Antonio},
  {Daguin}, {Dari}, {Dattilo}, {David}, {Davier}, {Day}, {Debreczeni}, \& {De
  Carolis}}]{2012JInst...7.3012A}
{Accadia}, T., {Acernese}, F., {Alshourbagy}, M., {et~al.} 2012,
  \href{http://dx.doi.org/10.1088/1748-0221/7/03/P03012}{\JournalTitle{Journal
  of Instrumentation}, 7, 3012}

\bibitem[{{Arcavi}(2018)}]{2018ApJ...855L..23A}
{Arcavi}, I. 2018,
  \href{http://dx.doi.org/10.3847/2041-8213/aab267}{\JournalTitle{\apj}, 855,
  L23}

\bibitem[{Barsotti {et~al.}(2018)Barsotti, McCuller, Evans, \&
  Fritschel}]{LIGOT1800042}
Barsotti, L., McCuller, L., Evans, M., {et~al.} 2018, {The A+ design curve},
  Tech. rep., LIGO

\bibitem[{{Bauswein} {et~al.}(2017){Bauswein}, {Just}, {Janka}, \&
  {Stergioulas}}]{Bauswein+17}
{Bauswein}, A., {Just}, O., {Janka}, H.-T., {et~al.} 2017,
  \href{http://dx.doi.org/10.3847/2041-8213/aa9994}{\JournalTitle{\apjl}, 850,
  L34}

\bibitem[{{Bauswein} {et~al.}(2014){Bauswein}, {Stergioulas}, \&
  {Janka}}]{Bauswein+14}
{Bauswein}, A., {Stergioulas}, N., \& {Janka}, H.-T. 2014,
  \href{http://dx.doi.org/10.1103/PhysRevD.90.023002}{\JournalTitle{\prd}, 90,
  023002}

\bibitem[{{Burbidge} {et~al.}(1957){Burbidge}, {Burbidge}, {Fowler}, \&
  {Hoyle}}]{Burbidge+57}
{Burbidge}, E.~M., {Burbidge}, G.~R., {Fowler}, W.~A., {et~al.} 1957,
  \href{http://dx.doi.org/10.1103/RevModPhys.29.547}{\JournalTitle{Reviews of
  Modern Physics}, 29, 547}

\bibitem[{{Cannon} {et~al.}(2012){Cannon}, {Cariou}, {Chapman},
  {Crispin-Ortuzar}, {Fotopoulos}, {Frei}, {Hanna}, {Kara}, {Keppel}, {Liao},
  {Privitera}, {Searle}, {Singer}, \& {Weinstein}}]{2012ApJ...748..136C}
{Cannon}, K., {Cariou}, R., {Chapman}, A., {et~al.} 2012,
  \href{http://dx.doi.org/10.1088/0004-637X/748/2/136}{\JournalTitle{\apj},
  748, 136}

\bibitem[{{Chan} {et~al.}(2018){Chan}, {Messenger}, {Heng}, \&
  {Hendry}}]{2018PhRvD..97l3014C}
{Chan}, M.~L., {Messenger}, C., {Heng}, I.~S., {et~al.} 2018,
  \href{http://dx.doi.org/10.1103/PhysRevD.97.123014}{\JournalTitle{\prd}, 97,
  123014}

\bibitem[{{Chen} {et~al.}(2018){Chen}, {Fishbach}, \&
  {Holz}}]{2018Natur.562..545C}
{Chen}, H.-Y., {Fishbach}, M., \& {Holz}, D.~E. 2018,
  \href{http://dx.doi.org/10.1038/s41586-018-0606-0}{\JournalTitle{\nat}, 562,
  545}

\bibitem[{{Chen} {et~al.}(2017){Chen}, {Holz}, {Miller}, {Evans}, {Vitale}, \&
  {Creighton}}]{hsinyu2017}
{Chen}, H.-Y., {Holz}, D.~E., {Miller}, J., {et~al.} 2017, \JournalTitle{arXiv
  e-prints}, arXiv:1709.08079

\bibitem[{{Chornock} {et~al.}(2019){Chornock}, {Cowperthwaite}, {Margutti}, \&
  {Milisavljevic}}]{chornockWP}
{Chornock}, R., {Cowperthwaite}, P.~S., {Margutti}, R., {et~al.} 2019,
  \JournalTitle{Astro2020 White Paper: ``Multi-Messenger Astronomy with
  Extremely Large Telescopes"}

\bibitem[{{Connaughton} {et~al.}(2016){Connaughton}, {Burns}, {Goldstein},
  {Blackburn}, {Briggs}, {Zhang}, {Camp}, {Christensen}, {Hui}, {Jenke},
  {Littenberg}, {McEnery}, {Racusin}, {Shawhan}, {Singer}, {Veitch},
  {Wilson-Hodge}, {Bhat}, {Bissaldi}, {Cleveland }, {Fitzpatrick}, {Giles},
  {Gibby}, {von Kienlin}, {Kippen}, {McBreen}, {Mailyan}, {Meegan}, {Paciesas},
  {Preece}, {Roberts}, {Sparke}, {Stanbro}, {Toelge}, \&
  {Veres}}]{2016ApJ...826L...6C}
{Connaughton}, V., {Burns}, E., {Goldstein}, A., {et~al.} 2016,
  \href{http://dx.doi.org/10.3847/2041-8205/826/1/L6}{\JournalTitle{\apj}, 826,
  L6}

\bibitem[{{Coughlin} {et~al.}(2018){Coughlin}, {Dietrich}, {Margalit}, \&
  {Metzger}}]{Coughlin+18}
{Coughlin}, M.~W., {Dietrich}, T., {Margalit}, B., {et~al.} 2018,
  \JournalTitle{arXiv e-prints},
  \href{http://arxiv.org/abs/1812.04803}{{\sffamily arXiv:1812.04803
  [astro-ph.HE]}}

\bibitem[{{Cowperthwaite} \& {Berger}(2015)}]{CowpBerger15}
{Cowperthwaite}, P.~S., \& {Berger}, E. 2015,
  \href{http://dx.doi.org/10.1088/0004-637X/814/1/25}{\JournalTitle{\apj}, 814,
  25}

\bibitem[{{Cowperthwaite} {et~al.}(2017){Cowperthwaite}, {Berger}, {Villar},
  {et~al.}}]{Cowperthwaite+17}
{Cowperthwaite}, P.~S., {Berger}, E., {Villar}, V.~A., {et~al.} 2017,
  \href{http://dx.doi.org/10.3847/2041-8213/aa8fc7}{\JournalTitle{\apjl}, 848,
  L17}

\bibitem[{{Cowperthwaite} {et~al.}(2018{\natexlab{a}}){Cowperthwaite},
  {Villar}, {Scolnic}, \& {Berger}}]{Cowp+18LSST}
{Cowperthwaite}, P.~S., {Villar}, V.~A., {Scolnic}, D.~M., {et~al.}
  2018{\natexlab{a}}, \JournalTitle{arXiv e-prints}, arXiv:1811.03098

\bibitem[{{Cowperthwaite} {et~al.}(2018{\natexlab{b}}){Cowperthwaite},
  {Berger}, {Rest}, {Chornock}, {Scolnic}, {Williams}, {Fong}, {Drout},
  {Foley}, {Margutti}, {Lunnan}, {Metzger}, \& {Quataert}}]{Cowp+18Cont}
{Cowperthwaite}, P.~S., {Berger}, E., {Rest}, A., {et~al.} 2018{\natexlab{b}},
  \href{http://dx.doi.org/10.3847/1538-4357/aabad9}{\JournalTitle{\apj}, 858,
  18}

\bibitem[{{De} {et~al.}(2018){De}, {Finstad}, {Lattimer}, {Brown}, {Berger}, \&
  {Biwer}}]{De+18}
{De}, S., {Finstad}, D., {Lattimer}, J.~M., {et~al.} 2018,
  \href{http://dx.doi.org/10.1103/PhysRevLett.121.091102}{\JournalTitle{Physical
  Review Letters}, 121, 091102}

\bibitem[{{Drout} {et~al.}(2017){Drout}, {Piro}, {Shappee},
  {et~al.}}]{Drout+17}
{Drout}, M.~R., {Piro}, A.~L., {Shappee}, B.~J., {et~al.} 2017,
  \href{http://dx.doi.org/10.1126/science.aaq0049}{\JournalTitle{Science}, 358,
  1570}

\bibitem[{{Fairhurst}(2018)}]{2018CQGra..35j5002F}
{Fairhurst}, S. 2018,
  \href{http://dx.doi.org/10.1088/1361-6382/aab675}{\JournalTitle{Classical and
  Quantum Gravity}, 35, 105002}

\bibitem[{{Farr} {et~al.}(2016){Farr}, {Berry}, {Farr}, {Haster}, {Middleton},
  {Cannon}, {Graff}, {Hanna}, {Mandel}, {Pankow}, {Price}, {Sidery}, {Singer},
  {Urban}, {Vecchio}, {Veitch}, \& {Vitale}}]{2016ApJ...825..116F}
{Farr}, B., {Berry}, C. P.~L., {Farr}, W.~M., {et~al.} 2016,
  \href{http://dx.doi.org/10.3847/0004-637X/825/2/116}{\JournalTitle{\apj},
  825, 116}

\bibitem[{{Farrow} {et~al.}(2019){Farrow}, {Zhu}, \&
  {Thrane}}]{2019arXiv190203300F}
{Farrow}, N., {Zhu}, X.-J., \& {Thrane}, E. 2019, \JournalTitle{arXiv
  e-prints}, arXiv:1902.03300

\bibitem[{{Foucart}(2012)}]{Foucart12}
{Foucart}, F. 2012,
  \href{http://dx.doi.org/10.1103/PhysRevD.86.124007}{\JournalTitle{\prd}, 86,
  124007}

\bibitem[{{Graham} {et~al.}(2019){Graham}, {Milisavljevic}, {Rest}, \&
  {Wheeler}}]{grahamWP}
{Graham}, M.~L., {Milisavljevic}, D., {Rest}, A., {et~al.} 2019,
  \JournalTitle{Astro2020 White Paper: ``Discovery Frontiers of Explosive
  Transients: An ELT \& LSST Perspectives"}

\bibitem[{{Guidorzi} {et~al.}(2017){Guidorzi}, {Margutti}, {Brout}, {Scolnic},
  {Fong}, {Alexander}, {Cowperthwaite}, {Annis}, {Berger}, {Blanchard},
  {Chornock}, {Coppejans}, {Eftekhari}, {Frieman}, {Huterer}, {Nicholl},
  {Soares-Santos}, {Terreran}, {Villar}, \& {Williams}}]{Guidorzi+17}
{Guidorzi}, C., {Margutti}, R., {Brout}, D., {et~al.} 2017,
  \href{http://dx.doi.org/10.3847/2041-8213/aaa009}{\JournalTitle{\apj}, 851,
  L36}

\bibitem[{{Holz} \& {Hughes}(2005)}]{2005ApJ...629...15H}
{Holz}, D.~E., \& {Hughes}, S.~A. 2005,
  \href{http://dx.doi.org/10.1086/431341}{\JournalTitle{\apj}, 629, 15}

\bibitem[{{Honda} {et~al.}(2006){Honda}, {Aoki}, {Ishimaru}, {Wanajo}, \&
  {Ryan}}]{Honda+06}
{Honda}, S., {Aoki}, W., {Ishimaru}, Y., {et~al.} 2006,
  \href{http://dx.doi.org/10.1086/503195}{\JournalTitle{\apj}, 643, 1180}

\bibitem[{{Horowitz} {et~al.}(2018){Horowitz}, {Arcones}, {C{\^o}t{\'e}},
  {Dillmann}, {Nazarewicz}, {Roederer}, {Schatz}, {Aprahamian}, {Atanasov},
  {Bauswein}, {Bliss}, {Brodeur}, {Clark}, {Frebel}, {Foucart}, {Hansen},
  {Just}, {Kankainen}, {McLaughlin}, {Kelly}, {Liddick}, {Lee}, {Lippuner},
  {Martin}, {Mendoza-Temis}, {Metzger}, {Mumpower}, {Perdikakis}, {Pereira},
  {O'Shea}, {Reifarth}, {Rogers}, {Siegel}, {Spyrou}, {Surman}, {Tang},
  {Uesaka}, \& {Wang}}]{Horowitz+18}
{Horowitz}, C.~J., {Arcones}, A., {C{\^o}t{\'e}}, B., {et~al.} 2018,
  \JournalTitle{arXiv e-prints},
  \href{http://arxiv.org/abs/1805.04637}{{\sffamily arXiv:1805.04637
  [astro-ph.SR]}}

\bibitem[{{IndIGO Collaboration}(2011)}]{ligoindia}
{IndIGO Collaboration}. 2011, LIGO-India, Proposal of the Consortium for Indian
  Initiative in Gravitational-wave Observations (IndIGO), Tech. Rep.
  LIGO-M1100296-v2, IndIGO,
  https://dcc.ligo.org/cgi-bin/DocDB/ShowDocument?docid=75988

\bibitem[{{Ivezi{\'c}} {et~al.}(2008){Ivezi{\'c}}, {Kahn}, {Tyson}, {Abel},
  {Acosta}, {Allsman}, {Alonso}, {AlSayyad}, {Anderson}, {Andrew}, {Angel},
  {Angeli}, {Ansari}, {Antilogus}, {Araujo}, {Armstrong}, {Arndt}, {Astier},
  {Aubourg}, {Auza}, {Axelrod}, {Bard}, {Barr}, {Barrau}, {Bartlett}, {Bauer},
  {Bauman}, {Baumont}, {Becker}, {Becla}, {Beldica}, {Bellavia}, {Bianco},
  {Biswas}, {Blanc}, {Blazek}, {Blandford}, {Bloom}, {Bogart}, {Bond},
  {Borgland}, {Borne}, {Bosch}, {Boutigny}, {Brackett}, {Bradshaw}, {Nielsen
  Brand t}, {Brown}, {Bullock}, {Burchat}, {Burke}, {Cagnoli}, {Calabrese},
  {Callahan}, \& {for the LSST Collaboration}}]{2008arXiv0805.2366I}
{Ivezi{\'c}}, {\v{Z}}., {Kahn}, S.~M., {Tyson}, J.~A., {et~al.} 2008,
  \JournalTitle{arXiv e-prints}, arXiv:0805.2366

\bibitem[{{Ji} {et~al.}(2016){Ji}, {Frebel}, {Chiti}, \& {Simon}}]{Ji+16}
{Ji}, A.~P., {Frebel}, A., {Chiti}, A., {et~al.} 2016,
  \href{http://dx.doi.org/10.1038/nature17425}{\JournalTitle{\nat}, 531, 610}

\bibitem[{{Kagra Collaboration} {et~al.}(2019){Kagra Collaboration}, {Akutsu},
  {Ando}, {Arai}, {Arai}, {Araki}, {Araya}, {Aritomi}, {Asada}, {Aso},
  {Atsuta}, {Awai}, {Bae}, {Baiotti}, {Barton}, {Cannon}, {Capocasa}, {Chen},
  {Chiu}, {Cho}, {Chu}, {Craig}, {Creus}, {Doi}, {Eda}, {Enomoto}, {Flaminio},
  {Fujii}, {Fujimoto}, {Fukunaga}, {Fukushima}, {Furuhata}, {Haino},
  {Hasegawa}, {Hashino}, {Hayama}, {Hirobayashi}, {Hirose}, {Hsieh}, {Huang},
  {Ikenoue}, {Inoue}, {Ioka}, {Itoh}, {Izumi}, {Kaji}, {Kajita}, {Kakizaki},
  {Kamiizumi}, {Kanbara}, {Kanda}, {Kanemura}, {Kaneyama}, {Kang}, {Kasuya},
  {Kataoka}, {Kawai}, {Kawamura}, {Kawasaki}, {Kim}, {Kim}, {Kim}, {Kim},
  {Kim}, {Kimura}, {Kinugawa}, {Kirii}, {Kitaoka}, {Kitazawa}, {Kojima},
  {Kokeyama}, {Komori}, {Kong}, {Kotake}, {Kozu}, {Kumar}, {Kuo}, {Kuroyanagi},
  {Lee}, {Lee}, {Lee}, {Leonardi}, {Lin}, {Lin}, {Liu}, {Liu}, {Majorana},
  {Mano}, {Marchio}, {Matsui}, {Matsushima}, {Michimura}, {Mio}, {Miyakawa},
  {Miyamoto}, {Miyamoto}, {Miyo}, {Miyoki}, {Morii}, {Morisaki}, {Moriwaki},
  {Morozumi}, {Musha}, {Nagano}, {Nagano}, {Nakamura}, {Nakamura}, {Nakano},
  {Nakano}, {Nakao}, {Narikawa}, {Naticchioni}, {Nguyen Quynh}, {Ni},
  {Nishizawa}, {Obuchi}, {Ochi}, {Oh}, {Oh}, {Ohashi}, {Ohishi}, {Ohkawa},
  {Okutomi}, {Ono}, {Oohara}, {Ooi}, {Pan}, {Park}, {Pe{\~n}a Arellano},
  {Pinto}, {Sago}, {Saijo}, {Saitou}, {Saito}, {Sakai}, {Sakai}, {Sakai},
  {Sasai}, {Sasaki}, {Sasaki}, {Sato}, {Sato}, {Sato}, {Sekiguchi}, {Seto},
  {Shibata}, {Shimoda}, {Shinkai}, {Shishido}, {Shoda}, {Somiya}, {Son},
  {Suemasa}, {Suzuki}, {Suzuki}, {Tagoshi}, {Tahara}, {Takahashi}, {Takahashi},
  {Takamori}, {Takeda}, {Tanaka}, {Tanaka}, {Tanaka}, {Tanioka}, {Tapia San
  Martin}, {Tatsumi}, {Tomaru}, {Tomura}, {Travasso}, {Tsubono}, {Tsuchida},
  {Uchikata}, {Uchiyama}, {Uehara}, {Ueki}, {Ueno}, {Uraguchi}, {Ushiba}, {van
  Putten}, {Vocca}, {Wada}, {Wakamatsu}, {Watanabe}, {Xu}, {Yamada},
  {Yamamoto}, {Yamamoto}, {Yamamoto}, {Yamamoto}, {Yamamoto}, {Yokogawa},
  {Yokoyama}, {Yokozawa}, {Yoon}, {Yoshioka}, {Yuzurihara}, {Zeidler}, \&
  {Zhu}}]{2019NatAs...3...35K}
{Kagra Collaboration}, {Akutsu}, T., {Ando}, M., {et~al.} 2019,
  \href{http://dx.doi.org/10.1038/s41550-018-0658-y}{\JournalTitle{Nature
  Astronomy}, 3, 35}

\bibitem[{{Kasen} {et~al.}(2017){Kasen}, {Metzger}, {Barnes}, {Quataert}, \&
  {Ramirez-Ruiz}}]{Kasen+17}
{Kasen}, D., {Metzger}, B., {Barnes}, J., {et~al.} 2017,
  \href{http://dx.doi.org/10.1038/nature24453}{\JournalTitle{\nat}, 551, 80}

\bibitem[{{Kim} {et~al.}(2015){Kim}, {Perera}, \&
  {McLaughlin}}]{2015MNRAS.448..928K}
{Kim}, C., {Perera}, B. B.~P., \& {McLaughlin}, M.~A. 2015,
  \href{http://dx.doi.org/10.1093/mnras/stu2729}{\JournalTitle{\mnras}, 448,
  928}

\bibitem[{{Lattimer} \& {Prakash}(2001)}]{Lattimer&Prakash01}
{Lattimer}, J.~M., \& {Prakash}, M. 2001,
  \href{http://dx.doi.org/10.1086/319702}{\JournalTitle{\apj}, 550, 426}

\bibitem[{{Lattimer} \& {Schramm}(1974)}]{Lattimer&Schramm74}
{Lattimer}, J.~M., \& {Schramm}, D.~N. 1974,
  \href{http://dx.doi.org/10.1086/181612}{\JournalTitle{\apjl}, 192, L145}

\bibitem[{{LIGO Scientific Collaboration} {et~al.}(2015){LIGO Scientific
  Collaboration}, {Aasi}, {Abbott}, {Abbott}, {Abbott}, {Abernathy}, {Ackley},
  {Adams}, {Adams}, {Addesso}, {Adhikari}, {Adya}, {Affeldt}, {Aggarwal},
  {Aguiar}, {Ain}, {Ajith}, {Alemic}, {Allen}, {Amariutei}, {Anderson},
  {Anderson}, {Arai}, {Araya}, {Arceneaux}, {Areeda}, {Ashton}, {Ast}, {Aston},
  {Aufmuth}, {Aulbert}, {Aylott}, {Babak}, {Baker}, {Ballmer}, {Barayoga},
  {Barbet}, {Barclay}, {Barish}, {Barker}, {Barr}, {Barsotti}, {Bartlett},
  {Barton}, {Bartos}, {Bassiri}, {Batch}, {Baune}, {Behnke}, {Bell}, {Bell},
  {Benacquista}, {Bergman}, {Bergmann}, {Berry}, {Betzwieser}, {Bhagwat},
  {Bhandare}, {Bilenko}, {Billingsley}, {Birch}, {Biscans}, {Biwer},
  {Blackburn}, {Blackburn}, {Blair}, {Blair}, {Bock}, {Bodiya}, {Bojtos},
  {Bond}, {Bork}, {Born}, {Bose}, {Brady}, {Braginsky}, {Brau}, {Bridges},
  {Brinkmann}, {Brooks}, {Brown}, {Brown}, {Brown}, {Buchman}, {Buikema},
  {Buonanno}, {Cadonati}, {Calder{\'o}n Bustillo}, {Camp}, {Cannon}, {Cao},
  {Capano}, {Caride}, {Caudill}, {Cavagli{\`a}}, {Cepeda}, {Chakraborty},
  {Chalermsongsak}, {Chamberlin}, {Chao}, {Charlton}, {Chen}, {Cho}, {Cho},
  {Chow}, {Christensen}, {Chu}, {Chung}, {Ciani}, {Clara}, {Clark}, {Collette},
  {Cominsky}, {Constancio}, {Cook}, {Corbitt}, {Cornish}, {Corsi}, {Costa},
  {Coughlin}, {Countryman}, {Couvares}, {Coward}, {Cowart}, {Coyne}, {Coyne},
  {Craig}, {Creighton}, {Creighton}, {Cripe}, {Crowder}, {Cumming},
  {Cunningham}, {Cutler}, {Dahl}, {Dal Canton}, {Damjanic}, {Danilishin},
  {Danzmann}, {Dartez}, {Dave}, {Daveloza}, {Davies}, {Daw}, {DeBra}, {Del
  Pozzo}, {Denker}, {Dent}, {Dergachev}, {DeRosa}, {DeSalvo}, {Dhurandhar},
  {D́{\i}az}, {Di Palma}, {Dojcinoski}, {Dominguez}, {Donovan}, {Dooley},
  {Doravari}, {Douglas}, {Downes}, {Driggers}, {Du}, {Dwyer}, {Eberle}, {Edo},
  {Edwards}, {Edwards}, {Effler}, {Eggenstein}, {Ehrens}, {Eichholz},
  {Eikenberry}, {Essick}, {Etzel}, {Evans}, {Evans}, {Factourovich},
  {Fairhurst}, {Fan}, {Fang}, {Farr}, {Farr}, {Favata}, {Fays}, {Fehrmann},
  {Fejer}, {Feldbaum}, {Ferreira}, {Fisher}, {Frei}, {Freise}, {Frey},
  {Fricke}, {Fritschel}, {Frolov}, {Fuentes-Tapia}, {Fulda}, {Fyffe}, {Gair},
  {Gaonkar}, {Gehrels}, {Gergely}, {Giaime}, {Giardina}, {Gleason}, {Goetz},
  {Goetz}, {Gondan}, {Gonz{\'a}lez}, {Gordon}, {Gorodetsky}, {Gossan},
  {Go{\ss}ler}, {Gr{\"a}f}, {Graff}, {Grant}, {Gras}, {Gray}, {Greenhalgh},
  {Gretarsson}, {Grote}, {Grunewald}, {Guido}, {Guo}, {Gushwa}, {Gustafson},
  {Gustafson}, {Hacker}, {Hall}, {Hammond}, {Hanke}, {Hanks}, {Hanna},
  {Hannam}, {Hanson}, {Hardwick}, {Harry}, {Harry}, {Hart}, {Hartman},
  {Haster}, {Haughian}, {Hee}, {Heintze}, {Heinzel}, {Hendry}, {Heng},
  {Heptonstall}, {Heurs}, {Hewitson}, {Hild}, {Hoak}, {Hodge}, {Hollitt},
  {Holt}, {Hopkins}, {Hosken}, {Hough}, {Houston}, {Howell}, {Hu}, {Huerta},
  {Hughey}, {Husa}, {Huttner}, {Huynh}, {Huynh-Dinh}, {Idrisy}, {Indik},
  {Ingram}, {Inta}, {Islas}, {Isler}, {Isogai}, {Iyer}, {Izumi}, {Jacobson},
  {Jang}, {Jawahar}, {Ji}, {Jim{\'e}nez-Forteza}, {Johnson}, {Jones}, {Jones},
  {Ju}, {Haris}, {Kalogera}, {Kand hasamy}, {Kang}, {Kanner}, {Katsavounidis},
  {Katzman}, {Kaufer}, {Kaufer}, {Kaur}, {Kawabe}, {Kawazoe}, {Keiser},
  {Keitel}, {Kelley}, {Kells}, {Keppel}, {Key}, {Khalaidovski}, {Khalili},
  {Khazanov}, {Kim}, {Kim}, {Kim}, {Kim}, {Kim}, {King}, {King}, {Kinzel},
  {Kissel}, {Klimenko}, {Kline}, {Koehlenbeck}, {Kokeyama}, {Kondrashov},
  {Korobko}, {Korth}, {Kozak}, {Kringel}, {Krishnan}, {Krueger}, {Kuehn},
  {Kumar}, {Kumar}, {Kuo}, {Landry}, {Lantz}, {Larson}, {Lasky}, {Lazzarini},
  {Lazzaro}, {Le}, {Leaci}, {Leavey}, {Lebigot}, {Lee}, {Lee}, {Lee}, {Leong},
  {Levin}, {Levine}, {Lewis}, {Li}, {Libbrecht}, {Libson}, {Lin}, {Littenberg},
  {Lockerbie}, {Lockett}, {Logue}, {Lombardi}, {Lormand}, {Lough}, {Lubinski},
  {L{\"u}ck}, {Lundgren}, {Lynch}, {Ma}, {Macarthur}, {MacDonald},
  {Machenschalk}, {MacInnis}, {Macleod}, {Maga{\~n}a-Sandoval}, {Magee},
  {Mageswaran}, {Maglione}, {Mailand}, {Mandel}, {Mandic}, {Mangano},
  {Mansell}, {M{\'a}rka}, {M{\'a}rka}, {Markosyan}, {Maros}, {Martin},
  {Martin}, {Martynov}, {Marx}, {Mason}, {Massinger}, {Matichard}, {Matone},
  {Mavalvala}, {Mazumder}, {Mazzolo}, {McCarthy}, {McClelland }, {McCormick},
  {McGuire}, {McIntyre}, {McIver}, {McLin}, {McWilliams}, {Meadors},
  {Meinders}, {Melatos}, {Mendell}, {Mercer}, {Meshkov}, {Messenger}, {Meyers},
  {Miao}, {Middleton}, {Mikhailov}, {Miller}, {Miller}, {Millhouse}, {Ming},
  {Mirshekari}, {Mishra}, {Mitra}, {Mitrofanov}, {Mitselmakher}, {Mittleman},
  {Moe}, {Mohanty}, {Mohapatra}, {Moore}, {Moraru}, {Moreno}, {Morriss},
  {Mossavi}, {Mow-Lowry}, {Mueller}, {Mueller}, {Mukherjee}, {Mullavey},
  {Munch}, {Murphy}, {Murray}, {Mytidis}, {Nash}, {Nayak}, {Necula}, {Nedkova},
  {Newton}, {Nguyen}, {Nielsen}, {Nissanke}, {Nitz}, {Nolting}, {Normandin},
  {Nuttall}, {Ochsner}, {O'Dell}, {Oelker}, {Ogin}, {Oh}, {Oh}, {Ohme},
  {Oppermann}, {Oram}, {O'Reilly}, {Ortega}, {O'Shaughnessy}, {Osthelder},
  {Ott}, {Ottaway}, {Ottens}, {Overmier}, {Owen}, {Padilla}, {Pai}, {Pai},
  {Palashov}, {Pal-Singh}, {Pan}, {Pankow}, {Pannarale}, {Pant}, {Papa},
  {Paris}, {Patrick}, {Pedraza}, {Pekowsky}, {Pele}, {Penn}, {Perreca},
  {Phelps}, {Pierro}, {Pinto}, {Pitkin}, {Poeld}, {Post}, {Poteomkin},
  {Powell}, {Prasad}, {Predoi}, {Premachandra}, {Prestegard}, {Price},
  {Principe}, {Privitera}, {Prix}, {Prokhorov}, {Puncken}, {P{\"u}rrer}, {Qin},
  {Quetschke}, {Quintero}, {Quiroga}, {Quitzow-James}, {Raab}, {Rabeling},
  {Radkins}, {Raffai}, {Raja}, {Rajalakshmi}, {Rakhmanov}, {Ramirez},
  {Raymond}, {Reed}, {Reid}, {Reitze}, {Reula}, {Riles}, {Robertson}, {Robie},
  {Rollins}, {Roma}, {Romano}, {Romanov}, {Romie}, {Rowan}, {R{\"u}diger},
  {Ryan}, {Sachdev}, {Sadecki}, {Sadeghian}, {Saleem}, {Salemi}, {Sammut},
  {Sandberg}, {Sanders}, {Sannibale}, {Santiago-Prieto}, {Sathyaprakash},
  {Saulson}, {Savage}, {Sawadsky}, {Scheuer}, {Schilling}, {Schmidt},
  {Schnabel}, {Schofield}, {Schreiber}, {Schuette}, {Schutz}, {Scott}, {Scott},
  {Sellers}, {Sengupta}, {Sergeev}, {Serna}, {Sevigny}, {Shaddock}, {Shahriar},
  {Shaltev}, {Shao}, {Shapiro}, {Shawhan}, {Shoemaker}, {Sidery}, {Siemens},
  {Sigg}, {Silva}, {Simakov}, {Singer}, {Singer}, {Singh}, {Sintes},
  {Slagmolen}, {Smith}, {Smith}, {Smith}, {Smith-Lefebvre}, {Son}, {Sorazu},
  {Souradeep}, {Staley}, {Stebbins}, {Steinke}, {Steinlechner}, {Steinlechner},
  {Steinmeyer}, {Stephens}, {Steplewski}, {Stevenson}, {Stone}, {Strain},
  {Strigin}, {Sturani}, {Stuver}, {Summerscales}, {Sutton}, {Szczepanczyk},
  {Szeifert}, {Talukder}, {Tanner}, {T{\'a}pai}, {Tarabrin}, {Taracchini},
  {Taylor}, {Tellez}, {Theeg}, {Thirugnanasambandam}, {Thomas}, {Thomas},
  {Thorne}, {Thorne}, {Thrane}, {Tiwari}, {Tomlinson}, {Torres}, {Torrie},
  {Traylor}, {Tse}, {Tshilumba}, {Ugolini}, {Unnikrishnan}, {Urban}, {Usman},
  {Vahlbruch}, {Vajente}, {Valdes}, {Vallisneri}, {van Veggel}, {Vass},
  {Vaulin}, {Vecchio}, {Veitch}, {Veitch}, {Venkateswara}, {Vincent-Finley},
  {Vitale}, {Vo}, {Vorvick}, {Vousden}, {Vyatchanin}, {Wade}, {Wade}, {Wade},
  {Walker}, {Wallace}, {Walsh}, {Wang}, {Wang}, {Wang}, {Ward}, {Warner},
  {Was}, {Weaver}, {Weinert}, {Weinstein}, {Weiss}, {Welborn}, {Wen},
  {Wessels}, {Westphal}, {Wette}, {Whelan}, {Whitcomb}, {White}, {Whiting},
  {Wilkinson}, {Williams}, {Williams}, {Williamson}, {Willis}, {Willke},
  {Wimmer}, {Winkler}, {Wipf}, {Wittel}, {Woan}, {Worden}, {Xie}, {Yablon},
  {Yakushin}, {Yam}, {Yamamoto}, {Yancey}, {Yang}, {Zanolin}, {Zhang}, {Zhang},
  {Zhang}, {Zhang}, {Zhao}, {Zhou}, {Zhu}, {Zucker}, {Zuraw}, \&
  {Zweizig}}]{2015CQGra..32g4001L}
{LIGO Scientific Collaboration}, {Aasi}, J., {Abbott}, B.~P., {et~al.} 2015,
  \href{http://dx.doi.org/10.1088/0264-9381/32/7/074001}{\JournalTitle{Classical
  and Quantum Gravity}, 32, 074001}

\bibitem[{{Margalit} \& {Metzger}(2017)}]{Margalit&Metzger17}
{Margalit}, B., \& {Metzger}, B.~D. 2017,
  \href{http://dx.doi.org/10.3847/2041-8213/aa991c}{\JournalTitle{\apjl}, 850,
  L19}

\bibitem[{{Margutti} {et~al.}(2018){Margutti}, {Cowperthwaite}, {Doctor},
  {Mortensen}, {Pankow}, {Salafia}, {Villar}, {Alexander}, {Annis}, {Andreoni},
  {Baldeschi}, {Balmaverde}, {Berger}, {Bernardini}, {Berry}, {Bianco},
  {Blanchard}, {Brocato}, {Carnerero}, {Cartier}, {Cenko}, {Chornock},
  {Chomiuk}, {Copperwheat}, {Coughlin}, {Coppejans}, {Corsi}, {D'Ammando},
  {Datrier}, {D'Avanzo}, {Dimitriadis}, {Drout}, {Foley}, {Fong}, {Fox},
  {Ghirlanda}, {Goldstein}, {Grindlay}, {Guidorzi}, {Haiman}, {Hendry}, {Holz},
  {Hung}, {Inserra}, {Jones}, {Kalogera}, {Kilpatrick}, {Lamb}, {Laskar},
  {Levan}, {Mason}, {Maguire}, {Melandri}, {Milisavljevic}, {Miller},
  {Narayan}, {Nielsen}, {Nicholl}, {Nissanke}, {Nugent}, {Pan}, {Pasham},
  {Paterson}, {Piranomonte}, {Racusin}, {Rest}, {Righi}, {Sand}, {Seaman},
  {Scolnic}, {Siellez}, {Singer}, {Szkody}, {Smith}, {Steeghs}, {Sullivan},
  {Tanvir}, {Terreran}, {Trimble}, {Valenti}, {LSST Transient}, \& {Variable
  Stars Collaboration}}]{MarguttiTOO}
{Margutti}, R., {Cowperthwaite}, P., {Doctor}, Z., {et~al.} 2018,
  \JournalTitle{arXiv e-prints}, arXiv:1812.04051

\bibitem[{{Metzger} {et~al.}(2015){Metzger}, {Bauswein}, {Goriely}, \&
  {Kasen}}]{Metzger+15}
{Metzger}, B.~D., {Bauswein}, A., {Goriely}, S., {et~al.} 2015,
  \href{http://dx.doi.org/10.1093/mnras/stu2225}{\JournalTitle{\mnras}, 446,
  1115}

\bibitem[{{Metzger} \& {Fern{\'a}ndez}(2014)}]{Metzger&Fernandez14}
{Metzger}, B.~D., \& {Fern{\'a}ndez}, R. 2014,
  \href{http://dx.doi.org/10.1093/mnras/stu802}{\JournalTitle{\mnras}, 441,
  3444}

\bibitem[{{Metzger} {et~al.}(2018){Metzger}, {Thompson}, \&
  {Quataert}}]{2018ApJ...856..101M}
{Metzger}, B.~D., {Thompson}, T.~A., \& {Quataert}, E. 2018,
  \href{http://dx.doi.org/10.3847/1538-4357/aab095}{\JournalTitle{\apj}, 856,
  101}

\bibitem[{{Most} {et~al.}(2018){Most}, {Weih}, {Rezzolla}, \&
  {Schaffner-Bielich}}]{Most+18}
{Most}, E.~R., {Weih}, L.~R., {Rezzolla}, L., {et~al.} 2018,
  \href{http://dx.doi.org/10.1103/PhysRevLett.120.261103}{\JournalTitle{Physical
  Review Letters}, 120, 261103}

\bibitem[{{Pankow}(2018)}]{2018ApJ...866...60P}
{Pankow}, C. 2018,
  \href{http://dx.doi.org/10.3847/1538-4357/aadc66}{\JournalTitle{\apj}, 866,
  60}

\bibitem[{{Pankow} {et~al.}(2018){Pankow}, {Chase}, {Coughlin}, {Zevin}, \&
  {Kalogera}}]{pankow2018}
{Pankow}, C., {Chase}, E.~A., {Coughlin}, S., {et~al.} 2018,
  \href{http://dx.doi.org/10.3847/2041-8213/aaacd4}{\JournalTitle{\apj}, 854,
  L25}

\bibitem[{{Piro} \& {Kollmeier}(2018)}]{Piro2018}
{Piro}, A.~L., \& {Kollmeier}, J.~A. 2018,
  \href{http://dx.doi.org/10.3847/1538-4357/aaaab3}{\JournalTitle{\apj}, 855,
  103}

\bibitem[{{Pol} {et~al.}(2019){Pol}, {McLaughlin}, \&
  {Lorimer}}]{2019ApJ...870...71P}
{Pol}, N., {McLaughlin}, M., \& {Lorimer}, D.~R. 2019,
  \href{http://dx.doi.org/10.3847/1538-4357/aaf006}{\JournalTitle{\apj}, 870,
  71}

\bibitem[{{Punturo} {et~al.}(2010){Punturo}, {Abernathy}, {Acernese}, {Allen},
  {Andersson}, {Arun}, {Barone}, {Barr}, {Barsuglia}, {Beker}, {Beveridge},
  {Birindelli}, {Bose}, {Bosi}, {Braccini}, {Bradaschia}, {Bulik}, {Calloni},
  {Cella}, {Chassande Mottin}, {Chelkowski}, {Chincarini}, {Clark}, {Coccia},
  {Colacino}, {Colas}, {Cumming}, {Cunningham}, {Cuoco}, {Danilishin},
  {Danzmann}, {De Luca}, {De Salvo}, {Dent}, {De Rosa}, {Di Fiore}, {Di
  Virgilio}, {Doets}, {Fafone}, {Falferi}, {Flaminio}, {Franc}, {Frasconi},
  {Freise}, {Fulda}, {Gair}, {Gemme}, {Gennai}, {Giazotto}, {Glampedakis},
  {Granata}, {Grote}, {Guidi}, {Hammond}, {Hannam}, {Harms}, {Heinert},
  {Hendry}, {Heng}, {Hennes}, {Hild}, {Hough}, {Husa}, {Huttner}, {Jones},
  {Khalili}, {Kokeyama}, {Kokkotas}, {Krishnan}, {Lorenzini}, {L{\"u}ck},
  {Majorana}, {Mandel}, {Mandic}, {Martin}, {Michel}, {Minenkov}, {Morgado},
  {Mosca}, {Mours}, {M{\"u}ller─Ebhardt}, {Murray}, {Nawrodt}, {Nelson},
  {Oshaughnessy}, {Ott}, {Palomba}, {Paoli}, {Parguez}, {Pasqualetti},
  {Passaquieti}, {Passuello}, {Pinard}, {Poggiani}, {Popolizio}, {Prato},
  {Puppo}, {Rabeling}, {Rapagnani}, {Read}, {Regimbau}, {Rehbein}, {Reid},
  {Rezzolla}, {Ricci}, {Richard}, {Rocchi}, {Rowan}, {R{\"u}diger}, {Sassolas},
  {Sathyaprakash}, {Schnabel}, {Schwarz}, {Seidel}, {Sintes}, {Somiya},
  {Speirits}, {Strain}, {Strigin}, {Sutton}, {Tarabrin}, {Th{\"u}ring}, {van
  den Brand}, {van Leewen}, {van Veggel}, {van den Broeck}, {Vecchio},
  {Veitch}, {Vetrano}, {Vicere}, {Vyatchanin}, {Willke}, {Woan}, {Wolfango}, \&
  {Yamamoto}}]{2010CQGra..27s4002P}
{Punturo}, M., {Abernathy}, M., {Acernese}, F., {et~al.} 2010,
  \href{http://dx.doi.org/10.1088/0264-9381/27/19/194002}{\JournalTitle{Classical
  and Quantum Gravity}, 27, 194002}

\bibitem[{{Radice} {et~al.}(2018){Radice}, {Perego}, {Zappa}, \&
  {Bernuzzi}}]{Radice+18}
{Radice}, D., {Perego}, A., {Zappa}, F., {et~al.} 2018,
  \href{http://dx.doi.org/10.3847/2041-8213/aaa402}{\JournalTitle{\apjl}, 852,
  L29}

\bibitem[{{Raithel} {et~al.}(2018){Raithel}, {{\"O}zel}, \&
  {Psaltis}}]{Raithel+18}
{Raithel}, C.~A., {{\"O}zel}, F., \& {Psaltis}, D. 2018,
  \href{http://dx.doi.org/10.3847/2041-8213/aabcbf}{\JournalTitle{\apjl}, 857,
  L23}

\bibitem[{{Rezzolla} {et~al.}(2018){Rezzolla}, {Most}, \& {Weih}}]{Rezzolla+18}
{Rezzolla}, L., {Most}, E.~R., \& {Weih}, L.~R. 2018,
  \href{http://dx.doi.org/10.3847/2041-8213/aaa401}{\JournalTitle{\apjl}, 852,
  L25}

\bibitem[{{Riess} {et~al.}(2018){Riess}, {Casertano}, {Yuan}, {Macri},
  {Bucciarelli}, {Lattanzi}, {MacKenty}, {Bowers}, {Zheng}, {Filippenko},
  {Huang}, \& {Anderson}}]{riess2018}
{Riess}, A.~G., {Casertano}, S., {Yuan}, W., {et~al.} 2018,
  \href{http://dx.doi.org/10.3847/1538-4357/aac82e}{\JournalTitle{\apj}, 861,
  126}

\bibitem[{{Ruiz} {et~al.}(2018){Ruiz}, {Shapiro}, \& {Tsokaros}}]{Ruiz+18}
{Ruiz}, M., {Shapiro}, S.~L., \& {Tsokaros}, A. 2018,
  \href{http://dx.doi.org/10.1103/PhysRevD.97.021501}{\JournalTitle{\prd}, 97,
  021501}

\bibitem[{{Sathyaprakash} {et~al.}(2012){Sathyaprakash}, {Abernathy},
  {Acernese}, {Ajith}, {Allen}, {Amaro-Seoane}, {Andersson}, {Aoudia}, {Arun},
  {Astone}, {Krishnan}, {Barack}, {Barone}, {Barr}, {Barsuglia}, {Bassan},
  {Bassiri}, {Beker}, {Beveridge}, {Bizouard}, {Bond}, {Bose}, {Bosi},
  {Braccini}, {Bradaschia}, {Britzger}, {Brueckner}, {Bulik}, {Bulten},
  {Burmeister}, {Calloni}, {Campsie}, {Carbone}, {Cella}, {Chalkley},
  {Chassande-Mottin}, {Chelkowski}, {Chincarini}, {Di Cintio}, {Clark},
  {Coccia}, {Colacino}, {Colas}, {Colla}, {Corsi}, {Cumming}, {Cunningham},
  {Cuoco}, {Danilishin}, {Danzmann}, {Daw}, {De Salvo}, {Del Pozzo}, {Dent},
  {De Rosa}, {Di Fiore}, {Emilio}, {Di Virgilio}, {Dietz}, {Doets}, {Dueck},
  {Edwards}, {Fafone}, {Fairhurst}, {Falferi}, {Favata}, {Ferrari}, {Ferrini},
  {Fidecaro}, {Flaminio}, {Franc}, {Frasconi}, {Freise}, {Friedrich}, {Fulda},
  {Gair}, {Galimberti}, {Gemme}, {Genin}, {Gennai}, {Giazotto}, {Glampedakis},
  {Gossan}, {Gouaty}, {Graef}, {Graham}, {Granata}, {Grote}, {Guidi}, {Hallam},
  {Hammond}, {Hannam}, {Harms}, {Haughian}, {Hawke}, {Heinert}, {Hendry},
  {Heng}, {Hennes}, {Hild}, {Hough}, {Huet}, {Husa}, {Huttner}, {Iyer},
  {Jones}, {Jones}, {Kamaretsos}, {Kant Mishra}, {Kawazoe}, {Khalili}, {Kley},
  {Kokeyama}, {Kokkotas}, {Kroker}, {Kumar}, {Kuroda}, {Lagrange}, {Lastzka},
  {Li}, {Lorenzini}, {Losurdo}, {L{\"u}ck}, {Majorana}, {Malvezzi}, {Mandel},
  {Mandic}, {Marka}, {Marin}, {Marion}, {Marque}, {Martin}, {McLeod},
  {Mckechan}, {Mehmet}, {Michel}, {Minenkov}, {Morgado}, {Morgia}, {Mosca},
  {Moscatelli}, {Mours}, {M{\"u}ller-Ebhardt}, {Murray}, {Naticchioni},
  {Nawrodt}, {Nelson}, {O' Shaughnessy}, {Ott}, {Palomba}, {Paoli}, {Parguez},
  {Pasqualetti}, {Passaquieti}, {Passuello}, {Perciballi}, {Piergiovanni},
  {Pinard}, {Pitkin}, {Plastino}, {Plissi}, {Poggiani}, {Popolizio}, {Porter},
  {Prato}, {Prodi}, {Punturo}, {Puppo}, {Rabeling}, {Racz}, {Rapagnani}, {Re},
  {Read}, {Regimbau}, {Rehbein}, {Reid}, {Ricci}, {Richard}, {Robinson},
  {Rocchi}, {Romano}, {Rowan}, {R{\"u}diger}, {Samblowski}, {Santamar{\'\i}a},
  {Sassolas}, {Schilling}, {Schmidt}, {Schnabel}, {Schutz}, {Schwarz}, {Scott},
  {Seidel}, {Sintes}, {Somiya}, {Sopuerta}, {Sorazu}, {Speirits}, {Storchi},
  {Strain}, {Strigin}, {Sutton}, {Tarabrin}, {Taylor}, {Th{\"u}rin},
  {Tokmakov}, {Tonelli}, {Tournefier}, {Vaccarone}, {Vahlbruch}, {van den
  Brand}, {Van Den Broeck}, {van der Putten}, {van Veggel}, {Vecchio},
  {Veitch}, {Vetrano}, {Vicere}, {Vyatchanin}, {We{\ss}els}, {Willke},
  {Winkler}, {Woan}, {Woodcraft}, \& {Yamamoto}}]{2012CQGra..29l4013S}
{Sathyaprakash}, B., {Abernathy}, M., {Acernese}, F., {et~al.} 2012,
  \href{http://dx.doi.org/10.1088/0264-9381/29/12/124013}{\JournalTitle{Classical
  and Quantum Gravity}, 29, 124013}

\bibitem[{{Schutz}(1986)}]{1986Natur.323..310S}
{Schutz}, B.~F. 1986,
  \href{http://dx.doi.org/10.1038/323310a0}{\JournalTitle{\nat}, 323, 310}

\bibitem[{{Scolnic} {et~al.}(2018){Scolnic}, {Kessler}, {Brout},
  {Cowperthwaite}, {Soares-Santos}, {Annis}, {Herner}, {Chen}, {Sako},
  {Doctor}, {Butler}, {Palmese}, {Diehl}, {Frieman}, {Holz}, {Berger},
  {Chornock}, {Villar}, {Nicholl}, {Biswas}, {Hounsell}, {Foley}, {Metzger},
  {Rest}, {Garc{\'\i}a-Bellido}, {M{\"o}ller}, {Nugent}, {Abbott}, {Abdalla},
  {Allam}, {Bechtol}, {Benoit-L{\'e}vy}, {Bertin}, {Brooks}, {Buckley-Geer},
  {Carnero Rosell}, {Carrasco Kind}, {Carretero}, {Castander}, {Cunha},
  {D'Andrea}, {da Costa}, {Davis}, {Doel}, {Drlica-Wagner}, {Eifler},
  {Flaugher}, {Fosalba}, {Gaztanaga}, {Gerdes}, {Gruen}, {Gruendl}, {Gschwend},
  {Gutierrez}, {Hartley}, {Honscheid}, {James}, {Johnson}, {Johnson}, {Krause},
  {Kuehn}, {Kuhlmann}, {Lahav}, {Li}, {Lima}, {Maia}, {March}, {Marshall},
  {Menanteau}, {Miquel}, {Neilsen}, {Plazas}, {Sanchez}, {Scarpine},
  {Schubnell}, {Sevilla-Noarbe}, {Smith}, {Smith}, {Sobreira}, {Suchyta},
  {Swanson}, {Tarle}, {Thomas}, {Tucker}, {Walker}, \& {DES
  Collaboration}}]{Scolnic+18}
{Scolnic}, D., {Kessler}, R., {Brout}, D., {et~al.} 2018,
  \href{http://dx.doi.org/10.3847/2041-8213/aa9d82}{\JournalTitle{\apj}, 852,
  L3}

\bibitem[{{Shibata} {et~al.}(2017){Shibata}, {Fujibayashi}, {Hotokezaka},
  {Kiuchi}, {Kyutoku}, {Sekiguchi}, \& {Tanaka}}]{Shibata+17}
{Shibata}, M., {Fujibayashi}, S., {Hotokezaka}, K., {et~al.} 2017,
  \href{http://dx.doi.org/10.1103/PhysRevD.96.123012}{\JournalTitle{\prd}, 96,
  123012}

\bibitem[{{Siegel} {et~al.}(2018){Siegel}, {Barnes}, \& {Metzger}}]{Siegel+17}
{Siegel}, D.~M., {Barnes}, J., \& {Metzger}, B.~D. 2018, \JournalTitle{arXiv
  e-prints}, \href{http://arxiv.org/abs/1810.00098}{{\sffamily arXiv:1810.00098
  [astro-ph.HE]}}

\bibitem[{{The LIGO Scientific Collaboration} {et~al.}(2018{\natexlab{a}}){The
  LIGO Scientific Collaboration}, {the Virgo Collaboration}, {Abbott},
  {Abbott}, {Abbott}, {Abraham}, {Acernese}, {Ackley}, {Adams}, {Adhikari},
  {Adya}, {Affeldt}, {Agathos}, {Agatsuma}, {Aggarwal}, {Aguiar}, {Aiello},
  {Ain}, {Ajith}, {Allen}, {Allocca}, {Aloy}, {Altin}, {Amato}, {Ananyeva},
  {Anderson}, {Anderson}, {Angelova}, {Antier}, {Appert}, {Arai}, {Araya},
  {Areeda}, {Ar{\`e}ne}, {Arnaud}, {Arun}, {Ascenzi}, {Ashton}, {Aston},
  {Astone}, {Aubin}, {Aufmuth}, {AultONeal}, {Austin}, {Avendano},
  {Avila-Alvarez}, {Babak}, {Bacon}, {Badaracco}, {Bader}, {Bae}, {Baker},
  {Baldaccini}, {Ballardin}, {Ballmer}, {Banagiri}, {Barayoga}, {Barclay},
  {Barish}, {Barker}, {Barkett}, {Barnum}, {Barone}, {Barr}, {Barsotti},
  {Barsuglia}, {Barta}, {Bartlett}, {Bartos}, {Bassiri}, {Basti}, {Bawaj},
  {Bayley}, {Bazzan}, {B{\'e}csy}, {Bejger}, {Belahcene}, {Bell}, {Beniwal},
  {Berger}, {Bergmann}, {Bernuzzi}, {Bero}, {Berry}, {Bersanetti}, {Bertolini},
  {Betzwieser}, {Bhand are}, {Bidler}, {Bilenko}, {Bilgili}, {Billingsley},
  {Birch}, {Birney}, {Birnholtz}, {Biscans}, {Biscoveanu}, {Bisht}, {Bitossi},
  {Bizouard}, {Blackburn}, {Blackman}, {Blair}, {Blair}, {Blair}, {Bloemen},
  {Bode}, {Boer}, {Boetzel}, {Bogaert}, {Bondu}, {Bonilla}, {Bonnand},
  {Booker}, {Boom}, {Booth}, {Bork}, {Boschi}, {Bose}, {Bossie}, {Bossilkov},
  {Bosveld}, {Bouffanais}, {Bozzi}, {Bradaschia}, {Brady}, {Bramley},
  {Branchesi}, {Brau}, {Briant}, {Briggs}, {Brighenti}, {Brillet}, {Brinkmann},
  {Brisson}, {Brockill}, {Brooks}, {Brown}, {Brunett}, {Buikema}, {Bulik},
  {Bulten}, {Buonanno}, {Buskulic}, {Bustamante Rosell}, {Buy}, {Byer},
  {Cabero}, {Cadonati}, {Cagnoli}, {Cahillane}, {Calder{\'o}n Bustillo},
  {Callister}, {Calloni}, {Camp}, {Campbell}, {Canepa}, {Cannon}, {Cao}, {Cao},
  {Capocasa}, {Carbognani}, {Caride}, {Carney}, {Carullo}, {Casanueva Diaz},
  {Casentini}, {Caudill}, {Cavagli{\`a}}, {Cavalier}, {Cavalieri}, {Cella},
  {Cerd{\'a}-Dur{\'a}n}, {Cerretani}, {Cesarini}, {Chaibi}, {Chakravarti},
  {Chamberlin}, {Chan}, {Chao}, {Charlton}, {Chase}, {Chassand e-Mottin},
  {Chatterjee}, {Chaturvedi}, {Chatziioannou}, {Cheeseboro}, {Chen}, {Chen},
  {Chen}, {Cheng}, {Cheong}, {Chia}, {Chincarini}, {Chiummo}, {Cho}, {Cho},
  {Cho}, {Christensen}, {Chu}, {Chua}, {Chung}, {Chung}, {Ciani}, {Ciobanu},
  {Ciolfi}, {Cipriano}, {Cirone}, {Clara}, {Clark}, {Clearwater}, {Cleva},
  {Cocchieri}, {Coccia}, {Cohadon}, {Cohen}, {Colgan}, {Colleoni}, {Collette},
  {Collins}, {Cominsky}, {Constancio}, {Conti}, {Cooper}, {Corban}, {Corbitt},
  {Cordero-Carri{\'o}n}, {Corley}, {Cornish}, {Corsi}, {Cortese}, {Costa},
  {Cotesta}, {Coughlin}, {Coughlin}, {Coulon}, {Countryman}, {Couvares},
  {Covas}, {Cowan}, {Coward}, {Cowart}, {Coyne}, {Coyne}, {Creighton},
  {Creighton}, {Cripe}, {Croquette}, {Crowder}, {Cullen}, {Cumming},
  {Cunningham}, {Cuoco}, {Dal Canton}, {D{\'a}lya}, {Danilishin}, {D'Antonio},
  {Danzmann}, {Dasgupta}, {Da Silva Costa}, {Datrier}, {Dattilo}, {Dave},
  {Davier}, {Davis}, {Daw}, {DeBra}, {Deenadayalan}, {Degallaix}, {De
  Laurentis}, {Del{\'e}glise}, {Del Pozzo}, {DeMarchi}, {Demos}, {Dent}, {De
  Pietri}, {Derby}, {De Rosa}, {De Rossi}, {DeSalvo}, {de Varona},
  {Dhurandhar}, {D{\'\i}az}, {Dietrich}, {Di Fiore}, {Di Giovanni}, {Di
  Girolamo}, {Di Lieto}, {Ding}, {Di Pace}, {Di Palma}, {Di Renzo}, {Dmitriev},
  {Doctor}, {Donovan}, {Dooley}, {Doravari}, {Dorrington}, {Downes}, {Drago},
  {Driggers}, {Du}, {Ducoin}, {Dupej}, {Dwyer}, {Easter}, {Edo}, {Edwards},
  {Effler}, {Ehrens}, {Eichholz}, {Eikenberry}, {Eisenmann}, {Eisenstein},
  {Essick}, {Estelles}, {Estevez}, {Etienne}, {Etzel}, {Evans}, {Evans},
  {Fafone}, {Fair}, {Fairhurst}, {Fan}, {Farinon}, {Farr}, {Farr},
  {Fauchon-Jones}, {Favata}, {Fays}, {Fazio}, {Fee}, {Feicht}, {Fejer}, {Feng},
  {Fernand ez-Galiana}, {Ferrante}, {Ferreira}, {Ferreira}, {Ferrini},
  {Fidecaro}, {Fiori}, {Fiorucci}, {Fishbach}, {Fisher}, {Fishner},
  {Fitz-Axen}, {Flaminio}, {Fletcher}, {Flynn}, {Fong}, {Font}, {Forsyth},
  {Fournier}, {Frasca}, {Frasconi}, {Frei}, {Freise}, {Frey}, {Frey},
  {Fritschel}, {Frolov}, {Fulda}, {Fyffe}, {Gabbard}, {Gadre}, {Gaebel},
  {Gair}, {Gammaitoni}, {Ganija}, {Gaonkar}, {Garcia},
  {Garc{\'\i}a-Quir{\'o}s}, {Garufi}, {Gateley}, {Gaudio}, {Gaur}, {Gayathri},
  {Gemme}, {Genin}, {Gennai}, {George}, {George}, {Gergely}, {Germain},
  {Ghonge}, {Ghosh}, {Ghosh}, {Ghosh}, {Giacomazzo}, {Giaime}, {Giardina},
  {Giazotto}, {Gill}, {Giordano}, {Glover}, {Godwin}, {Goetz}, {Goetz},
  {Goncharov}, {Gonz{\'a}lez}, {Gonzalez Castro}, {Gopakumar}, {Gorodetsky},
  {Gossan}, {Gosselin}, {Gouaty}, {Grado}, {Graef}, {Granata}, {Grant}, {Gras},
  {Grassia}, {Gray}, {Gray}, {Greco}, {Green}, {Green}, {Gretarsson}, {Groot},
  {Grote}, {Grunewald}, {Gruning}, {Guidi}, {Gulati}, {Guo}, {Gupta}, {Gupta},
  {Gustafson}, {Gustafson}, {Haegel}, {Halim}, {Hall}, {Hall}, {Hamilton},
  {Hammond}, {Haney}, {Hanke}, {Hanks}, {Hanna}, {Hannam}, {Hannuksela},
  {Hanson}, {Hardwick}, {Haris}, {Harms}, {Harry}, {Harry}, {Haster},
  {Haughian}, {Hayes}, {Healy}, {Heidmann}, {Heintze}, {Heitmann}, {Hello},
  {Hemming}, {Hendry}, {Heng}, {Hennig}, {Heptonstall}, {Hernandez Vivanco},
  {Heurs}, {Hild}, {Hinderer}, {Hoak}, {Hochheim}, {Hofman}, {Holgado},
  {Holland }, {Holt}, {Holz}, {Hopkins}, {Horst}, {Hough}, {Howell}, {Hoy},
  {Hreibi}, {Huang}, {Huerta}, {Huet}, {Hughey}, {Hulko}, {Husa}, {Huttner},
  {Huynh-Dinh}, {Idzkowski}, {Iess}, {Ingram}, {Inta}, {Intini}, {Irwin},
  {Isa}, {Isac}, {Isi}, {Iyer}, {Izumi}, {Jacqmin}, {Jadhav}, {Jani},
  {Janthalur}, {Jaranowski}, {Jenkins}, {Jiang}, {Johnson}, {Johnson-McDaniel},
  {Jones}, {Jones}, {Jones}, {Jonker}, {Ju}, {Junker}, {Kalaghatgi},
  {Kalogera}, {Kamai}, {Kand hasamy}, {Kang}, {Kanner}, {Kapadia}, {Karki},
  {Karvinen}, {Kashyap}, {Kasprzack}, {Katsanevas}, {Katsavounidis}, {Katzman},
  {Kaufer}, {Kawabe}, {Keerthana}, {K{\'e}f{\'e}lian}, {Keitel}, {Kennedy},
  {Key}, {Khalili}, {Khan}, {Khan}, {Khan}, {Khan}, {Khazanov}, {Khursheed},
  {Kijbunchoo}, {Kim}, {Kim}, {Kim}, {Kim}, {Kim}, {Kim}, {Kimball}, {King},
  {King}, {Kinley-Hanlon}, {Kirchhoff}, {Kissel}, {Kleybolte}, {Klika},
  {Klimenko}, {Knowles}, {Koch}, {Koehlenbeck}, {Koekoek}, {Koley},
  {Kondrashov}, {Kontos}, {Koper}, {Korobko}, {Korth}, {Kowalska}, {Kozak},
  {Kringel}, {Krishnendu}, {Kr{\'o}lak}, {Kuehn}, {Kumar}, {Kumar}, {Kumar},
  {Kumar}, {Kuo}, {Kutynia}, {Kwang}, {Lackey}, {Lai}, {Lam}, {Landry}, {Lane},
  {Lang}, {Lange}, {Lantz}, {Lanza}, {Lartaux-Vollard}, {Lasky}, {Laxen},
  {Lazzarini}, {Lazzaro}, {Leaci}, {Leavey}, {Lecoeuche}, {Lee}, {Lee}, {Lee},
  {Lee}, {Lee}, {Lee}, {Lehmann}, {Lenon}, {Leroy}, {Letendre}, {Levin}, {Li},
  {Li}, {Li}, {Li}, {Lin}, {Linde}, {Linker}, {Littenberg}, {Liu}, {Liu}, {Lo},
  {Lockerbie}, {London}, {Longo}, {Lorenzini}, {Loriette}, {Lormand},
  {Losurdo}, {Lough}, {Lousto}, {Lovelace}, {Lower}, {L{\"u}ck}, {Lumaca},
  {Lundgren}, {Lynch}, {Ma}, {Macas}, {Macfoy}, {MacInnis}, {Macleod},
  {Macquet}, {Maga{\~n}a-Sandoval}, {Maga{\~n}a Zertuche}, {Magee}, {Majorana},
  {Maksimovic}, {Malik}, {Man}, {Mandic}, {Mangano}, {Mansell}, {Manske},
  {Mantovani}, {Marchesoni}, {Marion}, {M{\'a}rka}, {M{\'a}rka}, {Markakis},
  {Markosyan}, {Markowitz}, {Maros}, {Marquina}, {Marsat}, {Martelli},
  {Martin}, {Martin}, {Martynov}, {Mason}, {Massera}, {Masserot}, {Massinger},
  {Masso-Reid}, {Mastrogiovanni}, {Matas}, {Matichard}, {Matone}, {Mavalvala},
  {Mazumder}, {McCann}, {McCarthy}, {McClelland }, {McCormick}, {McCuller},
  {McGuire}, {McIver}, {McManus}, {McRae}, {McWilliams}, {Meacher}, {Meadors},
  {Mehmet}, {Mehta}, {Meidam}, {Melatos}, {Mendell}, {Mercer}, {Mereni},
  {Merilh}, {Merzougui}, {Meshkov}, {Messenger}, {Messick}, {Metzdorff},
  {Meyers}, {Miao}, {Michel}, {Middleton}, {Mikhailov}, {Milano}, {Miller},
  {Miller}, {Millhouse}, {Mills}, {Milovich-Goff}, {Minazzoli}, {Minenkov},
  {Mishkin}, {Mishra}, {Mistry}, {Mitra}, {Mitrofanov}, {Mitselmakher},
  {Mittleman}, {Mo}, {Moffa}, {Mogushi}, {Mohapatra}, {Montani}, {Moore},
  {Moraru}, {Moreno}, {Morisaki}, {Mours}, {Mow-Lowry}, {Mukherjee},
  {Mukherjee}, {Mukherjee}, {Mukund}, {Mullavey}, {Munch}, {Mu{\~n}iz},
  {Muratore}, {Murray}, {Nagar}, {Nardecchia}, {Naticchioni}, {Nayak},
  {Neilson}, {Nelemans}, {Nelson}, {Nery}, {Neunzert}, {Ng}, {Ng}, {Nguyen},
  {Nichols}, {Nielsen}, {Nissanke}, {Nitz}, {Nocera}, {North}, {Nuttall},
  {Obergaulinger}, {Oberling}, {O'Brien}, {O'Dea}, {Ogin}, {Oh}, {Oh}, {Ohme},
  {Ohta}, {Okada}, {Oliver}, {Oppermann}, {Oram}, {O'Reilly}, {Ormiston},
  {Ortega}, {O'Shaughnessy}, {Ossokine}, {Ottaway}, {Overmier}, {Owen}, {Pace},
  {Pagano}, {Page}, {Pai}, {Pai}, {Palamos}, {Palashov}, {Palomba},
  {Pal-Singh}, {Pan}, {Pang}, {Pang}, {Pankow}, {Pannarale}, {Pant},
  {Paoletti}, {Paoli}, {Papa}, {Parida}, {Parker}, {Pascucci}, {Pasqualetti},
  {Passaquieti}, {Passuello}, {Patil}, {Patricelli}, {Pearlstone}, {Pedersen},
  {Pedraza}, {Pedurand}, {Pele}, {Penn}, {Perego}, {Perez}, {Perreca},
  {Pfeiffer}, {Phelps}, {Phukon}, {Piccinni}, {Pichot}, {Piergiovanni},
  {Pillant}, {Pinard}, {Pirello}, {Pitkin}, {Poggiani}, {Pong}, {Ponrathnam},
  {Popolizio}, {Porter}, {Powell}, {Prajapati}, {Prasad}, {Prasai}, {Prasanna},
  {Pratten}, {Prestegard}, {Privitera}, {Prodi}, {Prokhorov}, {Puncken},
  {Punturo}, {Puppo}, {P{\"u}rrer}, {Qi}, {Quetschke}, {Quinonez}, {Quintero},
  {Quitzow-James}, {Raab}, {Radkins}, {Radulescu}, {Raffai}, {Raja}, {Rajan},
  {Rajbhandari}, {Rakhmanov}, {Ramirez}, {Ramos-Buades}, {Rana}, {Rao},
  {Rapagnani}, {Raymond}, {Razzano}, {Read}, {Regimbau}, {Rei}, {Reid},
  {Reitze}, {Ren}, {Ricci}, {Richardson}, {Richardson}, {Ricker},
  {Riemenschneider}, {Riles}, {Rizzo}, {Robertson}, {Robie}, {Robinet},
  {Rocchi}, {Rolland}, {Rollins}, {Roma}, {Romanelli}, {Romano}, {Romel},
  {Romie}, {Rose}, {Rosi{\'n}ska}, {Rosofsky}, {Ross}, {Rowan}, {R{\"u}diger},
  {Ruggi}, {Rutins}, {Ryan}, {Sachdev}, {Sadecki}, {Sakellariadou}, {Salafia},
  {Salconi}, {Saleem}, {Salemi}, {Samajdar}, {Sammut}, {Sanchez}, {Sanchez},
  {Sanchis-Gual}, {Sandberg}, {Sand ers}, {Santiago}, {Sarin}, {Sassolas},
  {Sathyaprakash}, {Saulson}, {Sauter}, {Savage}, {Schale}, {Scheel},
  {Scheuer}, {Schmidt}, {Schnabel}, {Schofield}, {Sch{\"o}nbeck}, {Schreiber},
  {Schulte}, {Schutz}, {Schwalbe}, {Scott}, {Scott}, {Seidel}, {Sellers},
  {Sengupta}, {Sennett}, {Sentenac}, {Sequino}, {Sergeev}, {Setyawati},
  {Shaddock}, {Shaffer}, {Shahriar}, {Shaner}, {Shao}, {Sharma}, {Shawhan},
  {Shen}, {Shink}, {Shoemaker}, {Shoemaker}, {ShyamSundar}, {Siellez},
  {Sieniawska}, {Sigg}, {Silva}, {Singer}, {Singh}, {Singhal}, {Sintes},
  {Sitmukhambetov}, {Skliris}, {Slagmolen}, {Slaven-Blair}, {Smith}, {Smith},
  {Somala}, {Son}, {Sorazu}, {Sorrentino}, {Souradeep}, {Sowell}, {Spencer},
  {Srivastava}, {Srivastava}, {Staats}, {Stachie}, {Standke}, {Steer},
  {Steinke}, {Steinlechner}, {Steinlechner}, {Steinmeyer}, {Stevenson},
  {Stocks}, {Stone}, {Stops}, {Strain}, {Stratta}, {Strigin}, {Strunk},
  {Sturani}, {Stuver}, {Sudhir}, {Summerscales}, {Sun}, {Sunil}, {Suresh},
  {Sutton}, {Swinkels}, {Szczepa{\'n}czyk}, {Tacca}, {Tait}, {Talbot},
  {Talukder}, {Tanner}, {T{\'a}pai}, {Taracchini}, {Tasson}, {Taylor}, {Thies},
  {Thomas}, {Thomas}, {Thondapu}, {Thorne}, {Thrane}, {Tiwari}, {Tiwari},
  {Tiwari}, {Toland}, {Tonelli}, {Tornasi}, {Torres-Forn{\'e}}, {Torrie},
  {T{\"o}yr{\"a}}, {Travasso}, {Traylor}, {Tringali}, {Trovato}, {Trozzo},
  {Trudeau}, {Tsang}, {Tse}, {Tso}, {Tsukada}, {Tsuna}, {Tuyenbayev}, {Ueno},
  {Ugolini}, {Unnikrishnan}, {Urban}, {Usman}, {Vahlbruch}, {Vajente},
  {Valdes}, {van Bakel}, {van Beuzekom}, {van den Brand}, {Van Den Broeck}, \&
  {Vander-Hyde}}]{2018arXiv181112907T}
{The LIGO Scientific Collaboration}, {the Virgo Collaboration}, {Abbott},
  B.~P., {et~al.} 2018{\natexlab{a}}, \JournalTitle{arXiv e-prints},
  arXiv:1811.12907

\bibitem[{{The LIGO Scientific Collaboration} {et~al.}(2018{\natexlab{b}}){The
  LIGO Scientific Collaboration}, {the Virgo Collaboration}, {Abbott},
  {Abbott}, {Abbott}, {Acernese}, {Ackley}, {Adams}, {Adams}, {Addesso},
  {Adhikari}, {Adya}, {Affeldt}, {Agarwal}, {Agathos}, {Agatsuma}, {Aggarwal},
  {Aguiar}, {Aiello}, {Ain}, {Ajith}, {Allen}, {Allen}, {Allocca}, {Aloy},
  {Altin}, {Amato}, {Ananyeva}, {Anderson}, {Anderson}, {Angelova}, {Antier},
  {Appert}, {Arai}, {Araya}, {Areeda}, {Ar{\`e}ne}, {Arnaud}, {Arun},
  {Ascenzi}, {Ashton}, {Ast}, {Aston}, {Astone}, {Atallah}, {Aubin}, {Aufmuth},
  {Aulbert}, {AultONeal}, {Austin}, {Avila-Alvarez}, {Babak}, {Bacon},
  {Badaracco}, {Bader}, {Bae}, {Baker}, {Baldaccini}, {Ballardin}, {Ballmer},
  {Banagiri}, {Barayoga}, {Barclay}, {Barish}, {Barker}, {Barkett}, {Barnum},
  {Barone}, {Barr}, {Barsotti}, {Barsuglia}, {Barta}, {Bartlett}, {Bartos},
  {Bassiri}, {Basti}, {Batch}, {Bawaj}, {Bayley}, {Bazzan}, {B{\'e}csy},
  {Beer}, {Bejger}, {Belahcene}, {Bell}, {Beniwal}, {Bensch}, {Berger},
  {Bergmann}, {Bernuzzi}, {Bero}, {Berry}, {Bersanetti}, {Bertolini},
  {Betzwieser}, {Bhandare}, {Bilenko}, {Bilgili}, {Billingsley}, {Billman},
  {Birch}, {Birney}, {Birnholtz}, {Biscans}, {Biscoveanu}, {Bisht}, {Bitossi},
  {Bizouard}, {Blackburn}, {Blackman}, {Blair}, {Blair}, {Blair}, {Bloemen},
  {Bock}, {Bode}, {Boer}, {Boetzel}, {Bogaert}, {Bohe}, {Bondu}, {Bonilla},
  {Bonnand}, {Booker}, {Boom}, {Booth}, {Bork}, {Boschi}, {Bose}, {Bossie},
  {Bossilkov}, {Bosveld}, {Bouffanais}, {Bozzi}, {Bradaschia}, {Brady},
  {Bramley}, {Branchesi}, {Brau}, {Briant}, {Brighenti}, {Brillet},
  {Brinkmann}, {Brisson}, {Brockill}, {Brooks}, {Brown}, {Brunett}, {Buchanan},
  {Buikema}, {Bulik}, {Bulten}, {Buonanno}, {Buskulic}, {Buy}, {Byer},
  {Cabero}, {Cadonati}, {Cagnoli}, {Cahillane}, {Calder{\'o}n Bustillo},
  {Callister}, {Calloni}, {Camp}, {Canepa}, {Canizares}, {Cannon}, {Cao},
  {Cao}, {Capano}, {Capocasa}, {Carbognani}, {Caride}, {Carney}, {Carullo},
  {Casanueva Diaz}, {Casentini}, {Caudill}, {Cavagli{\`a}}, {Cavalier},
  {Cavalieri}, {Cella}, {Cepeda}, {Cerd{\'a}-Dur{\'a}n}, {Cerretani},
  {Cesarini}, {Chaibi}, {Chamberlin}, {Chan}, {Chao}, {Charlton}, {Chase},
  {Chassande-Mottin}, {Chatterjee}, {Chatziioannou}, {Cheeseboro}, {Chen},
  {Chen}, {Chen}, {Cheng}, {Chia}, {Chincarini}, {Chiummo}, {Chmiel}, {Cho},
  {Cho}, {Chow}, {Christensen}, {Chu}, {Chua}, {Chua}, {Chung}, {Chung},
  {Ciani}, {Ciobanu}, {Ciolfi}, {Cipriano}, {Cirelli}, {Cirone}, {Clara},
  {Clark}, {Clearwater}, {Cleva}, {Cocchieri}, {Coccia}, {Cohadon}, {Cohen},
  {Colla}, {Collette}, {Collins}, {Cominsky}, {Constancio}, {Conti}, {Cooper},
  {Corban}, {Corbitt}, {Cordero-Carri{\'o}n}, {Corley}, {Cornish}, {Corsi},
  {Cortese}, {Costa}, {Cotesta}, {Coughlin}, {Coughlin}, {Coulon},
  {Countryman}, {Couvares}, {Covas}, {Cowan}, {Coward}, {Cowart}, {Coyne},
  {Coyne}, {Creighton}, {Creighton}, {Cripe}, {Crowder}, {Cullen}, {Cumming},
  {Cunningham}, {Cuoco}, {Dal Canton}, {D{\'a}lya}, {Danilishin}, {D'Antonio},
  {Danzmann}, {Dasgupta}, {Da Silva Costa}, {Dattilo}, {Dave}, {Davier},
  {Davis}, {Daw}, {Day}, {DeBra}, {Deenadayalan}, {Degallaix}, {De Laurentis},
  {Del{\'e}glise}, {Del Pozzo}, {Demos}, {Denker}, {Dent}, {De Pietri},
  {Derby}, {Dergachev}, {De Rosa}, {De Rossi}, {DeSalvo}, {de Varona},
  {Dhurandhar}, {D{\'\i}az}, {Dietrich}, {Di Fiore}, {Di Giovanni}, {Di
  Girolamo}, {Di Lieto}, {Ding}, {Di Pace}, {Di Palma}, {Di Renzo}, {Dmitriev},
  {Doctor}, {Dolique}, {Donovan}, {Dooley}, {Doravari}, {Dorrington}, {Dovale
  {\'A}lvarez}, {Downes}, {Drago}, {Dreissigacker}, {Driggers}, {Du}, {Dupej},
  {Dwyer}, {Easter}, {Edo}, {Edwards}, {Effler}, {Eggenstein}, {Ehrens},
  {Eichholz}, {Eikenberry}, {Eisenmann}, {Eisenstein}, {Essick}, {Estelles},
  {Estevez}, {Etienne}, {Etzel}, {Evans}, {Evans}, {Fafone}, {Fair},
  {Fairhurst}, {Fan}, {Farinon}, {Farr}, {Farr}, {Fauchon-Jones}, {Favata},
  {Fays}, {Fee}, {Fehrmann}, {Feicht}, {Fejer}, {Feng}, {Fernandez-Galiana},
  {Ferrante}, {Ferreira}, {Ferrini}, {Fidecaro}, {Fiori}, {Fiorucci},
  {Fishbach}, {Fisher}, {Fishner}, {Fitz-Axen}, {Flaminio}, {Fletcher}, {Fong},
  {Font}, {Forsyth}, {Forsyth}, {Fournier}, {Frasca}, {Frasconi}, {Frei},
  {Freise}, {Frey}, {Frey}, {Fritschel}, {Frolov}, {Fulda}, {Fyffe}, {Gabbard},
  {Gadre}, {Gaebel}, {Gair}, {Gammaitoni}, {Ganija}, {Gaonkar}, {Garcia},
  {Garc{\'\i}a-Quir{\'o}s}, {Garufi}, {Gateley}, {Gaudio}, {Gaur}, {Gayathri},
  {Gemme}, {Genin}, {Gennai}, {George}, {George}, {Gergely}, {Germain},
  {Ghonge}, {Ghosh}, {Ghosh}, {Ghosh}, {Giacomazzo}, {Giaime}, {Giardina},
  {Giazotto}, {Gill}, {Giordano}, {Glover}, {Goetz}, {Goetz}, {Goncharov},
  {Gonz{\'a}lez}, {Gonzalez Castro}, {Gopakumar}, {Gorodetsky}, {Gossan},
  {Gosselin}, {Gouaty}, {Grado}, {Graef}, {Granata}, {Grant}, {Gras}, {Gray},
  {Greco}, {Green}, {Green}, {Gretarsson}, {Groot}, {Grote}, {Grunewald},
  {Gruning}, {Guidi}, {Gulati}, {Guo}, {Gupta}, {Gupta}, {Gushwa}, {Gustafson},
  {Gustafson}, {Halim}, {Hall}, {Hall}, {Hamilton}, {Hamilton}, {Hammond},
  {Haney}, {Hanke}, {Hanks}, {Hanna}, {Hannam}, {Hannuksela}, {Hanson},
  {Hardwick}, {Harms}, {Harry}, {Harry}, {Hart}, {Haster}, {Haughian}, {Healy},
  {Heidmann}, {Heintze}, {Heitmann}, {Hello}, {Hemming}, {Hendry}, {Heng},
  {Hennig}, {Heptonstall}, {Hernandez}, {Heurs}, {Hild}, {Hinderer}, {Hoak},
  {Hochheim}, {Hofman}, {Holland}, {Holt}, {Holz}, {Hopkins}, {Horst}, {Hough},
  {Houston}, {Howell}, {Hreibi}, {Huerta}, {Huet}, {Hughey}, {Hulko}, {Husa},
  {Huttner}, {Huynh-Dinh}, {Iess}, {Indik}, {Ingram}, {Inta}, {Intini}, {Isa},
  {Isac}, {Isi}, {Iyer}, {Izumi}, {Jacqmin}, {Jani}, {Jaranowski}, {Johnson},
  {Johnson}, {Jones}, {Jones}, {Jonker}, {Ju}, {Junker}, {Kalaghatgi},
  {Kalogera}, {Kamai}, {Kand hasamy}, {Kang}, {Kanner}, {Kapadia}, {Karki},
  {Karvinen}, {Kasprzack}, {Katolik}, {Katsanevas}, {Katsavounidis}, {Katzman},
  {Kaufer}, {Kawabe}, {Keerthana}, {K{\'e}f{\'e}lian}, {Keitel}, {Kemball},
  {Kennedy}, {Key}, {Khalili}, {Khamesra}, {Khan}, {Khan}, {Khan}, {Khan},
  {Khazanov}, {Kijbunchoo}, {Kim}, {Kim}, {Kim}, {Kim}, {Kim}, {Kim}, {King},
  {King}, {Kinley-Hanlon}, {Kirchhoff}, {Kissel}, {Kleybolte}, {Klimenko},
  {Knowles}, {Koch}, {Koehlenbeck}, {Koley}, {Kondrashov}, {Kontos}, {Korobko},
  {Korth}, {Kowalska}, {Kozak}, {Kr{\"a}mer}, {Kringel}, {Krishnan},
  {Kr{\'o}lak}, {Kuehn}, {Kumar}, {Kumar}, {Kumar}, {Kuo}, {Kutynia}, {Kwang},
  {Lackey}, {Lai}, {Landry}, {Lang}, {Lange}, {Lantz}, {Lanza},
  {Lartaux-Vollard}, {Lasky}, {Laxen}, {Lazzarini}, {Lazzaro}, {Leaci},
  {Leavey}, {Lee}, {Lee}, {Lee}, {Lee}, {Lee}, {Lehmann}, {Lenon}, {Leonardi},
  {Leroy}, {Letendre}, {Levin}, {Li}, {Li}, {Li}, {Linker}, {Littenberg},
  {Liu}, {Liu}, {Lo}, {Lockerbie}, {London}, {Longo}, {Lorenzini}, {Loriette},
  {Lormand}, {Losurdo}, {Lough}, {Lousto}, {Lovelace}, {L{\"u}ck}, {Lumaca},
  {Lundgren}, {Lynch}, {Ma}, {Macas}, {Macfoy}, {Machenschalk}, {MacInnis},
  {Macleod}, {Maga{\~n}a Hernandez}, {Maga{\~n}a-Sandoval}, {Maga{\~n}a
  Zertuche}, {Magee}, {Majorana}, {Maksimovic}, {Man}, {Mandic}, {Mangano},
  {Mansell}, {Manske}, {Mantovani}, {Marchesoni}, {Marion}, {M{\'a}rka},
  {M{\'a}rka}, {Markakis}, {Markosyan}, {Markowitz}, {Maros}, {Marquina},
  {Marsat}, {Martelli}, {Martellini}, {Martin}, {Martin}, {Martynov}, {Mason},
  {Massera}, {Masserot}, {Massinger}, {Masso-Reid}, {Mastrogiovanni}, {Matas},
  {Matichard}, {Matone}, {Mavalvala}, {Mazumder}, {McCann}, {McCarthy},
  {McClelland }, {McCormick}, {McCuller}, {McGuire}, {McIver}, {McManus},
  {McRae}, {McWilliams}, {Meacher}, {Meadors}, {Mehmet}, {Meidam},
  {Mejuto-Villa}, {Melatos}, {Mendell}, {Mendoza-Gandara}, {Mercer}, {Mereni},
  {Merilh}, {Merzougui}, {Meshkov}, {Messenger}, {Messick}, {Metzdorff},
  {Meyers}, {Miao}, {Michel}, {Middleton}, {Mikhailov}, {Milano}, {Miller},
  {Miller}, {Miller}, {Miller}, {Millhouse}, {Mills}, {Milovich-Goff},
  {Minazzoli}, {Minenkov}, {Ming}, {Mishra}, {Mitra}, {Mitrofanov},
  {Mitselmakher}, {Mittleman}, {Moffa}, {Mogushi}, {Mohan}, {Mohapatra},
  {Montani}, {Moore}, {Moraru}, {Moreno}, {Morisaki}, {Mours}, {Mow-Lowry},
  {Mueller}, {Muir}, {Mukherjee}, {Mukherjee}, {Mukherjee}, {Mukund},
  {Mullavey}, {Munch}, {Mu{\~n}iz}, {Muratore}, {Murray}, {Nagar}, {Napier},
  {Nardecchia}, {Naticchioni}, {Nayak}, {Neilson}, {Nelemans}, {Nelson},
  {Nery}, {Neunzert}, {Nevin}, {Newport}, {Ng}, {Ng}, {Nguyen}, {Nguyen},
  {Nichols}, {Nielsen}, {Nissanke}, {Nitz}, {Nocera}, {Nolting}, {North},
  {Nuttall}, {Obergaulinger}, {Oberling}, {O'Brien}, {O'Dea}, {Ogin}, {Oh},
  {Oh}, {Ohme}, {Ohta}, {Okada}, {Oliver}, {Oppermann}, {Oram}, {O'Reilly},
  {Ormiston}, {Ortega}, {O'Shaughnessy}, {Ossokine}, {Ottaway}, {Overmier},
  {Owen}, {Pace}, {Pagano}, {Page}, {Page}, {Pai}, {Pai}, {Palamos},
  {Palashov}, {Palomba}, {Pal-Singh}, {Pan}, {Pan}, {Pang}, {Pang}, {Pankow},
  {Pannarale}, {Pant}, {Paoletti}, {Paoli}, {Papa}, {Parida}, {Parker},
  {Pascucci}, {Pasqualetti}, {Passaquieti}, {Passuello}, {Patil}, {Patricelli},
  {Pearlstone}, {Pedersen}, {Pedraza}, {Pedurand}, {Pekowsky}, {Pele}, {Penn},
  {Perez}, {Perreca}, {Perri}, {Pfeiffer}, {Phelps}, {Phukon}, {Piccinni},
  {Pichot}, {Piergiovanni}, {Pierro}, {Pillant}, {Pinard}, {Pinto}, {Pirello},
  {Pitkin}, {Poggiani}, {Popolizio}, {Porter}, {Possenti}, {Post}, {Powell},
  {Prasad}, {Pratt}, {Pratten}, {Predoi}, {Prestegard}, {Principe},
  {Privitera}, {Prodi}, {Prokhorov}, {Puncken}, {Punturo}, {Puppo},
  {P{\"u}rrer}, {Qi}, {Quetschke}, {Quintero}, {Quitzow-James}, {Raab},
  {Rabeling}, {Radkins}, {Raffai}, {Raja}, {Rajan}, {Rajbhandari}, {Rakhmanov},
  {Ramirez}, {Ramos-Buades}, {Rana}, {Rapagnani}, {Raymond}, {Razzano}, {Read},
  {Regimbau}, {Rei}, {Reid}, {Reitze}, {Ren}, {Ricci}, {Ricker},
  {Riemenschneider}, {Riles}, {Rizzo}, {Robertson}, {Robie}, {Robinet},
  {Robson}, {Rocchi}, {Rolland}, {Rollins}, {Roma}, {Romano}, {Romel}, {Romie},
  {Rosi{\'n}ska}, {Ross}, {Rowan}, {R{\"u}diger}, {Ruggi}, {Rutins}, {Ryan},
  {Sachdev}, {Sadecki}, {Sakellariadou}, {Salconi}, {Saleem}, {Salemi},
  {Samajdar}, {Sammut}, {Sampson}, {Sanchez}, {Sanchez}, {Sanchis-Gual},
  {Sandberg}, {Sand ers}, {Sarin}, {Sassolas}, {Sathyaprakash}, {Saulson},
  {Sauter}, {Savage}, {Sawadsky}, {Schale}, {Scheel}, {Scheuer}, {Schmidt},
  {Schnabel}, {Schofield}, {Sch{\"o}nbeck}, {Schreiber}, {Schuette}, {Schulte},
  {Schutz}, {Schwalbe}, {Scott}, {Scott}, {Seidel}, {Sellers}, {Sengupta},
  {Sennett}, {Sentenac}, {Sequino}, {Sergeev}, {Setyawati}, {Shaddock},
  {Shaffer}, {Shah}, {Shahriar}, {Shaner}, {Shao}, {Shapiro}, {Shawhan},
  {Shen}, {Shoemaker}, {Shoemaker}, {Siellez}, {Siemens}, {Sieniawska}, {Sigg},
  {Silva}, {Singer}, {Singh}, {Singhal}, {Sintes}, {Slagmolen}, {Slaven-Blair},
  {Smith}, {Smith}, {Smith}, {Somala}, {Son}, {Sorazu}, {Sorrentino},
  {Souradeep}, {Spencer}, {Srivastava}, {Staats}, {Steer}, {Steinke},
  {Steinlechner}, {Steinlechner}, {Steinmeyer}, {Steltner}, {Stevenson},
  {Stocks}, {Stone}, {Stops}, {Strain}, {Stratta}, {Strigin}, {Strunk},
  {Sturani}, {Stuver}, {Summerscales}, {Sun}, {Sunil}, {Suresh}, {Sutton},
  {Swinkels}, {Szczepa{\'n}czyk}, {Tacca}, {Tait}, {Talbot}, {Talukder},
  {Tamanini}, {Tanner}, {T{\'a}pai}, {Taracchini}, {Tasson}, {Taylor},
  {Taylor}, {Tewari}, {Theeg}, {Thies}, {Thomas}, {Thomas}, {Thomas}, {Thorne},
  {Thrane}, {Tiwari}, {Tiwari}, {Tokmakov}, {Toland}, {Tonelli}, \&
  {Tornasi}}]{2018arXiv181100364T}
---. 2018{\natexlab{b}}, \JournalTitle{arXiv e-prints}, arXiv:1811.00364

\bibitem[{{Van Den Broeck}(2014)}]{2014JPhCS.484a2008V}
{Van Den Broeck}, C. 2014,
  \href{http://dx.doi.org/10.1088/1742-6596/484/1/012008}{in Journal of Physics
  Conference Series, Vol. 484, Journal of Physics Conference Series}, 012008

\bibitem[{{Vitale} \& {Chen}(2018)}]{2018PhRvL.121b1303V}
{Vitale}, S., \& {Chen}, H.-Y. 2018,
  \href{http://dx.doi.org/10.1103/PhysRevLett.121.021303}{\JournalTitle{\prl},
  121, 021303}

\bibitem[{{Woosley} {et~al.}(1994){Woosley}, {Wilson}, {Mathews}, {Hoffman}, \&
  {Meyer}}]{Woosley+94}
{Woosley}, S.~E., {Wilson}, J.~R., {Mathews}, G.~J., {et~al.} 1994,
  \href{http://dx.doi.org/10.1086/174638}{\JournalTitle{\apj}, 433, 229}

\end{thebibliography}
}
\end{document}